\documentclass[a4paper, 11pt]{article}
\pdfoutput=1

\usepackage{jheppub}
\usepackage{caption}
\usepackage{subcaption}
\usepackage{graphicx}
\usepackage{xcolor}
\usepackage{amsmath,amssymb}
\usepackage{mathtools}

\title{First saturation correction in high energy proton-nucleus
collisions: Part III. Ensemble averaging}
\author[a]{Ming Li}
\affiliation[a]{Department of Physics, North Carolina State University, Raleigh, NC 27695, USA}
\author[a,b]{and Vladimir V. Skokov}
\affiliation[b]{RIKEN/BNL Research Center, Brookhaven National Laboratory, Upton, NY 11973}

\emailAdd{mli48@ncsu.edu}
\emailAdd{vskokov@ncsu.edu}

\abstract{
In high energy proton-nucleus collisions, the gluon saturation effects from the nucleus are fully incorporated into the light-like  Wilson lines. The gluon saturation effects from the proton, which are anticipated to be important either in the extreme high energy limit or towards the dense-dense (nucleus-nucleus) collision regimes, have been studied perturbatively within the Color Glass Condensate effective theory  in previous papers of this series. A configuration-by-configuration  expression for the single inclusive semi-hard gluon production including the first saturation correction was obtained. In this paper, we perform ensemble averaging in the McLerran-Venugopalan model and the Dipole Approximation. We find that, in the saturation correction, the effects of the initial state interactions are negligible while the final state interactions play most important role and give a positive-valued contribution  to the semi-hard gluon spectrum. Furthermore,  we show that the single gluon spectrum scales approximately $1/k_{\perp}^{4}$ at small $k_{\perp}$, suggesting that a resummation of higher order saturation corrections is required to regulate the infrared region of the gluon spectrum. 
}

\begin{document}

\maketitle
\flushbottom

\section{Introduction}
In previous two papers \cite{Li:2021zmf, Li:2021yiv}, we have studied the gluon saturation effects in the proton in high energy proton-nucleus collisions within the Color Glass Condensate effective theory \cite{McLerran:2001sr,Iancu:2003xm,Kovchegov:2012mbw}.  Specifically, using the next-to-leading order solutions of the classical Yang-Mills equations in the dilute-dense regime, we derived the complete first saturation correction to single inclusive semi-hard gluon production. The final result is a semi-analytic expression which has an explicit dependence on the color charge densities of the projectile (proton) and, through the light-like Wilson lines, on the target (nucleus) color charge densities.  As a functional of the color charge densities of the projectile and the target, this result describes the configuration-by-configuration  single inclusive gluon production. It thus can be used to construct event-by-event double/multiple gluon productions~\cite{Kovner:2010xk}. Experiments at RHIC and LHC, however, record particle productions that are averaged results over many events. To relate theoretical predictions to experimental data, one needs to average over all possible configurations of the color charge densities -- the ensemble averaging. 

The most widely used model to perform ensemble averaging is the McLerran-Venugopalan (MV) model \cite{McLerran:1993ni,McLerran:1993ka}. This model assumes that  the color charge densities, as random variables, are distributed according to  the Gaussian distribution. It is expected to be a good approximation when the color charge densities are parametrically large, $\rho\sim 1/g$ with $g$ being the coupling constant for strong interactions. In the regime when the color charge densities are parametrically small $\rho \sim g$, their distributions are far from Gaussian  and their correlation functions were explicitly computed using a model wavefunction for the proton~\cite{Dumitru:2020fdh,Dumitru:2020gla, Dumitru:2021tvw}. Since we are interested in the gluon saturation effects in proton, the MV model will be used to average over both the proton and the nucleus color charge configurations. 

 While averaging over the proton color charge densities can be carried out straightforwardly owing to the polynomial dependence, averaging over the nucleus is more involved as it requires computing Wilson line correlators in the adjoint representation. A two-Wilson-line correlator, proportional to the dipole operator, has a closed form expression in the MV model. General adjoint representation Wilson line correlators  that contain more than two Wilson lines do not have closed form expressions except for  the large-$N_c$ limit. In this paper, we will use the Dipole Approximation proposed by Kovner and Rezaeian~\cite{Kovner:2017ssr} that approximates Wilson line correlators in terms of products of dipole operators. It is only applied to correlators involving even numbers of independent Wilson lines. It is also argued to be applicable in the black disk limit when the target is extremely dense so that the number of domains with transverse area of order $1/Q_s^2$ is large, $N_{\perp} = S_{\perp}Q_s^2\gg 1$. Here $S_{\perp}$ is the transverse area of the target and $Q_{s}$ denotes the gluon saturation scale.  For four-Wilson-Line correlators, we will present numerical evidence that the Dipole Approximation works quite well and that the correlators factorize into products of the dipole correlators.

The parametric expansion of the single inclusive gluon spectrum in terms of projectile saturation corrections is expected to be 
\begin{equation}
\label{Eq:Expan}
\frac{dN}{d^2\mathbf{k}} = \frac{1}{\alpha_s}\sum_{n = 1}^{\infty} \left[\frac{Q_{s,P}^2}{k_{\perp}^2}\right]^n f_{(n)} \left(\frac{Q_{s,T}^2}{k_{\perp}^2}\right).
\end{equation}
Here $\alpha_s = g^2/4\pi$ is the coupling constant for strong interactions. For the leading order result ($n=1$), the function $f_{(1)}(Q_{s,T}^2/k_{\perp}^2)$ has a closed form expression. It approaches  a constant independent of $k_{\perp}$,  $f_{(1)}  \rightarrow \mathrm{const}$,  for $k\ll Q_{s,T}$. In the opposite limit,  $k_{\perp}\gg Q_{s,T}$, $f_{(1)} (Q_{s,T}^2/k_{\perp}^2) \rightarrow 1/k_{\perp}^2$. As a result, the leading order gluon spectrum scales like $1/k_{\perp}^2$ at small $k_{\perp}$ ($k_{\perp}\ll Q_{s,T}$) and $1/k_{\perp}^4$ at large $k_{\perp}$ ($k_{\perp}\gg Q_{s,T}$). Therefore the total number of gluons $N_{\mathrm{tot}} = \int d^2\mathbf{k} dN/d^2\mathbf{k}$ is logarithmically divergent in the infrared.  It is believed that including all order saturation corrections in the projectile would render the total number of gluons infrared safe~\cite{Krasnitz:2000gz,Krasnitz:2001qu,Lappi:2003bi, Blaizot:2010kh, Jalilian-Marian:2005ccm}.

 The goal of this paper is to analyze the first saturation correction, the second term $n=2$ in the sum~\eqref{Eq:Expan}.  We expect  the parametric dependence of $f_{(2)}(Q_{s,T}^2/k_{\perp}^2)$ on $k_{\perp}$ to be the same as for $f_{(1)}(Q_{s,T}^2/k_{\perp}^2)$.  Thus the contribution to the first saturation correction term to the gluon spectrum scales like $1/k_{\perp}^4$ at small $k_{\perp}$, which leads to worse divergence of $N_{\rm tot}$ compared to that at the leading order. Apparently, at any fixed order, the  saturation corrections to the gluon spectrum would diverge faster than $1/k_{\perp}^2$ at small $k_{\perp}$; this signifies the need for all order resummation  in the fully developed  saturation region $k_{\perp}\ll Q_{s,P}, Q_{s,T}$.  Nevertheless, a fixed order saturation correction can provide valuable information 
in the semi-hard momentum region $Q_{s,P} \ll k_{\perp} \ll Q_{s,T}$.  In this momentum region, the first saturation correction serves as a perturbative correction,  controlled by the small parameter $Q_{s,P}^2/k^2$.  For large momentum $k_{\perp}\gg Q_{s,T}$, the leading order order expansion coefficient of  $f_{(2)}(Q_{s,T}^2/k_{\perp}^2)$ is of the order $Q_{s,T}^2/k_{\perp}^2$; this can be obtained either by direct perturbative calculations or by expanding the target Wilson lines in the dilute/high-momentum regime. Therefore, at large $k_{\perp}$, the first saturation contribution to the spectrum behaves as  $1/k_{\perp}^6$.

Within the first saturation correction, one can separate effects from initial state interactions and effects from final state interactions.  Here the initial state interactions and final state interactions are identified by whether the interactions happen before or after the eikonal scattering of the projectile off the target. We found that the magnitude of the initial state interactions are negligible compared to the leading order result.  Interestingly, one can derive an all order resummed expression purely for the initial state interactions. This all order result does not change the $1/k_{\perp}^2$ scaling for the gluon spectrum at small $k_{\perp}$, suggesting that the {final state interactions must be included} to tame the infrared behavior of the spectrum. We will analyze the various terms from the final state interactions within the first saturation correction, singling out the dominant contribution and demonstrating how the $1/k_{\perp}^4$ scaling at small $k_{\perp}$ and the $1/k_{\perp}^6$ scaling at large $k_{\perp}$ are interpolated numerically. \\

The paper is organized as follows. In Sec.~\ref{sec:review}, a brief review of the main analytic results obtained in previous papers is given. This is followed by performing ensemble averaging over the projectile using the MV model in Sec.~\ref{sec:average_projectile}.  The main challenge is to average over target Wilson line correlators within the MV model. We discuss the Dipole Approximation in Sec.~\ref{sec:average_target} and provide numerical evidences that support factorization relations of adjoint Wilson line correlators in momentum space.  In Sec.~\ref{sec:final_results}, numerical results for  the single inclusive gluon spectrums are presented and discussed, focusing on effects due to final state interactions.

\section{Event-by-event expressions: a brief recap}
\label{sec:review}
We summarize the main results obtained in previous papers \cite{Li:2021zmf, Li:2021yiv}. Unlike in previous paper \cite{Li:2021yiv} in which the projectile color charge density is defined in the  light-cone gauge, the results presented in this paper are in terms of the covariant gauge color charge density. Working with the covariant gauge color charge density is more convenient to make comparison with existing computations in the literature, especially when performing ensemble averaging. The single inclusive gluon production amplitude in high energy proton-nucleus collisions can be formally expanded as 
\begin{equation}
M(\mathbf{k}) = M_{(1)}(\mathbf{k}) + M_{(3)}(\mathbf{k}) + M_{(5)}(\mathbf{k})+\ldots
\end{equation}
The expansion is not simply a perturbative expansion in terms of coupling constant $g$. At each order, it only contains terms enhanced by the color charge density of the proton. To be specific, $M_{(1)} \sim g \rho_P$, $M_{(3)}\sim g^3\rho_P^2$ and $M_{(5)} \sim g^5\rho_P^3$.  The spectrum of single inclusive gluon production then follows 
\begin{equation}
\begin{split}
\frac{dN}{d^2\mathbf{k}} =& |M_{(1)}(\mathbf{k})|^2+ M_{(1)}^{\ast}(\mathbf{k})M_{(3)}(\mathbf{k}) + M_{(1)}(\mathbf{k})M_{(3)}^{\ast}(\mathbf{k}) \\
&+ |M_{(3)}(\mathbf{k})|^2 + M_{(1)}^{\ast}(\mathbf{k})M_{(5)}(\mathbf{k}) + M_{(1)}(\mathbf{k})M_{(5)}^{\ast}(\mathbf{k})+ \ldots
\end{split}
\end{equation}
Among the various terms, one identifies the leading order term
\begin{equation}
\frac{dN}{d^2\mathbf{k}}\Big|_{\mathrm{LO}} = |M_{(1)}(\mathbf{k})|^2
\end{equation}
and the first saturation correction
\begin{equation}\label{eq:fsc_m3m3+m1m5}
\frac{dN}{d^2\mathbf{k}}\Big|_{\mathrm{FSC}} = |M_{(3)}(\mathbf{k})|^2 + M_{(1)}^{\ast}(\mathbf{k})M_{(5)}(\mathbf{k}) + M_{(1)}(\mathbf{k})M_{(5)}^{\ast}(\mathbf{k}).
\end{equation}
The interference terms $M_{(1)}^{\ast}(\mathbf{k})M_{(3)}(\mathbf{k}) + M_{(1)}(\mathbf{k})M_{(3)}^{\ast}(\mathbf{k}) $ are odd power in $\rho_P$ and thus vanish after ensemble averaging using Gaussian distributions. \\

The leading order result has been rederived  using a variety  of different approaches in the last two decades~\cite{Kovchegov:1998bi, Kopeliovich:1998nw, Kovner:2001vi, Dumitru:2001ux, Blaizot:2004wu}. Within the CGC framework, its explicit expression is
\begin{equation}\label{eq:LO_ebye}
\frac{dN}{d^2\mathbf{k}}\Big|_{\mathrm{LO}} = - \frac{g^2}{(2\pi)^2\pi} \int_{\mathbf{p}, \mathbf{q}} L^i(\mathbf{p}, \mathbf{k}) L^i(\mathbf{q}, -\mathbf{k}) \rho_P^a(\mathbf{p})\rho_P^b(\mathbf{q})\Big[U(\mathbf{k}-\mathbf{p})U^{T}(-\mathbf{k}-\mathbf{q})\Big]^{ab}. 
\end{equation}
Here the Lipatov vertex is $
L^i(\mathbf{p},\mathbf{k}) =\left(\frac{\mathbf{p}^i}{p_{\perp}^2}- \frac{\mathbf{k}^i}{k_{\perp}^2}\right)$ and  $\int_{\mathbf{p}} = \int \frac{d^2\mathbf{p}}{(2\pi)^2}$ -- the shorthand notation for transverse momentum integration measure. The projectile color charge density $\rho_P^a(\mathbf{x}) = \int_{-\infty}^{\infty} dx^- \rho_P^a(x^-, \mathbf{x})$ has been integrated over longitudinal coordinate and it is understood as being in the covariant gauge. The target Wilson line in the adjoint representation is defined by 
\begin{equation}\label{eq:def_WL}
U(\mathbf{x}) = \mathcal{P}\mathrm{exp}\left\{ig\int_{-\infty}^{+\infty} dx^- \Phi(x^-, \mathbf{x})\right\}
\end{equation}
with the gauge potential field related to the target color charge density $\Phi(x^-,\mathbf{x}_{\perp}) =\int d^2\mathbf{z} G_0(\mathbf{x} -\mathbf{z})\rho_T(x^-,\mathbf{z})$. The Green function in momentum space is simply $G_0(\mathbf{p}) = 1/p^2_{\perp}$.  \\

For the first saturation correction eq.~\eqref{eq:fsc_m3m3+m1m5}, there are two contributions. The order-$g^3$ gluon production amplitude squared has the following form 
\begin{equation}\label{eq:M3M3_ebye}
\begin{split}
&|M_{(3)}(\mathbf{k})|^2\\
=&-\frac{g^6}{(2\pi)^2\pi}\bigg\{\int_{\mathbf{p},\mathbf{q}} \mathcal{H}_1^{ij}(\mathbf{p},\mathbf{q},\mathbf{k}) \alpha_{P,(3)}^{b,i}(\mathbf{p})\alpha_{P,(3)}^{d,j}(\mathbf{q})  \Big[U(\mathbf{k}-\mathbf{p})U^{T}(-\mathbf{k}-\mathbf{q})\Big]^{bd}\\
&\qquad+\int_{\mathbf{p},\mathbf{p}_1,\mathbf{p}_2, \mathbf{q},\mathbf{q}_1}\mathcal{H}_2(\mathbf{p},\mathbf{p}_1,\mathbf{p}_2, \mathbf{q},\mathbf{q}_1, \mathbf{k})  \rho_{P}^{b_1}(\mathbf{p}_1)\rho_P^{b_2}(\mathbf{p}_2) \rho_P^{d_1}(\mathbf{q}-\mathbf{q}_1)T^d_{d_1d_2}\rho_P^{d_2}(\mathbf{q}_1)\\
&\qquad\qquad  \times  \Big[U(\mathbf{k}-\mathbf{p}-\mathbf{p}_1)T^aU^{T}(\mathbf{p}-\mathbf{p}_2)\Big]^{b_1b_2}U^{da}(-\mathbf{k}-\mathbf{q})+c.c.\\
&\qquad+\int_{\mathbf{p},\mathbf{p}_1,\mathbf{p}_2, \mathbf{q},\mathbf{q}_1,\mathbf{q}_2}\mathcal{H}_3(\mathbf{p},\mathbf{p}_1, \mathbf{p}_2,\mathbf{q},  \mathbf{q}_1, \mathbf{q}_2, \mathbf{k})   \rho_P^{b_1}(\mathbf{p}_1)\rho_P^{b_2}(\mathbf{p}_2)\rho_P^{d_1}(\mathbf{q}_1)  \rho_P^{d_2}(\mathbf{q}_2)\\
&\qquad\qquad \times \Big[U(\mathbf{k}-\mathbf{p}-\mathbf{p}_1)T^aU^{T}(\mathbf{p}-\mathbf{p}_2)\Big]^{b_1b_2} \Big[U(-\mathbf{k}-\mathbf{q}-\mathbf{q}_1) T^aU^T(\mathbf{q}-\mathbf{q}_2)\Big]^{d_1d_2}\bigg\}.\\
\end{split}
\end{equation}
The integrand functions are 
\begin{equation}
\mathcal{H}_1^{ij}(\mathbf{p},\mathbf{q},\mathbf{k}) = \frac{1}{k_{\perp}^2} \left[ (\mathbf{k}-\mathbf{p})^i(\mathbf{k}+\mathbf{q})^j + k_{\perp}^2\delta^{ij} -\mathbf{k}^i\mathbf{k}^j\right],
\end{equation}
\begin{equation}
\mathcal{H}_2(\mathbf{p}, \mathbf{p}_1, \mathbf{p}_2,\mathbf{q}, \mathbf{q}_1, \mathbf{k})=\Gamma^j_{(3),\mathrm{II}}(\mathbf{p},\mathbf{p}_1,\mathbf{p}_2, \mathbf{k})\Gamma_{(3),\mathrm{I}}^j(\mathbf{q},\mathbf{q}_1, -\mathbf{k}),
\end{equation}
\begin{equation}
\mathcal{H}_3(\mathbf{p},\mathbf{p}_1, \mathbf{p}_2,\mathbf{q},  \mathbf{q}_1, \mathbf{q}_2, \mathbf{k}) =-\Gamma^j_{(3),\mathrm{II}}(\mathbf{p},\mathbf{p}_1,\mathbf{p}_2, \mathbf{k})\Gamma^{j,\ast}_{(3),\mathrm{II}}(\mathbf{q},\mathbf{q}_1,\mathbf{q}_2, -\mathbf{k}).
\end{equation}
Here, we  introduced the effective vertices 
\begin{equation}
\Gamma^j_{(3),I}(\mathbf{p}_1,\mathbf{p}, \mathbf{k}) = -\frac{1}{2}\left[\frac{\mathbf{k}\times \mathbf{p}_1}{|\mathbf{p}-\mathbf{p}_1|^2p_1^2}\frac{\epsilon^{ij}\mathbf{k}^i}{k_{\perp}^2}+ \frac{(\mathbf{k}-\mathbf{p})\cdot\mathbf{p}_1}{|\mathbf{p}-\mathbf{p}_1|^2p_1^2}\frac{\mathbf{k}^j}{k_{\perp}^2}\right],
\end{equation}
\begin{equation}
\begin{split}
&\Gamma^j_{(3),\mathrm{II}}(\mathbf{p},\mathbf{p}_1,\mathbf{p}_2, \mathbf{k})\\
 = &-\left[\frac{(\mathbf{p}-\mathbf{p}_2)\cdot\mathbf{p}_2}{p_2^2} \frac{(\mathbf{k}-\mathbf{p})\cdot\mathbf{p}_1}{|\mathbf{k}-\mathbf{p}|^2 p_1^2} +\frac{\mathbf{p} \times \mathbf{p}_2}{p_2^2}\frac{(\mathbf{k}-\mathbf{p}-\mathbf{p}_1)\cdot\mathbf{p}_1}{|\mathbf{k}-\mathbf{p}|^2p_1^2} \frac{(\mathbf{k}\times \mathbf{p})}{p_{\perp}^2 }\left(i \frac{\mathbf{k}\cdot(\mathbf{k}-\mathbf{p})}{|\mathbf{k}\times \mathbf{p}|} -1\right) \right]\frac{\mathbf{k}^j}{k_{\perp}^2}\\
&-\Bigg[\frac{(\mathbf{k}\cdot\mathbf{p})(\mathbf{p} \times \mathbf{p}_2)}{p_{\perp}^2p_2^2}  \frac{(\mathbf{k}-\mathbf{p})\cdot\mathbf{p}_1}{|\mathbf{k}-\mathbf{p}|^2p_1^2} +\frac{1}{2}\frac{(\mathbf{k}\times \mathbf{p})(\mathbf{p}\cdot\mathbf{p}_2)}{p_{\perp}^2p_2^2}  \frac{(\mathbf{k}-\mathbf{p})\cdot\mathbf{p}_1}{|\mathbf{k}-\mathbf{p}|^2p_1^2} \\
&\qquad+\frac{1}{2}\frac{(\mathbf{p}-\mathbf{p}_2)\cdot\mathbf{p}_2}{p_2^2}\frac{(\mathbf{k}-\mathbf{p}-\mathbf{p}_1)\cdot\mathbf{p}_1}{|\mathbf{k}-\mathbf{p}|^2p_1^2} \frac{(\mathbf{k}\times\mathbf{p}) }{p_{\perp}^2}\left(i\frac{\mathbf{p}\cdot(\mathbf{p}-\mathbf{k})}{|\mathbf{p}\times \mathbf{k}|} +1\right) \\
&\qquad+\frac{1}{2}\frac{\mathbf{p}\times \mathbf{p}_2}{p_2^2}\frac{(\mathbf{k}-\mathbf{p})\times \mathbf{p}_1}{|\mathbf{k}-\mathbf{p}|^2p_1^2} \frac{(k_{\perp}^2+p_{\perp}^2-\mathbf{k}\cdot\mathbf{p})}{p_{\perp}^2}  i\frac{\mathbf{k}\times \mathbf{p}}{|\mathbf{k}\times \mathbf{p}|} \Bigg] \frac{\epsilon^{ij}\mathbf{k}^i}{k_{\perp}^2}.\\
\end{split}
\end{equation}
In eq.~\eqref{eq:M3M3_ebye}, the first line is expressed in terms of perturbative solutions of the projectile Weizs\"acker-Williams (WW) field $\alpha_P^{a,i}(\mathbf{x})$ while the other terms are expressed in the projectile color charge density directly. When performing ensemble averaging, the path ordering along the longitudinal direction matters and one needs to keep track of the orderings inside the WW field. That is why we used the WW field instead of the color charge density.  In details, this will be discussed in the following sections. \\

In eq.~\eqref{eq:fsc_m3m3+m1m5}, the interference terms from the order-$g$ and order-$g^5$ gluon production amplitudes are
\begin{equation}\label{eq:M1M5_ebye}
\begin{split}
&M_{(1)}^{\ast} (\mathbf{k})M_{(5)}(\mathbf{k}) + M_{(1)}(\mathbf{k})M_{(5)} ^{\ast} (\mathbf{k})\\
=&-\frac{g^6}{(2\pi)^2\pi}\Big\{\int_{\mathbf{p},\mathbf{q}}\mathcal{F}_1^{ij}(\mathbf{q}, \mathbf{p},  \mathbf{k}) \alpha_{P,(5)}^{b,i} (\mathbf{p})\alpha_{P,(1)}^{d,j}(\mathbf{q})  \Big[U(\mathbf{k}-\mathbf{p})U^T(-\mathbf{k}-\mathbf{q})\Big]^{bd}\\
&\qquad+\int_{\mathbf{p},\mathbf{q},\mathbf{p}_1,\mathbf{p}_3,\mathbf{p}_4}\mathcal{F}_2(\mathbf{p}, \mathbf{q}, \mathbf{p}_1, \mathbf{p}_3, \mathbf{p}_4, \mathbf{k}) \rho_P^{b_1}(\mathbf{p}-\mathbf{p_1})T^b_{b_1b_2}\rho_P^{b_2}(\mathbf{p}_1)\rho_P^{b_3}(\mathbf{p}_3)\rho_P^{b_4}(\mathbf{p}_4)\\
&\qquad\qquad \times U^{ba}(\mathbf{q}-\mathbf{p}) \Big[U(\mathbf{k}-\mathbf{q}-\mathbf{p}_3)T^aU^{T}(-\mathbf{k}-\mathbf{p}_4)\Big]^{b_3b_4}\\
&\qquad+\int_{\mathbf{q},\mathbf{p},\mathbf{p}_1,\mathbf{p}_2,\mathbf{p}_3,\mathbf{p}_4}\mathcal{F}_3(\mathbf{p}, \mathbf{q}, \mathbf{p}_1, \mathbf{p}_2, \mathbf{p}_3, \mathbf{p}_4, \mathbf{k})\rho^{b_1}_P(\mathbf{p}_1)\rho^{b_2}_P(\mathbf{p}_2)\rho^{b_3}_P(\mathbf{p}_3)\rho^{b_4}_P(\mathbf{p}_4)\\
&\qquad\qquad\times\Big[ U(\mathbf{p}-\mathbf{p}_1)T^aU^{T}(\mathbf{q}-\mathbf{p}-\mathbf{p}_2)\Big]^{b_1b_2} \Big[U(\mathbf{k}-\mathbf{q}-\mathbf{p}_3) T^aU^{T}(-\mathbf{k}-\mathbf{p}_4)\Big]^{b_3b_4}\Big\}\\
&\qquad+c.c.
\end{split}
\end{equation}
Here $c.c.$ represents the corresponding complex conjugate terms. 
The integrand functions are 
\begin{equation}
\mathcal{F}_1^{ij}(\mathbf{p}, \mathbf{q},  \mathbf{k}) =\mathcal{H}_1^{ij}(\mathbf{p},\mathbf{q},\mathbf{k}) = \frac{1}{k_{\perp}^2} \left[ (\mathbf{k}-\mathbf{p})^i(\mathbf{k}+\mathbf{q})^j + k_{\perp}^2\delta^{ij} -\mathbf{k}^i\mathbf{k}^j\right],
\end{equation}
\begin{equation}
\mathcal{F}_2(\mathbf{p},\mathbf{q},\mathbf{p}_1, \mathbf{p}_3, \mathbf{p}_4, \mathbf{k}) = \Upsilon_2^j(\mathbf{p}, \mathbf{q}, \mathbf{p}_1, \mathbf{p}_3, \mathbf{k}) L^j(\mathbf{p}_4 , -\mathbf{k}),
\end{equation}
\begin{equation}
\mathcal{F}_3(\mathbf{p}, \mathbf{q}, \mathbf{p}_1, \mathbf{p}_2, \mathbf{p}_3, \mathbf{p}_4, \mathbf{k}) = \Upsilon_3^{j }(\mathbf{p}, \mathbf{q}, \mathbf{p}_1, \mathbf{p}_2, \mathbf{p}_3,  \mathbf{k})  L^j(\mathbf{p}_4, -\mathbf{k}).
\end{equation}
We introduced the  effective vertex, 
\begin{equation}
\Upsilon_2^j(\mathbf{p}, \mathbf{q}, \mathbf{p}_1, \mathbf{p}_3, \mathbf{k}) = \Upsilon^{\parallel }_2(\mathbf{p}, \mathbf{q}, \mathbf{p}_1, \mathbf{p}_3, \mathbf{k})\frac{\mathbf{k}^j}{k_{\perp}^2} + \Upsilon_2^{\perp}(\mathbf{p}, \mathbf{q}, \mathbf{p}_1, \mathbf{p}_3, \mathbf{k}) \frac{\epsilon^{ij} \mathbf{k}^i}{k_{\perp}^2}
\end{equation}
with the  components 
\begin{equation}
\begin{split}
&\Upsilon^{\parallel }_2(\mathbf{p}, \mathbf{q}, \mathbf{p}_1, \mathbf{p}_3, \mathbf{k})\\
=& -\frac{(\mathbf{q}-\mathbf{p})\cdot\mathbf{p}_1}{2|\mathbf{p}-\mathbf{p}_1|^2p_1^2 }\frac{(\mathbf{k}-\mathbf{q})\cdot \mathbf{p}_3}{|\mathbf{k}-\mathbf{q}|^2p_3^2} +\frac{\mathbf{q}\cdot\mathbf{p}_1}{2q_{\perp}^2|\mathbf{p}-\mathbf{p}_1|^2 p_1^2}\frac{(\mathbf{k}-\mathbf{q}-\mathbf{p}_3) \cdot\mathbf{p}_3}{p_3^2} \\
&-i\frac{(\mathbf{q}-\mathbf{p})\cdot\mathbf{p}_1}{2|\mathbf{p}-\mathbf{p}_1|^2 p_1^2} \frac{(\mathbf{k}-\mathbf{q})\times \mathbf{p}_3}{|\mathbf{k}-\mathbf{q}|^2p_3^2} \frac{\mathbf{k}\times \mathbf{q}}{q_{\perp}^2}\left(\frac{\mathbf{k}\cdot\mathbf{q}}{|\mathbf{k}\times \mathbf{q}| }+i\right)\\
&+i\frac{(\mathbf{p}_1\times \mathbf{q})}{2 p_1^2|\mathbf{p}-\mathbf{p}_1|^2}\frac{(\mathbf{k}-\mathbf{q}-\mathbf{p}_3)\cdot\mathbf{p}_3}{|\mathbf{k}-\mathbf{q}|^2p_3^2} \frac{\mathbf{k}\times \mathbf{q}}{q_{\perp}^2} \left(\frac{\mathbf{k}\cdot(\mathbf{k}-\mathbf{q})}{|\mathbf{k}\times \mathbf{q}|} + i \right)
\end{split}
\end{equation}
and 
\begin{equation}
\begin{split}
&\Upsilon_2^{\perp}(\mathbf{p}, \mathbf{q}, \mathbf{p}_1, \mathbf{p}_3, \mathbf{k})\\
=
&+\frac{(\mathbf{p}_1\times \mathbf{k})}{2|\mathbf{p}-\mathbf{p}_1|^2p_1^2}\frac{(\mathbf{k}-\mathbf{q})\cdot \mathbf{p}_3}{|\mathbf{k}-\mathbf{q}|^2p_3^2} +\frac{\mathbf{k}\cdot(\mathbf{k}-\mathbf{q})}{ |\mathbf{k}-\mathbf{q}|^2}\frac{\mathbf{q}\cdot\mathbf{p}_1}{2q_{\perp}^2 |\mathbf{p}-\mathbf{p}_1|^2 p_1^2}\frac{(\mathbf{k}-\mathbf{q})\times \mathbf{p}_3}{p_3^2} \\
&-i\frac{(\mathbf{q}-\mathbf{p})\cdot\mathbf{p}_1}{2|\mathbf{p}-\mathbf{p}_1|^2 p_1^2}\frac{(\mathbf{k}-\mathbf{q}-\mathbf{p}_3)\cdot\mathbf{p}_3}{ |\mathbf{k}-\mathbf{q}|^2p_3^2} \frac{\mathbf{k}\times \mathbf{q}}{q_{\perp}^2}\left(\frac{\mathbf{q}\cdot(\mathbf{q}-\mathbf{k})}{|\mathbf{q}\times \mathbf{k}|} -i\right)\\
&+i\frac{(\mathbf{p}_1\times \mathbf{q})}{2p_1^2|\mathbf{p}-\mathbf{p}_1|^2}\frac{(\mathbf{k}-\mathbf{q})\times\mathbf{p}_3}{|\mathbf{k}-\mathbf{q}|^2p_3^2} \frac{(\mathbf{k}\times \mathbf{q})}{q^2_{\perp} }\frac{(k_{\perp}^2+q_{\perp}^2-\mathbf{k}\cdot\mathbf{q})}{|\mathbf{k}\times \mathbf{q}|}.\\
\end{split}
\end{equation}
Another effective vertex is
\begin{equation}
\Upsilon_3^{j }(\mathbf{p}, \mathbf{q}, \mathbf{p}_1, \mathbf{p}_2, \mathbf{p}_3,  \mathbf{k}) = \Upsilon_3^{\| }(\mathbf{p}, \mathbf{q}, \mathbf{p}_1, \mathbf{p}_2, \mathbf{p}_3,  \mathbf{k}) \frac{\mathbf{k}^j}{k_{\perp}^2} + \Upsilon_3^{\perp }(\mathbf{p}, \mathbf{q}, \mathbf{p}_1, \mathbf{p}_2, \mathbf{p}_3,  \mathbf{k}) \frac{\epsilon^{ij}\mathbf{k}^i}{k_{\perp}^2} 
\end{equation}
Here the parallel component is 
\begin{equation}\label{eq:upsilon3parallel}
\begin{split}
&\Upsilon_3^{\| }(\mathbf{p}, \mathbf{q}, \mathbf{p}_1, \mathbf{p}_2, \mathbf{p}_3,  \mathbf{k}) \\
=&+\frac{\mathbf{p}\times\mathbf{p}_1}{p_{\perp}^2p_1^2} \frac{(\mathbf{q}-\mathbf{p}-\mathbf{p}_2)\cdot\mathbf{p}_2}{|\mathbf{q}-\mathbf{p}|^2p_2^2} \frac{(\mathbf{k}-\mathbf{q})\times \mathbf{p}_3}{|\mathbf{k}-\mathbf{q}|^2p_3^2} i\mathcal{I}_1 (\mathbf{p}, \mathbf{q}, \mathbf{k}) \\
&+ \frac{(\mathbf{p}-\mathbf{p}_1)\cdot\mathbf{p}_1}{p^2_{\perp}p_1^2} \frac{(\mathbf{q}-\mathbf{p}-\mathbf{p}_2)\cdot\mathbf{p}_2}{|\mathbf{q}-\mathbf{p}|^2p_2^2}\frac{(\mathbf{k}-\mathbf{q}-\mathbf{p}_3)\cdot\mathbf{p}_3}{|\mathbf{k}-\mathbf{q}|^2p_3^2}i\mathcal{I}_2(\mathbf{p}, \mathbf{q}, \mathbf{k})\\
& + \frac{\mathbf{p}\times\mathbf{p}_1}{p^2_{\perp}p_1^2} \frac{(\mathbf{q}-\mathbf{p})\times\mathbf{p}_2}{|\mathbf{q}-\mathbf{p}|^2p_2^2}\frac{(\mathbf{k}-\mathbf{q}-\mathbf{p}_3)\cdot\mathbf{p}_3}{|\mathbf{k}-\mathbf{q}|^2p_3^2}i\mathcal{I}_3(\mathbf{p},\mathbf{q}, \mathbf{k})\\
&+\left[-\frac{\mathbf{p}\times \mathbf{p}_1}{p_1^2}\frac{\mathbf{q}\times\mathbf{p}}{p_{\perp}^2}  +\frac{1}{2}\frac{\mathbf{p}\cdot\mathbf{p}_1}{p_1^2} \frac{\mathbf{q}\cdot\mathbf{p}}{p_{\perp}^2}\right]\frac{(\mathbf{q}-\mathbf{p})\cdot\mathbf{p}_2}{|\mathbf{q}-\mathbf{p}|^2p_2^2} \frac{(\mathbf{k}-\mathbf{q}-\mathbf{p}_3)\cdot\mathbf{p}_3}{q_{\perp}^2p_3^2} \\
&-\frac{(\mathbf{p}-\mathbf{p}_1)\cdot\mathbf{p}_1}{p_1^2} \frac{(\mathbf{q}-\mathbf{p})\cdot\mathbf{p}_2}{ |\mathbf{q}-\mathbf{p}|^2p_2^2} \left[\frac{1}{2}\frac{(\mathbf{k}-\mathbf{q})\cdot \mathbf{p}_3}{|\mathbf{k}-\mathbf{q}|^2p_3^2} +\frac{(\mathbf{k}-\mathbf{q})\times \mathbf{p}_3}{|\mathbf{k}-\mathbf{q}|^2p_3^2}  \frac{\mathbf{k}\times \mathbf{q}}{q_{\perp}^2}\left(i \frac{(\mathbf{k}\cdot \mathbf{q})}{|\mathbf{k}\times \mathbf{q}|}-1\right)\right]\\
&+\left[- \frac{(\mathbf{p}\times \mathbf{p}_1)}{p_1^2} \frac{\mathbf{q}\cdot\mathbf{p}}{p_{\perp}^2} +\frac{1}{2 }\frac{\mathbf{p}\cdot\mathbf{p}_1}{p_1^2} \frac{ (\mathbf{p}\times \mathbf{q})}{p_{\perp}^2} \right]\frac{(\mathbf{q}-\mathbf{p})\cdot\mathbf{p}_2}{|\mathbf{q}-\mathbf{p}|^2p_2^2} \frac{(\mathbf{k}-\mathbf{q}-\mathbf{p}_3)\cdot\mathbf{p}_3}{|\mathbf{k}-\mathbf{q}|^2p_3^2} \frac{\mathbf{k}\times \mathbf{q}}{q_{\perp}^2} \left(i\frac{\mathbf{k}\cdot(\mathbf{k}-\mathbf{q})}{|\mathbf{k}\times \mathbf{q}|} -1  \right)\\
\end{split}
\end{equation}
while the perpendicular part is 
\begin{equation}\label{eq:upsilon3perpendicular}
\begin{split}
&\Upsilon_3^{\perp }(\mathbf{p}, \mathbf{q}, \mathbf{p}_1, \mathbf{p}_2, \mathbf{p}_3,  \mathbf{k})\\
=&\frac{\mathbf{p}\times \mathbf{p}_1}{p_{\perp}^2p_1^2} \frac{(\mathbf{q}-\mathbf{p}-\mathbf{p}_2)\cdot\mathbf{p}_2}{|\mathbf{q}-\mathbf{p}|^2p_2^2} \frac{(\mathbf{k}-\mathbf{q}-\mathbf{p}_3)\cdot\mathbf{p}_3}{|\mathbf{k}-\mathbf{q}|^2p_3^2} i \mathcal{G}_1(\mathbf{p}, \mathbf{q}, \mathbf{k})\\
&+\frac{(\mathbf{p}-\mathbf{p}_1)\cdot\mathbf{p}_1}{p_{\perp}^2p_1^2} \frac{(\mathbf{q}-\mathbf{p}-\mathbf{p}_2)\cdot\mathbf{p}_2}{|\mathbf{q}-\mathbf{p}|^2p_2^2} \frac{(\mathbf{k}-\mathbf{q})\times \mathbf{p}_3}{|\mathbf{k}-\mathbf{q}|^2p_3^2}i\mathcal{G}_2(\mathbf{p}, \mathbf{q}, \mathbf{k})\\
&+\frac{\mathbf{p}\times \mathbf{p}_1}{p_{\perp}^2p_1^2} \frac{(\mathbf{q}-\mathbf{p})\times \mathbf{p}_2}{|\mathbf{q}-\mathbf{p}|^2p_2^2} \frac{(\mathbf{k}-\mathbf{q})\times \mathbf{p}_3}{|\mathbf{k}-\mathbf{q}|^2p_3^2}  i\mathcal{G}_3(\mathbf{p}, \mathbf{q}, \mathbf{k})\\
&+\left[-\frac{\mathbf{p}\times \mathbf{p}_1}{p_1^2}\frac{\mathbf{q}\times \mathbf{p}}{p_{\perp}^2} +\frac{1}{2}\frac{\mathbf{p}\cdot\mathbf{p}_1}{p_1^2} \frac{\mathbf{q}\cdot\mathbf{p}}{p_{\perp}^2}\right]\frac{(\mathbf{q}-\mathbf{p})\cdot\mathbf{p}_2}{|\mathbf{q}-\mathbf{p}|^2p_2^2}  \frac{(\mathbf{k}-\mathbf{q})\times \mathbf{p}_3}{ |\mathbf{k}-\mathbf{q}|^2p_3^2}\frac{\mathbf{k}\cdot(\mathbf{k}-\mathbf{q})}{q_{\perp}^2}\\
&+\left[-\frac{1}{2 }\frac{\mathbf{p}\times\mathbf{p}_1}{p_1^2}\frac{\mathbf{k}\cdot\mathbf{p}}{p_{\perp}^2} +\frac{1}{3 } \frac{\mathbf{p}\cdot\mathbf{p}_1}{p_1^2} \frac{\mathbf{p}\times \mathbf{k}}{p_{\perp}^2}\right]\frac{(\mathbf{q}-\mathbf{p})\cdot\mathbf{p}_2}{|\mathbf{q}-\mathbf{p}|^2 p_2^2}\frac{(\mathbf{k}-\mathbf{q})\cdot\mathbf{p}_3}{|\mathbf{k}-\mathbf{q}|^2p_3^2}\\
& +i\left[-\frac{\mathbf{p}\times \mathbf{p}_1}{p_1^2} \frac{\mathbf{q}\cdot\mathbf{p}}{p_{\perp}^2} + \frac{1}{2 }\frac{\mathbf{p}\cdot\mathbf{p}_1}{p_1^2}\frac{(\mathbf{p}\times \mathbf{q})}{p_{\perp}^2}\right]\frac{(\mathbf{q}-\mathbf{p})\cdot\mathbf{p}_2}{|\mathbf{q}-\mathbf{p}|^2\mathbf{p}_2^2} \frac{(\mathbf{k}-\mathbf{q})\times \mathbf{p}_3}{|\mathbf{k}-\mathbf{q}|^2p_3^2}  \frac{(\mathbf{k}\times \mathbf{q})}{q^2_{\perp} }\frac{(k_{\perp}^2+q_{\perp}^2-\mathbf{k}\cdot\mathbf{q})}{|\mathbf{k}\times \mathbf{q}|} \\
& -\frac{(\mathbf{p}-\mathbf{p}_1)\cdot\mathbf{p}_1}{p_1^2} \frac{(\mathbf{q}-\mathbf{p})\cdot\mathbf{p}_2}{|\mathbf{q}-\mathbf{p}|^2p_2^2} \frac{(\mathbf{k}-\mathbf{q}-\mathbf{p}_3)\cdot\mathbf{p}_3}{|\mathbf{k}-\mathbf{q}|^2p_3^2}  \frac{\mathbf{k}\times \mathbf{q}}{q_{\perp}^2 }\left(i\frac{\mathbf{q}\cdot(\mathbf{q}-\mathbf{k})}{|\mathbf{q}\times \mathbf{k}|} +1\right).\\
\end{split}
\end{equation}
The explicit expressions for the six auxiliary functions $\mathcal{I}_1,\mathcal{I}_2,\mathcal{I}_3, \mathcal{G}_1,\mathcal{G}_2,\mathcal{G}_3$ are given in \cite{Li:2021yiv}. Again, in eq.~\eqref{eq:M1M5_ebye}, the pure initial state interaction term is expressed in terms of order-$g$ and order-$g^5$ WW field while other terms are expressed in terms of the color charge density directly.

\section{Ensemble averaging over the projectile}
\label{sec:average_projectile}
In the MV model, the color charges  are distributed according to  Gaussian distributions independently for different spatial positions and  colors. The two point correlation function is
\begin{equation}\label{eq:MV_gaussian_correlator}
\langle \rho^a(x^-, \mathbf{x}) \rho^b(y^-, \mathbf{y})\rangle = \delta^{ab}\mu^2(x^-, \mathbf{x}) \delta(x^--y^-)\delta^{(2)}(\mathbf{x}-\mathbf{y}).
\end{equation}
Here $\mu^2(x^-, \mathbf{x})$ represents the color charge density squared and it is ultimately related to the gluon saturation scale $Q_{s}^2$. The spatial profile of $\mu^2(x^-, \mathbf{x})$ can be modeled to describe experimental data~\cite{Kowalski:2003hm,Mantysaari:2016ykx}. For the purpose of this paper, it is sufficient to consider homogeneous distribution/translational invariance on the transverse plane and, additionally, integrate out the longitudinal coordinate. 
For  $x^{-}$-integrated color charge density,  the momentum space two-point correlator for the projectile simplifies to
\begin{equation}
\left\langle \rho_P^a(\mathbf{p})\rho_P^b(\mathbf{q})\right\rangle =\bar{\mu}_P^2 \delta^{ab}(2\pi)^2\delta^{(2)}(\mathbf{p}+\mathbf{q}).
\end{equation}
Here $\bar{\mu}_P^2 = \int dx^- \mu_P^2(x^-)$. High order correlation functions can be readily computed using Wick contractions. Terms in the first saturation correction to single gluon productions due to final state interactions all contain four-point correlators and they are computed by 
\begin{equation}
\begin{split}
&\left\langle \rho_P^{b_1}(\mathbf{p}_1)\rho_P^{b_2}(\mathbf{p}_2)\rho_P^{d_1}(\mathbf{q}_1)  \rho_P^{d_2}(\mathbf{q}_2)\right\rangle\\
=&(2\pi)^4\bar{\mu}_P^4 \delta^{b_1b_2}\delta^{d_1d_2}\delta^{(2)}(\mathbf{p}_1+\mathbf{p}_2)\delta^{(2)}(\mathbf{q}_1+\mathbf{q}_2)  + (2\pi)^4\bar{\mu}_P^4\delta^{b_1d_1}\delta^{b_2d_2}\delta^{(2)}(\mathbf{p}_1+\mathbf{q}_1)\delta^{(2)}(\mathbf{p}_2+\mathbf{q}_2) \\
&\quad + (2\pi)^4\bar{\mu}_P^4\delta^{b_1d_2}\delta^{b_2d_1}\delta^{(2)}(\mathbf{p}_1+\mathbf{q}_2)\delta^{(2)}(\mathbf{p}_2+\mathbf{q}_1).\\
\end{split}
\end{equation}
We also need to average over the perturbative solutions of the WW field correlators 
\begin{equation}
\left\langle \alpha_{P,(3)}^{a,i}(\mathbf{p}) \alpha_{P,(3)}^{b,j}(\mathbf{q})\right\rangle, \quad \left\langle \alpha_{P,(1)}^{a,i}(\mathbf{p}) \alpha_{P,(5)}^{b,j}(\mathbf{q})\right\rangle. 
\end{equation}
These correlators can be computed straightforwardly using the explicit expressions of the WW field, see Appendix.~\ref{appendixB}. Alternatively, they can be obtained by expanding the full WW correlator $\langle \alpha_{P}^{a,i}(\mathbf{x}) \alpha_{P}^{b,j}(\mathbf{y})\rangle$
whose closed form expression is well-known, see Appendix.~\ref{appendixA}. \\

We carry out ensemble averaging over the projectile color charge densities in the  leading order result eq.~\eqref{eq:LO_ebye} and in the first saturation corrections eq.~\eqref{eq:M3M3_ebye} and eq.~\eqref{eq:M1M5_ebye}.  The leading order result reduces to 
\begin{equation}
\Big\langle \frac{dN}{d^2\mathbf{k}} \Big|_{\mathrm{LO}}(\rho_P, \rho_T) \Big\rangle_{\rho_P}=\frac{g^2 \bar{\mu}_P^2}{(2\pi)^2 \pi} \frac{1}{k_{\perp}^2} \int_{\mathbf{p}} \frac{|\mathbf{k}-\mathbf{p}|^2}{p_{\perp}^2} \mathrm{Tr}[ U(\mathbf{k}-\mathbf{p})U^{T}(-\mathbf{k}+\mathbf{p})].
\end{equation}
For the first saturation correction, eq.~\eqref{eq:M3M3_ebye} becomes
\begin{equation}\label{eq:M3M3_average_rhoP}
\begin{split}
&\Big\langle |M_{(3)}(\mathbf{k})|^2 (\rho_P, \rho_T)\Big\rangle_{\rho_P}\\
=&\frac{g^6\bar{\mu}_P^4}{(2\pi)^2\pi}\Big\{N_c\int_{\mathbf{p},\mathbf{p}_1}\widetilde{\mathcal{H}}_1(\mathbf{p},\mathbf{p}_1, \mathbf{k})\mathrm{Tr}\Big[U(\mathbf{k}-\mathbf{p})U^{T}(-\mathbf{k}+\mathbf{p})\Big]\\
&+\int_{\mathbf{p},\mathbf{p}_1,\mathbf{p}_2} \widetilde{\mathcal{H}}_2(\mathbf{p}, \mathbf{p}_1, \mathbf{p}_2,\mathbf{k})\mathrm{Tr}\Big[ U(\mathbf{k}-\mathbf{p}-\mathbf{p}_1) T^aU^{T}(\mathbf{p}-\mathbf{p}_2)T^d\Big]U^{da}(-\mathbf{k}+\mathbf{p}_1+\mathbf{p}_2)+c.c\\
&+\int_{\mathbf{p},\mathbf{p}_1, \mathbf{q},\mathbf{q}_1} \widetilde{\mathcal{H}}_{3}^{(A)}(\mathbf{p},\mathbf{p}_1, \mathbf{q},\mathbf{q}_1, \mathbf{k})\mathrm{Tr}\Big[U(\mathbf{k}-\mathbf{p}-\mathbf{p}_1)T^aU^T(\mathbf{p}+\mathbf{p}_1)\Big]\mathrm{Tr}\Big[ U(-\mathbf{k}-\mathbf{q}-\mathbf{q}_1) T^aU^T(\mathbf{q}+\mathbf{q}_1)\Big]\\
&+\int_{\mathbf{p},\mathbf{q},\mathbf{p}_1,\mathbf{p}_2} \widetilde{\mathcal{H}}_{3}^{(B)}(\mathbf{p},\mathbf{q},\mathbf{p}_1,\mathbf{p}_2,\mathbf{k}) \mathrm{Tr}\Big[ U(\mathbf{k}-\mathbf{p}-\mathbf{p}_1)T^aU^T(\mathbf{p}-\mathbf{p}_2) U(\mathbf{q}+\mathbf{p}_2)T^aU^T(-\mathbf{k}-\mathbf{q}+\mathbf{p}_1) \Big]\Big\}.\\
\end{split}
\end{equation}
The integrand functions after averaging are related to the original ones by 
\begin{equation}
\widetilde{\mathcal{H}}_1(\mathbf{p},\mathbf{p}_1, \mathbf{k}) = \frac{[(\mathbf{k}-\mathbf{p})\cdot(\mathbf{p}-\mathbf{p}_1)]^2 + [\mathbf{k}\times (\mathbf{p}-\mathbf{p}_1)]^2}{2k_{\perp}^2 |\mathbf{p}-\mathbf{p}_1|^4 p_1^4}\, ,
\end{equation}
\begin{equation}
\begin{split}
\widetilde{\mathcal{H}}_2(\mathbf{p},\mathbf{p}_1,\mathbf{p}_2,\mathbf{k})=&\mathcal{H}_2(\mathbf{p},\mathbf{p}_1,\mathbf{p}_2, \mathbf{q}=-\mathbf{p}_1-\mathbf{p}_2,\mathbf{q}_1=-\mathbf{p}_2, \mathbf{k})  \\
&- \mathcal{H}_2(\mathbf{p},\mathbf{p}_1,\mathbf{p}_2, \mathbf{q}=-\mathbf{p}_1-\mathbf{p}_2,\mathbf{q}_1=-\mathbf{p}_1,\mathbf{k}),\\
\end{split}
\end{equation}
\begin{equation}
\widetilde{\mathcal{H}}_{3}^{(A)}(\mathbf{p},\mathbf{p}_1, \mathbf{q},\mathbf{q}_1, \mathbf{k}) =-\mathcal{H}_3(\mathbf{p},\mathbf{p}_1, \mathbf{p}_2=-\mathbf{p}_1,\mathbf{q},  \mathbf{q}_1, \mathbf{q}_2=-\mathbf{q}_1, \mathbf{k}),
\end{equation}
\begin{equation}
\begin{split}
\widetilde{\mathcal{H}}_{3}^{(B)}(\mathbf{p},\mathbf{q},\mathbf{p}_1,\mathbf{p}_2,\mathbf{k})
=&\mathcal{H}_3(\mathbf{p},\mathbf{p}_1, \mathbf{p}_2,\mathbf{q},  \mathbf{q}_1 = -\mathbf{p}_1, \mathbf{q}_2=-\mathbf{p}_2, \mathbf{k})\\
&-\mathcal{H}_3(\mathbf{p},\mathbf{p}_1, \mathbf{p}_2,\mathbf{q} = -\mathbf{k}-\mathbf{q},  \mathbf{q}_1 = -\mathbf{p}_2, \mathbf{q}_2=-\mathbf{p}_1, \mathbf{k}). \\
\end{split}
\end{equation}
Eq.~\eqref{eq:M1M5_ebye} after ensemble averaging over the projectile color charge densities becomes
\begin{equation}\label{eq:M1M5_average_rhoP}
\begin{split}
&\Big\langle \left[M_{(1)}^{\ast}(\mathbf{k})M_{(5)}(\mathbf{k}) +M_{(1)}(\mathbf{k})M_{(5)}^{\ast}(\mathbf{k}) \right](\rho_P,\rho_T)\Big\rangle_{\rho_P}\\
=&\frac{g^6\bar{\mu}_P^4}{(2\pi)^2\pi}\Big\{N_c\int_{\mathbf{p},\mathbf{q}} \widetilde{\mathcal{F}}_1(\mathbf{p},\mathbf{q},\mathbf{k}) \mathrm{Tr}[U(\mathbf{k}-\mathbf{q})U^{T}(-\mathbf{k}+\mathbf{q})]\\
&+ \int_{\mathbf{p}_3,\mathbf{p}_4, \mathbf{q}} \widetilde{\mathcal{F}}_2(\mathbf{p}_3,\mathbf{p}_4,\mathbf{q},\mathbf{k})\mathrm{Tr}\Big[U(\mathbf{k}-\mathbf{q}-\mathbf{p}_3)T^eU^{T}(\mathbf{q}+\mathbf{p}_3+\mathbf{p}_4)T^{d}\Big]U^{de}(-\mathbf{k}-\mathbf{p}_4)\\
&+\int_{\mathbf{q},\mathbf{p},\mathbf{p}_1,\mathbf{p}_3}\widetilde{\mathcal{F}}^{(A)}_3(\mathbf{p}, \mathbf{q}, \mathbf{p}_1,  \mathbf{p}_3, \mathbf{k})\mathrm{Tr}[U(\mathbf{p}-\mathbf{p}_1)T^aU^{T}(\mathbf{q}-\mathbf{p}+\mathbf{p}_1)]\mathrm{Tr}[U(\mathbf{k}-\mathbf{q}-\mathbf{p}_3)T^aU^{T}(-\mathbf{k}+\mathbf{p}_3)]\\
&+\int_{\mathbf{q},\mathbf{p}, \mathbf{p}_3, \mathbf{p}_4}\widetilde{\mathcal{F}}_3^{(B)}(\mathbf{p},\mathbf{q},\mathbf{p}_3,\mathbf{p}_4,\mathbf{k})\mathrm{Tr}\Big[U(\mathbf{k}-\mathbf{q}-\mathbf{p}_3)T^{a}U^{T}(-\mathbf{k}-\mathbf{p}_4)U(\mathbf{q}-\mathbf{p}+\mathbf{p}_4) T^aU^{T}(\mathbf{p}+\mathbf{p}_3)\Big]\Big\}\\
&+c.c.
\end{split}
\end{equation}
The new integrand functions are related to the original ones by
\begin{equation}\label{eq:new_integrands_Fs}
\widetilde{\mathcal{F}}_1(\mathbf{p},\mathbf{q},\mathbf{k})=-\frac{(\mathbf{k}-\mathbf{q})^2q^2_{\perp}}{4k_{\perp}^2q_{\perp}^4p_{\perp}^4},
\end{equation}
\begin{equation}
\begin{split}
\widetilde{\mathcal{F}}_2(\mathbf{p}_3,\mathbf{p}_4,\mathbf{q},\mathbf{k})
= &\mathcal{F}_2(\mathbf{p}=-\mathbf{p}_3-\mathbf{p}_4, \mathbf{q}, \mathbf{p}_1=-\mathbf{p}_4,\mathbf{p}_3,\mathbf{p}_4,\mathbf{k})\\
&-\mathcal{F}_2(\mathbf{p}=-\mathbf{p}_3-\mathbf{p}_4, \mathbf{q}, \mathbf{p}_1=-\mathbf{p}_3, \mathbf{p}_3, \mathbf{p}_4,\mathbf{k}),\\
\end{split}
\end{equation}
\begin{equation}
\widetilde{\mathcal{F}}_3^{(A)}(\mathbf{p},\mathbf{q},\mathbf{p}_1,\mathbf{p}_3,\mathbf{k})=-\mathcal{F}_3(\mathbf{p}, \mathbf{q}, \mathbf{p}_1, \mathbf{p}_2=-\mathbf{p}_1, \mathbf{p}_3, \mathbf{p}_4=-\mathbf{p}_3, \mathbf{k}),
\end{equation}
\begin{equation}
\begin{split}
\widetilde{\mathcal{F}}^{(B)}_3(\mathbf{p},\mathbf{q}, \mathbf{p}_3, \mathbf{p}_4,\mathbf{k})
=&\mathcal{F}_3(\mathbf{p}, \mathbf{q}, \mathbf{p}_1=-\mathbf{p}_3, \mathbf{p}_2=-\mathbf{p}_4, \mathbf{p}_3, \mathbf{p}_4, \mathbf{k}) \\
&- \mathcal{F}_3(\mathbf{p}= \mathbf{q}-\mathbf{p}, \mathbf{q}, \mathbf{p}_1=-\mathbf{p}_4, \mathbf{p}_2=-\mathbf{p}_3, \mathbf{p}_3, \mathbf{p}_4, \mathbf{k}).\\
\end{split}
\end{equation}
Comparing the expressions after ensemble averaging over the projectile 
 eq.~\eqref{eq:M3M3_average_rhoP} and eq.~\eqref{eq:M1M5_average_rhoP} with the configuration-by-configuration expressions eq.~\eqref{eq:M3M3_ebye} and eq.~\eqref{eq:M1M5_ebye}, the total number of terms roughly doubles due to Wick contractions. This also defines the relation between the new integrand functions to the original. The functional dependence on the target Wilson lines looks the same in eq.~\eqref{eq:M3M3_average_rhoP} and eq.~\eqref{eq:M1M5_average_rhoP}.  Indeed,  
there are four distinct structures for the adjoint Wilson lines in these expressions:
\begin{equation}\label{eq:four_products_WL}
\begin{split}
&\mathrm{Tr}\left[U(\mathbf{k}_1)U^T(\mathbf{k}_2)\right], \\
&\mathrm{Tr}\left[ U(\mathbf{k}_1)T^aU^{T}(\mathbf{k}_2)T^d\right] U^{da}(\mathbf{k}_3), \\
&\mathrm{Tr}\left[U(\mathbf{k}_1)T^aU^{T}(\mathbf{k}_2)\right] \mathrm{Tr}\left[U(\mathbf{k}_3)T^aU^{T}(\mathbf{k}_4)\right],\\
&\mathrm{Tr}\left[ U(\mathbf{k}_1)T^aU^{T}(\mathbf{k}_2)U(\mathbf{k}_3)T^aU^{T}(\mathbf{k}_4)\right].\\
\end{split}
\end{equation}
It is important to note that there are constraints on the momentum arguments: the net momentum in all of them have to be zero. That is  $\mathbf{k}_1+\mathbf{k}_2 =0$ for the term with two Wilson lines, $\mathbf{k}_1+\mathbf{k}_2+\mathbf{k}_3=0$ for the term with three Wilson lines and $\mathbf{k}_1+\mathbf{k}_2+\mathbf{k}_3 + \mathbf{k}_4=0$ for the terms with four Wilson lines. This zero net momentum constraint is a reflect of the translational invariance on the transverse plane assumed for the projectile color charge fluctuations in the MV model.  Moreover, there is one exception that the momentum arguments of the third Wilson line factor are different.  In eq.~\eqref{eq:M3M3_average_rhoP}, one has $\mathbf{k}_1+\mathbf{k}_2 = \mathbf{k} = -(\mathbf{k}_3+\mathbf{k}_4)$ with $\mathbf{k}$ the external momentum of the gluon probed. On the othe hand, in eq.~\eqref{eq:M1M5_average_rhoP}, one has $\mathbf{k}_1+\mathbf{k}_2 = \mathbf{q} = -(\mathbf{k}_3+\mathbf{k}_4)$ with $\mathbf{q}$ being a dummy integration variable. 

Relabeling the transverse momentum and collecting terms in eq.~\eqref{eq:M3M3_average_rhoP} and eq.~\eqref{eq:M1M5_average_rhoP} that have the same Wilson line structures, one obtains 
\begin{equation}\label{eq:combined_M1M5_M3M3_average_rhoP}
\begin{split}
&\Big\langle \frac{dN}{d^2\mathbf{k}}\Big|_{FSC}(\rho_P, \rho_T)\Big\rangle_{\rho_P}=\frac{g^6\bar{\mu}_P^4}{(2\pi)^2\pi}\Big\{N_c\int_{\mathbf{p},\mathbf{k}_1} \mathsf{G}_1(\mathbf{p}, \mathbf{k}_1, \mathbf{k}) \mathrm{Tr}[U(\mathbf{k}_1)U^{T}(-\mathbf{k}_1)]\\
&+\int_{\mathbf{k}_1,\mathbf{k}_2, \mathbf{q}} \mathsf{G}_2(\mathbf{q}, \mathbf{k}_1, \mathbf{k}_2, \mathbf{k}) \mathrm{Tr}\Big[U(\mathbf{k}_1)T^eU^{T}(\mathbf{k}_2)T^{d}\Big]U^{de}(-\mathbf{k}_1-\mathbf{k}_2)\\
&+\int_{\mathbf{q},\mathbf{p},\mathbf{k}_1,\mathbf{k}_4}\mathsf{G}_3(\mathbf{p}, \mathbf{q}, \mathbf{k}_1, \mathbf{k}_4, \mathbf{k})\mathrm{Tr}\Big[U(\mathbf{k}_1)T^aU^{T}(\mathbf{q}-\mathbf{k}_1)\Big]\mathrm{Tr}\Big[U(-\mathbf{q}-\mathbf{k}_4)T^aU^{T}(\mathbf{k}_4)\Big]\\
&+\int_{\mathbf{p},\mathbf{q},\mathbf{k}_1,\mathbf{k}_4} \mathsf{G}_4(\mathbf{p}, \mathbf{q}, \mathbf{k}_1, \mathbf{k}_4, \mathbf{k})\mathrm{Tr}\Big[U(\mathbf{k}_1)T^aU^T(\mathbf{k}-\mathbf{k}_1)\Big]\mathrm{Tr}\Big[ U(-\mathbf{k}-\mathbf{k}_4) T^aU^T(\mathbf{k}_4)\Big]\\
&+ \int_{\mathbf{p},\mathbf{k}_1, \mathbf{k}_2, \mathbf{k}_3}\mathsf{G}_5(\mathbf{p}, \mathbf{k}_1, \mathbf{k}_2, \mathbf{k}_3, \mathbf{k}) \mathrm{Tr}\Big[U(\mathbf{k}_1)T^{a}U^{T}(\mathbf{k}_2)U(\mathbf{k}_3) T^aU^{T}(-\mathbf{k}_1-\mathbf{k}_2-\mathbf{k}_3)\Big]\Big\}\\
&+c.c.
\end{split}
\end{equation}
Here the integrand functions are related to the previously defined by
\begin{equation}
\mathsf{G}_1(\mathbf{p}, \mathbf{k}_1, \mathbf{k}) = \widetilde{\mathcal{F}}_1(\mathbf{p},\mathbf{k}-\mathbf{k}_1,\mathbf{k}) +\frac{1}{2}\widetilde{\mathcal{H}}_1(\mathbf{k}-\mathbf{k}_1,\mathbf{p}, \mathbf{k}),
\end{equation}
\begin{equation}
\begin{split}
\mathsf{G}_2(\mathbf{q}, \mathbf{k}_1, \mathbf{k}_2, \mathbf{k})=&\widetilde{\mathcal{F}}_2(\mathbf{k}-\mathbf{q}-\mathbf{k}_1,\,-\mathbf{k}+\mathbf{k}_1+\mathbf{k}_2,\,\mathbf{q},\,\mathbf{k})\\
& +\widetilde{\mathcal{H}}_2(\mathbf{q},\, \mathbf{k}-\mathbf{q}-\mathbf{k}_1, \,\mathbf{q}-\mathbf{k}_2,\,\mathbf{k}),\\
\end{split}
\end{equation}
\begin{equation}
\mathsf{G}_3(\mathbf{p}, \mathbf{q}, \mathbf{k}_1, \mathbf{k}_4, \mathbf{k}) = \widetilde{\mathcal{F}}^{(A)}_3(\mathbf{p}, \,\mathbf{q}, \,\mathbf{p}-\mathbf{k}_1, \, \mathbf{k}+\mathbf{k}_4, \,\mathbf{k}),
\end{equation}
\begin{equation}
\mathsf{G}_4(\mathbf{p}, \mathbf{q}, \mathbf{k}_1, \mathbf{k}_4, \mathbf{k})=-\frac{1}{2}\widetilde{\mathcal{H}}_{3}^{(A)}(\mathbf{p},\,\mathbf{k}_1-\mathbf{p}, \,\mathbf{q},\,\mathbf{k}_4-\mathbf{q}, \,\mathbf{k}),
\end{equation}
\begin{equation}
\begin{split}
\mathsf{G}_5(\mathbf{p}, \mathbf{k}_1, \mathbf{k}_2,\mathbf{k}_3, \mathbf{k})=&\widetilde{\mathcal{F}}_3^{(B)}(\mathbf{p},\,\mathbf{k}+\mathbf{k}_2+\mathbf{k}_3+\mathbf{p},\,-\mathbf{k}_1-\mathbf{k}_2-\mathbf{k}_3+\mathbf{p},\,-\mathbf{k}-\mathbf{k}_2,\,\mathbf{k})\\
& +\frac{1}{2} \widetilde{\mathcal{H}}_{3}^{(B)}(\mathbf{p},\,\mathbf{k}_2+\mathbf{k}_3-\mathbf{p},\,\mathbf{k}-\mathbf{k}_1-\mathbf{p},\,\mathbf{p}-\mathbf{k}_2,\,\mathbf{k})\,.\\
\end{split}
\end{equation}
The numeric factor $\frac{1}{2}$ is due to the complex conjugate terms added in eq.~\eqref{eq:combined_M1M5_M3M3_average_rhoP}. Note that the Wilson line in momentum space is in general complex. Eq.~\eqref{eq:combined_M1M5_M3M3_average_rhoP} is the main result of this section. 
It makes its dependence on the target Wilson lines manifest while sweeping all the complications into the integrand functions. 
This form is superior for analysis and, especially, for numerical computations 
as potential cancelations occur on the level of simple and analytic kinematic factors rather than on the level of the numerically evaluated target Wilson line correlators.

\section{Ensemble averaging over the target}
\label{sec:average_target}
The next step is to do ensemble averaging over the target Wilson lines within the MV model. We only need to compute the four terms in eq.~\eqref{eq:four_products_WL}. Keep in mind that all the Wilson lines are in the adjoint representation.  The first term is simply related to the gluon dipole correlator 
\begin{equation}
D(\mathbf{k}_1) = \frac{1}{S_{\perp}(N_c^2-1)} \Big\langle \mathrm{Tr}[U(\mathbf{k}_1)U^T(-\mathbf{k}_1)]\Big\rangle\,.
\end{equation}
It has a closed form analytic expression whose derivations and phenomenological applications have been extensively studied \cite{Kovner:2001vi, Fujii:2002vh, Blaizot:2004wv, Fukushima:2007dy, Fukushima:2017mko, Albacete:2010sy, Albacete:2012xq, Lappi:2013zma}.  

For the other three correlators in eq.~\eqref{eq:four_products_WL}, one first notes that the three-Wilson-line correlator turns out to be related to one of the four-Wilson-line correlators
\begin{equation}
\left\langle \mathrm{Tr}[U(\mathbf{k}_1)T^eU^T(\mathbf{k}_2)T^d]U^{de}(\mathbf{k}_3)\right\rangle =\int_{\mathbf{p}}\left\langle \mathrm{Tr}[U(\mathbf{k}_1) T^e U^T(\mathbf{k}_2)U(\mathbf{p})T^eU^T(\mathbf{k}_3-\mathbf{p})]\right\rangle \,.
\end{equation}
As a result, one only need to figure out how to obtain the following two four-Wilson-line correlators
\begin{equation}\label{eq:4WL_correlators}
\begin{split}
&\left\langle \mathrm{Tr}\left[U(\mathbf{k}_1)T^aU^{T}(\mathbf{k}_2)\right] \mathrm{Tr}\left[U(\mathbf{k}_3)T^aU^{T}(\mathbf{k}_4)\right]\right\rangle ,\\
&\left\langle \mathrm{Tr}\left[ U(\mathbf{k}_1)T^aU^{T}(\mathbf{k}_2)U(\mathbf{k}_3)T^aU^{T}(\mathbf{k}_4)\right]\right\rangle .\\
\end{split}
\end{equation}
For the adjoint Wilson line correlators that contain more than two Wilson lines, there are several ways to compute them. 

On the analytic side, closed form expressions in the \textit{coordinate space} can be systematically deduced in the large-$N_c$ limit \cite{JalilianMarian:2004da, Dominguez:2012ad, Kovchegov:2012nd}.  In this limit,  eq.~\eqref{eq:4WL_correlators} could also be expressed in terms of dipole and quadrupole of Wilson lines in the fundamental representation for which  closed form expressions are known. However, our analytic result for the single inclusive gluon production is given in momentum space. 
The coordinate-space expression can be obtained in principal, but in practice it is not useful as it will lead to many two-dimensional Fourier transforms of rather complicated expressions which cannot be computed reliably numerically owing to the presence of the sign-alternating (and in general complex) phases.  
 

Recently, a new approximation scheme was proposed by Kovner and Rezaeian in \cite{Kovner:2017ssr} and has been applied to study the multiparticle production in high energy nuclear collisions \cite{Altinoluk:2018ogz, Agostini:2021xca}.

\subsection{The Dipole Approximation}
The idea behind this approximation is to express any Wilson line correlator with arbitrary color indices in terms of the dipole correlators. Physically, the approximation projects out the color singlet states of the projectile and is applicable for a very dense target, i.e. in the vicinity of the black disk limit. See discussions in Refs.~\cite{Kovner:2017ssr, Altinoluk:2018ogz}.  Mathematically, it is implemented in coordinate space by
\begin{equation}
\Big\langle U_{ab}(\mathbf{x})U_{cd}(\mathbf{y})\Big\rangle = \frac{1}{N_c^2-1} \delta_{ac}\delta_{bd} \,D(\mathbf{x},\mathbf{y})
\end{equation}
with the coordinate space dipole correlator
\begin{equation}
D(\mathbf{x},\mathbf{y}) = \frac{1}{N_c^2-1} \Big\langle \mathrm{Tr}[U(\mathbf{x})U^{\dagger}(\mathbf{y})]\Big\rangle.
\end{equation}
For expressions containing more than two adjoint Wilson lines, the Wick contraction is first carried out before taking the average. To understand the Wick contraction procedure, imagine the nucleus is composed of domains of the transverse size $1/Q_s^2$, determined by the saturation scale. The total number of domains is $N_{\perp} = S_{\perp} Q_s^2$. For the domains in the nucleus spatially separated larger than $1/Q_s$, it is reasonable to assume that color fluctuations are independent and uncorrelated.  In general, all the possible configurations contribute to the final observable. The dipole approximation, however, only takes into account configurations that pairs of gluons are largely separated (larger than $1/Q_s$) so that one can identify each pair as an independent dipole. The configurations, when more than two gluons simultaneously locate within area $1/Q^2_s$,   are  suppressed by $1/N_{\perp}$ compared with configurations in which pairs of gluons are widely separated. The dipole approximation is expected to be exact in the limit $N_{\perp} \gg 1$.  
One consequence of this approximation is that  the expectation value of  the correlators with an odd number of Wilson lines is zero. 

 In a sense, the dipole approximation should be applied  directly to observables that are expressed as integral of Wilson line correlators and other factors. These other kinematic factors might restrict range of  the validity of the approximation. 



Using the dipole approximation, one proceeds to calculate the four-Wilson-line correlators as follows 
\begin{equation}\label{eq:fourWL_dipole_App}
\begin{split}
&\big\langle \mathrm{Tr}[U(\mathbf{x}_1)T^aU^{\dagger}(\mathbf{x}_2)U(\mathbf{x}_3)T^aU^{\dagger}(\mathbf{x}_4)]\big\rangle\\
\simeq &N_c(N_c^2-1)D(\mathbf{x}_1, \mathbf{x}_4) D(\mathbf{x}_2,\mathbf{x}_3) -N_c D(\mathbf{x}_1, \mathbf{x}_3)D(\mathbf{x}_2, \mathbf{x}_4).\\
\end{split}
\end{equation}
Note that the second term is subleasing in the  large-$N_c$ limit.  Now consider two different extreme spatial configurations. First, taking $\mathbf{x}_2=\mathbf{x}_3$ or $\mathbf{x}_1=\mathbf{x}_4$ on the left hand side of the equation, the four-Wilson-line correlator reduces to a gluon dipole correlator. The first term on the right hand side correctly captures these limits while the second term seems violate these limits. 
Second, taking the limit $\mathbf{x}_1 =\mathbf{x}_2$ or $\mathbf{x}_3=\mathbf{x}_4$, the four-Wilson-line correlator reduces to the three-Wilson-line correlator 
\begin{equation}\label{eq:DA_3WL_correlator}
\big\langle \mathrm{Tr}[U(\mathbf{x}_1)T^aU^{\dagger}(\mathbf{x}_2)T^d]U^{da}(\mathbf{x}_3)\big\rangle\simeq N_c(N_c^2-2)D(\mathbf{x}_1, \mathbf{x}_3) D(\mathbf{x}_2,\mathbf{x}_3)
\end{equation}
This three-Wilson-line correlator is exactly the one present in the first saturation correction to single inclusive gluon production, eq.~\eqref{eq:combined_M1M5_M3M3_average_rhoP}. However, according to the dipole approximation, the left hand side of eq.~\eqref{eq:DA_3WL_correlator} contains odd numbers of Wilson lines and thus should be set to zero. 

How to reconcile these apparent contradictions? As was alluded to before the dipole approximation is not generally valid for any fixed configuration. It only accounts for configurations with widely separated pairs of gluons, as they dominate in the large $N_{\perp}$ limit. Taking the limits $\mathbf{x}_2=\mathbf{x}_3$ or $\mathbf{x}_1 = \mathbf{x}_4$ in the first case brings the two pairs of gluons close and thus goes beyond  the region of the applicability  of the  dipole approximation. In the second case, the two pairs of gluons share one common point and thus cannot be properly distinguished as two independent dipoles. In other words, one is not allowed to freely change the spatial configurations when using the dipole approximation.

Performing Fourier transformations of eq.~\eqref{eq:fourWL_dipole_App}, 
we obtain 
\begin{equation}\label{eq:DA_result_4WL_two}
\begin{split}
&\left\langle \mathrm{Tr}\left[ U(\mathbf{k}_1)T^aU^{T}(\mathbf{k}_2)U(\mathbf{k}_3)T^aU^{T}(\mathbf{k}_4)\right]\right\rangle\\
\simeq &S_{\perp}N_c\Big[(N_c^2-1) (2\pi)^2\delta(\mathbf{k}_2+\mathbf{k}_3)- (2\pi)^2\delta(\mathbf{k}_1+\mathbf{k}_3)\Big]D(\mathbf{k}_1) D(\mathbf{k}_2). \\
\end{split}
\end{equation}
Here we  accounted to  the momentum constraint $\mathbf{k}_1+\mathbf{k}_2+\mathbf{k}_3+\mathbf{k}_4 =0$ and introduced the transverse area $S_{\perp} = (2\pi)^2 \delta^{(2)}(\mathbf{k}=\mathbf{0})$. Note that the four-Wilson-line correlator as a function of $\mathbf{k}_1, \mathbf{k}_2, \mathbf{k}_3$ is in general complex-valued  but the dipole approximation predicts that its dominant part is real as given by the right hand side of eq.~\eqref{eq:DA_result_4WL_two}.

 The other four-Wilson-line correlator in eq.~\eqref{eq:4WL_correlators} can be similarly computed 
\begin{equation}\label{eq:DA_result_4WL_one}
\begin{split}
&\left\langle \mathrm{Tr}\left[U(\mathbf{k}_1)T^aU^{T}(\mathbf{k}_2)\right] \mathrm{Tr}\left[U(\mathbf{k}_3)T^aU^{T}(\mathbf{k}_4)\right]\right\rangle\\
\simeq &S_{\perp}N_c\Big[(2\pi)^2 \delta(\mathbf{k}_2+\mathbf{k}_3) -(2\pi)^2\delta(\mathbf{k}_1+\mathbf{k}_3)\Big] D(\mathbf{k}_1)D(\mathbf{k}_2).\\
\end{split}
\end{equation}
Indeed, under the dipole approximation, Wilson line correlators are expressed as  products of dipole correlators. They have simplified factorization forms in momentum space as shown in eqs.~\eqref{eq:DA_result_4WL_two} and ~\eqref{eq:DA_result_4WL_one}.

In the following sections, we numerically compute the momentum space Wilson line correlators within the MV model and  validate  eqs.~\eqref{eq:DA_result_4WL_two} and ~\eqref{eq:DA_result_4WL_one}. We also discuss in detail the problem associated with the three-Wilson-line correlator eq. \eqref{eq:DA_3WL_correlator}.

\subsection{Wilson Line correlators in momentum space: factorization relations }
In this section, we compute the Wilson line correlators numerically. We follow the procedures explained in \cite{Lappi:2007ku}. First computing the coordinate space Wilson line on a 2d lattice through
\begin{equation}
U(\mathbf{x}) = \prod_{i=1}^{N_y} \mathrm{Exp}\left\{ig \frac{\rho_T(x^-_i, \mathbf{x})}{\nabla^2-m^2}\right\} .
\end{equation}
Here longitudinal path integral in eq.~\eqref{eq:def_WL} is expressed as a product of a series of discrete exponentials at different longitudinal coordinates. The target color charge densities $\rho_T(x^-_i, \mathbf{x})$ are random variables at given position $(x_i^-, \mathbf{x})$ satisfying Gaussian distributions eq. ~\eqref{eq:MV_gaussian_correlator}.  
The next step is to use a Fast Fourier Transformation to obtain the momentum space Wilson line $U(\mathbf{k})$ on a 2d lattice. Nearest neighbor prescription is used to extract values of $U(\mathbf{k})$ if the momentum $\mathbf{k}$ does not precisely fall on a lattice site; this forces us to use the lattice parameters that correspond to fine lattice spacings in the momentum space. 
The ensemble averaging is then obtained by computing  a large number of configurations. 

The input and lattice  parameters are as follows. The longitudinal extension is characterized by $N_y = 50$. Here the infrared behavior is regularized but the cutoff scale $m=0.2\,\mathrm{GeV}$.  The nucleus is represented by a square sheet with side length $L_x = 24 \, \mathrm{fm}$. The number of lattice sites along each direction is $N_x = 512$.  This corresponds to momentum step $\Delta k = 2\pi/L_x \simeq 0.052 \, \mathrm{GeV}$. The momentum range covered is $-k_{\mathrm{max}} \leq k_x \leq k_{\mathrm{max}}$ with $k_{\mathrm{max}} = ( N_x/2)\Delta k = 13.2\, \mathrm{GeV}$. The other input parameter is the Gaussian width $g^2\mu_T(x^-, \mathbf{x})$, which is known to be linearly related to the gluon saturation scale $g^2\mu_T \simeq c\, Q_{s,T}$ \cite{Lappi:2007ku}, where  the proportionality factor $c = 1.43$ \cite{Schlichting:2019bvy, Schenke:2012wb, Schenke:2012hg}.  In our calculations we choose $Q_{s,T} = 1.5\, \mathrm{GeV}$.\\

 According to eq.~\eqref{eq:DA_result_4WL_two}, the four-Wilson-line correlator is peaked
at $\mathbf{k}_3= -\mathbf{k}_2$ and at $\mathbf{k}_3 = -\mathbf{k}_1$. The former peak is $N_c^2-1$ higher than the latter.  Furthermore, the peak values are   $N_c(N_c^2-1)S_{\perp}^2D(\mathbf{k}_1)D(\mathbf{k}_2)$ and $-N_cS_{\perp}^2D(\mathbf{k}_1)D(\mathbf{k}_2)$, respectively.  To demonstrate this prediction,  we randomly picked $\mathbf{k}_2 = (-5.48\mathrm{GeV}, 0)$ and $\mathbf{k}_1 = (-8.06\mathrm{GeV},0)$ as an example.  We then vary $\mathbf{k}_3 = (k_{3x}, k_{3y})$ along different trajectories passing through $\mathbf{k}_2$. 

\begin{figure}[htp]
\centering 
\begin{subfigure}[b]{0.65\textwidth}
\includegraphics[width = \textwidth]{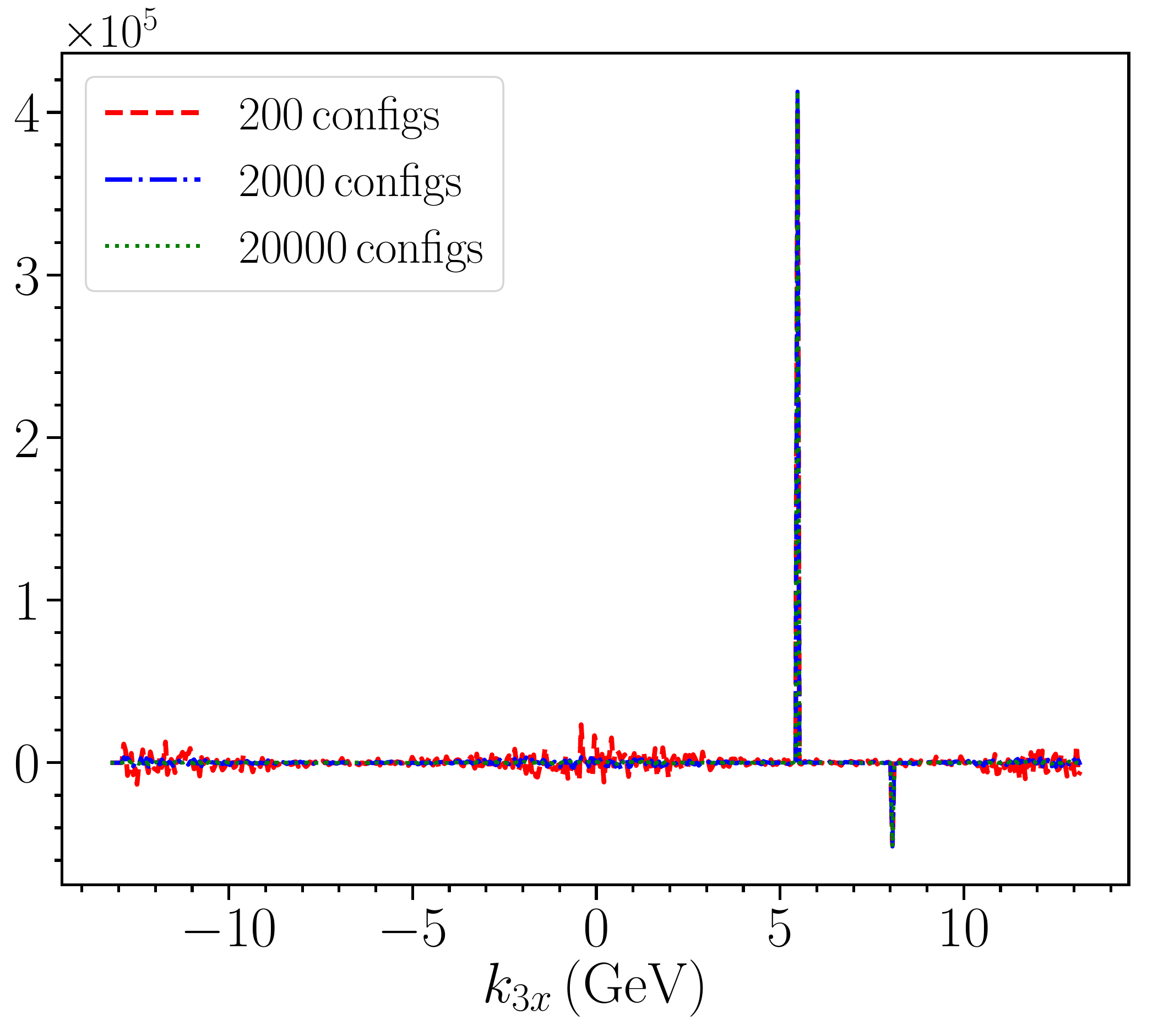}
\caption{Linear scale}
\label{fig:tj1_linear}
\end{subfigure}
\begin{subfigure}[b]{0.65\textwidth}
\includegraphics[width=\textwidth]{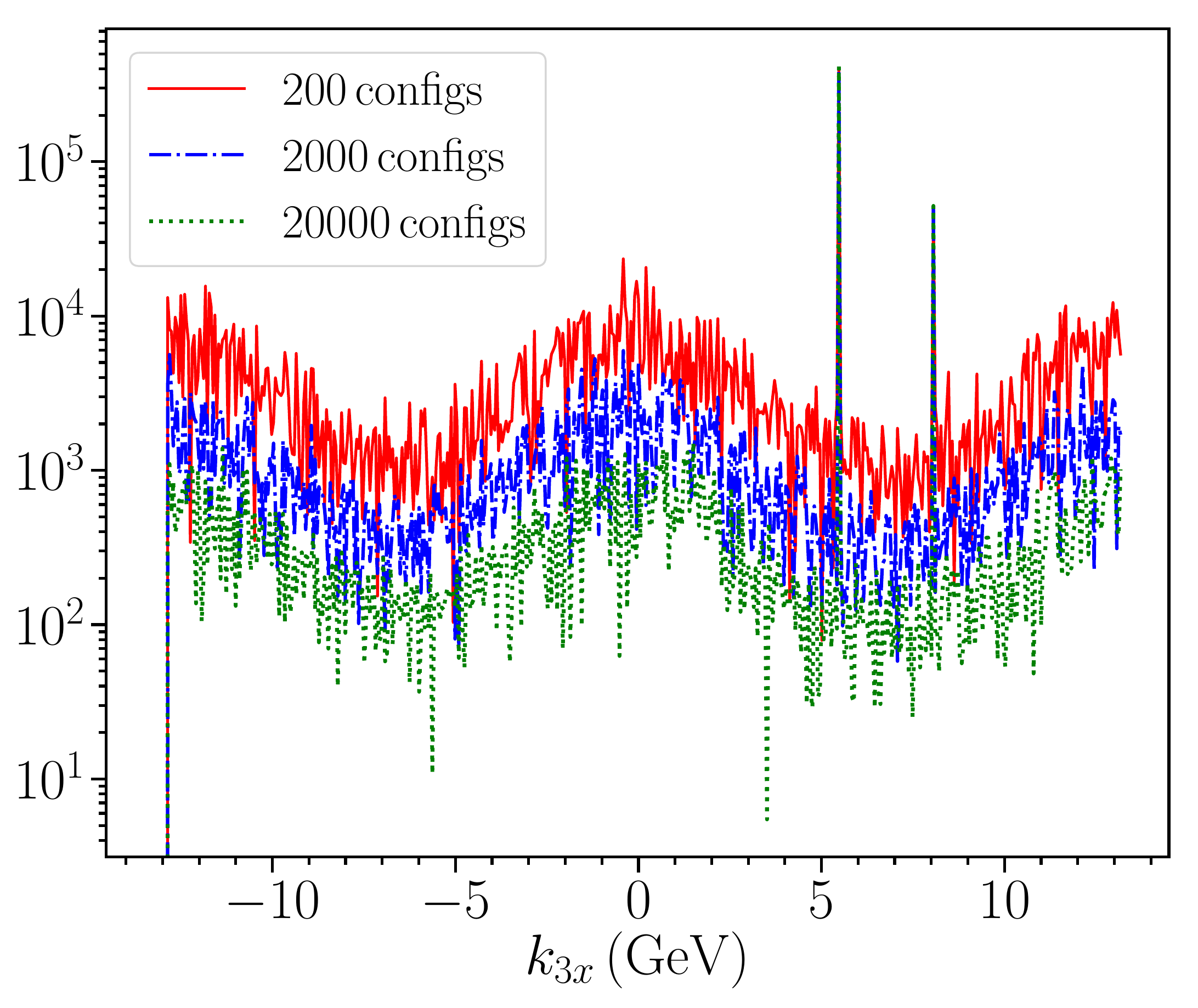}
\caption{Log scale}
\label{fig:tj1_log}
\end{subfigure}
\caption{Four-Wilson-line correlator $\langle \mathrm{Tr}[ U(\mathbf{k}_1)T^aU^{T}(\mathbf{k}_2)U(\mathbf{k}_3)T^aU^{T}(\mathbf{k}_4)]\rangle$ peaks at $\mathbf{k}_3=-\mathbf{k}_2$ and $\mathbf{k}_3=-\mathbf{k}_1$. The figures are plotted by choosing $\mathbf{k}_1 = (-8.06\mathrm{GeV},0)$ and $\mathbf{k}_2 = (-5.48\mathrm{GeV}, 0) $ and $\mathbf{k}_3=(k_{3x},0)$ is varied along $x$-direction. With increasing the number of averaging configurations, the peak values are unchanged while values at other momenta are reduced.}
\label{fig:4U_peaked}
\end{figure}

In the first situation, we choose $\mathbf{k}_3 =(k_{3x},0)$. The four-Wilson-line correlator is in general complex; the real part of it  can have either sign. However, the right hand side of eq.~\eqref{eq:DA_result_4WL_two} is real.  In Fig.~\ref{fig:tj1_linear} the real part of the four-Wilson-line correlator is plotted on a linear scale. In Fig.~\ref{fig:tj1_log}, the norm of the four-Wilson-line correlator is plotted on a log scale. In each figure,  curves from increasing number of averaging configurations are presented.  In Fig.~\ref{fig:tj1_linear}, two peaks at $k_{3x} = 5.48\, \mathrm{GeV}$ and $k_{3x} = 8.06\,\mathrm{GeV}$ have ratio $N_c^2-1$ and being opposite in sign. Note that these two peaks are not smeared Dirac delta function but are peaks at isolated points. Increasing the number of averaging configurations do not change the peak values but reduces values at other momentum. Fig.~\ref{fig:tj1_linear} might be a little bit disguising. The relative magnitudes of the four-Wilson-line correlator at different momentum $\mathbf{k}_3$ can be more clearly seen on the log scale in Fig.~\ref{fig:tj1_log}.
 For 20000 configurations, the peak values at $\mathbf{k}_3=-\mathbf{k}_2$ is at least more than two orders of magnitude larger compared with values at other momenta. It seems that by  increasing the number of configurations from $N_{\mathrm{conf}} = 200$ to $N_{\mathrm{conf}} = 20000$, fluctuations at other momenta are reduced by $1/10$, that is they scale like $1/\sqrt{N_{\mathrm{conf}}}$ demonstrating that these peaks are likely artifacts of finite statistics. 
 
\begin{figure}[htp]
\centering 
\begin{subfigure}[b]{0.65\textwidth}
\includegraphics[width = \textwidth]{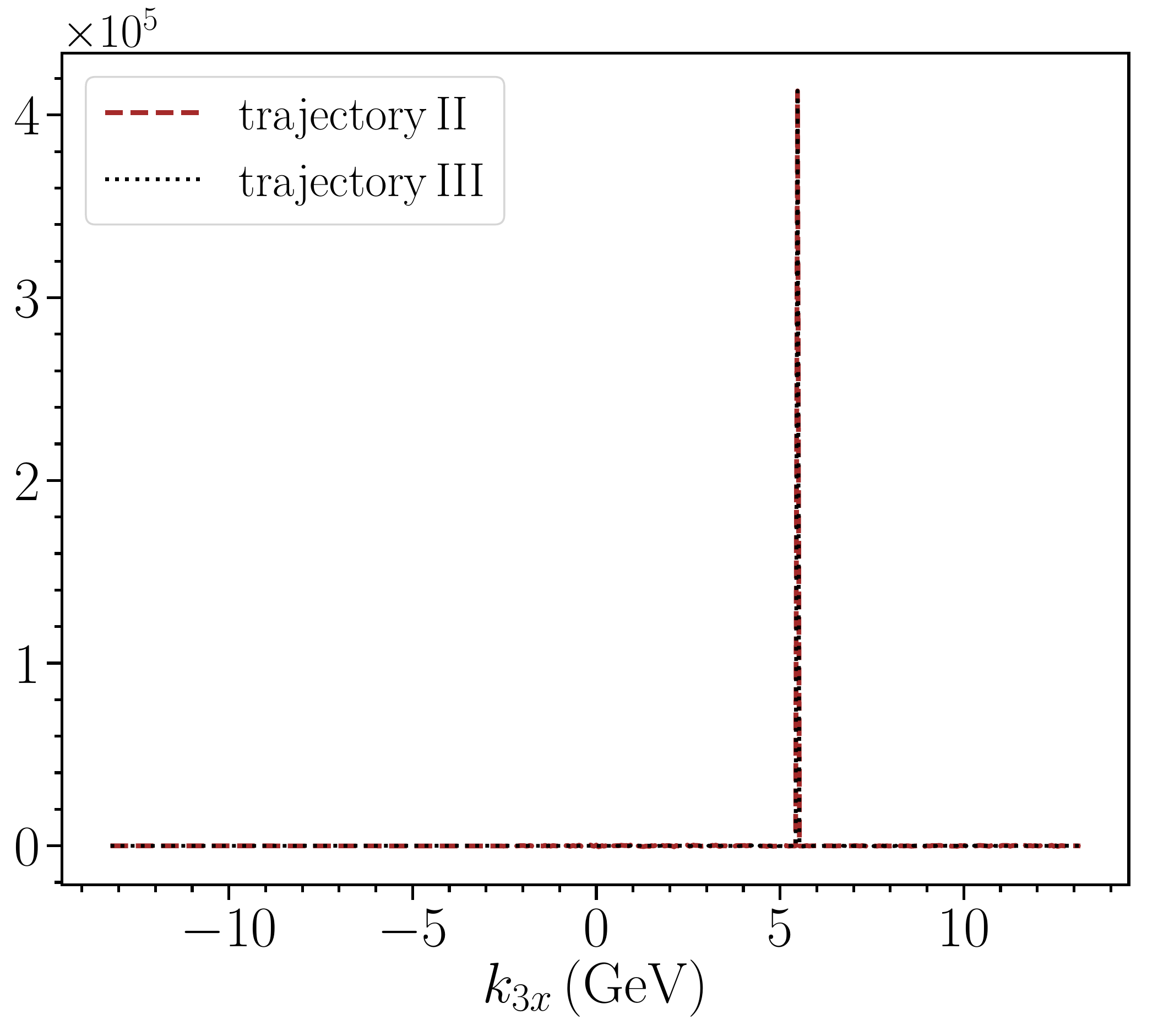}
\caption{Linear scale}
\label{fig:2tjs_linear}
\end{subfigure}
\begin{subfigure}[b]{0.65\textwidth}
\includegraphics[width=\textwidth]{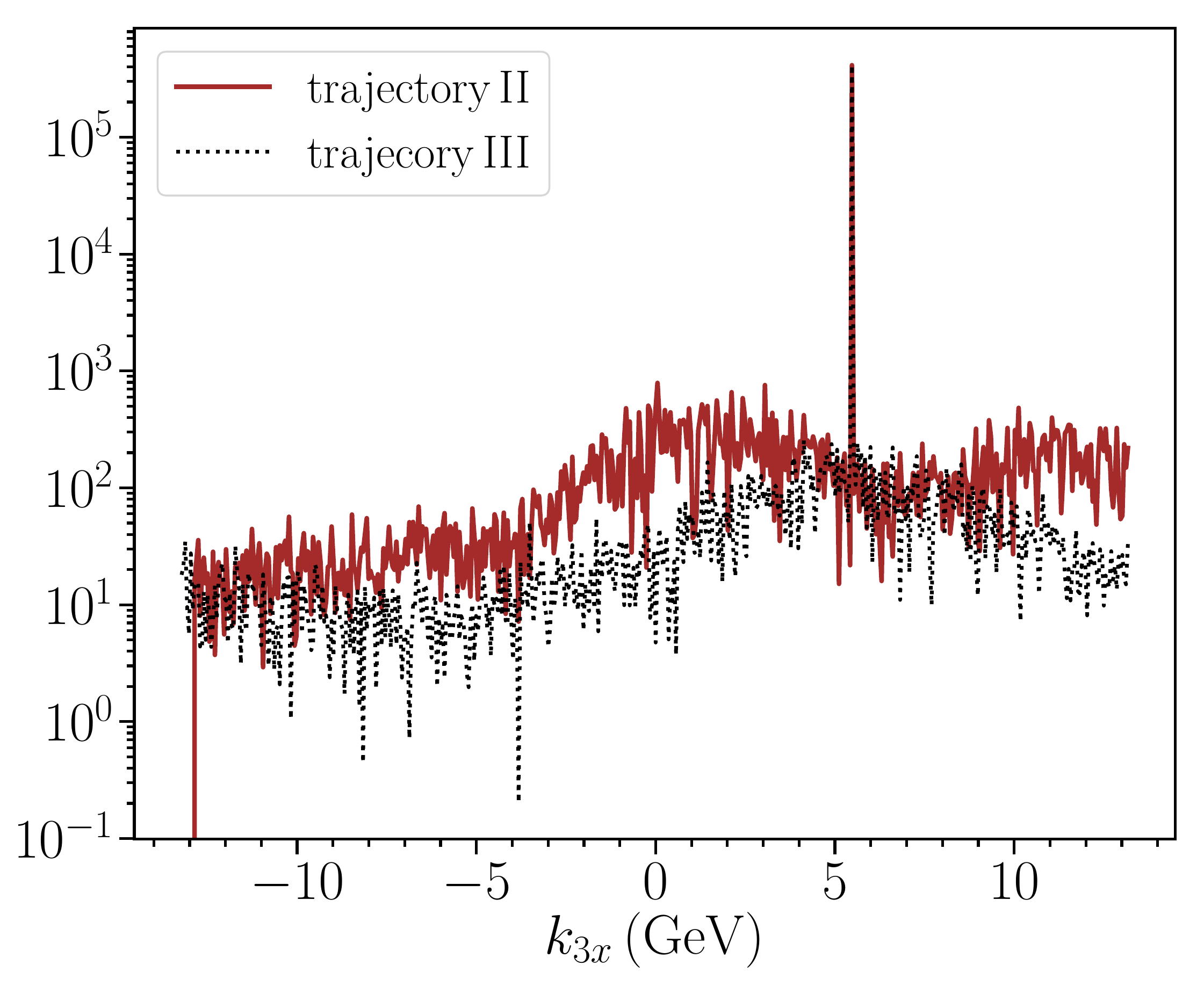}
\caption{Log scale}
\label{fig:2tjs_log}
\end{subfigure}
\caption{Four-Wilson-line correlator $\langle \mathrm{Tr}[ U(\mathbf{k}_1)T^aU^{T}(\mathbf{k}_2)U(\mathbf{k}_3)T^aU^{T}(\mathbf{k}_4)]\rangle$ peaks at $\mathbf{k}_3=-\mathbf{k}_2$.  The figures are exampled by choosing $\mathbf{k}_2 = (-5.48\mathrm{GeV}, 0)$. When varying $\mathbf{k}_3$, two different trajectories passing through $-\mathbf{k}_2$ on the transverse plane are plotted.}
\label{fig:4U_peaked_twodirections}
\end{figure}

In Fig.~\ref{fig:4U_peaked_twodirections}, we continue computing the four-Wilson-line correlator. This time we vary $\mathbf{k}_3$ on the transverse plane along two different trajectories passing through the point $-\mathbf{k}_2=(5.48\,\mathrm{GeV},0)$ but do not pass through the point $-\mathbf{k}_1 = (8.06\,\mathrm{GeV},0)$. For trajectory II, $\mathbf{k}_3 = (k_{3x}, -0.5(k_{3x}+k_{2x}))$ and for trajectory III, $\mathbf{k}_3 = (-k_{2x}, k_{3x}+k_{2x})$. Both figures are results from averaging 20000 configurations. On the linear scale plot in Fig.~\ref{fig:2tjs_linear}, one can see the same peak at $\mathbf{k}_3 = -\mathbf{k}_2$ no matter how $\mathbf{k}_3$ approaches $-\mathbf{k}_2$ on the 2d plane. On the log scale plot in Fig.~\ref{fig:2tjs_log}, the relative magnitude of the peak value and values at other momenta is demonstrated. The peak value is roughly three orders of magnitude larger than values at other momenta. \\
 
We now turn to investigating the next two factorization relations (assuming $\mathbf{k}_1\neq \mathbf{k}_2$)
\begin{equation}\label{eq:factorization_one}
\big\langle\mathrm{Tr}[U(\mathbf{k}_1)T^aU^T(\mathbf{k}_2)U(-\mathbf{k}_2)T^aU^{T}(-\mathbf{k}_1)]\big\rangle =N_c(N_c^2-1) S_{\perp}^2D(\mathbf{k}_1)D(\mathbf{k}_2),
\end{equation} 
\begin{equation}\label{eq:factorization_two}
\big\langle\mathrm{Tr}[U(\mathbf{k}_1)T^aU^T(\mathbf{k}_2)U(-\mathbf{k}_1)T^aU^{T}(-\mathbf{k}_2)]\big\rangle = -N_cS_{\perp}^2 D(\mathbf{k}_1)D(\mathbf{k}_2).
\end{equation} 
\begin{figure}[htp]
\centering 
\begin{subfigure}[b]{0.65\textwidth}
\includegraphics[width = \textwidth]{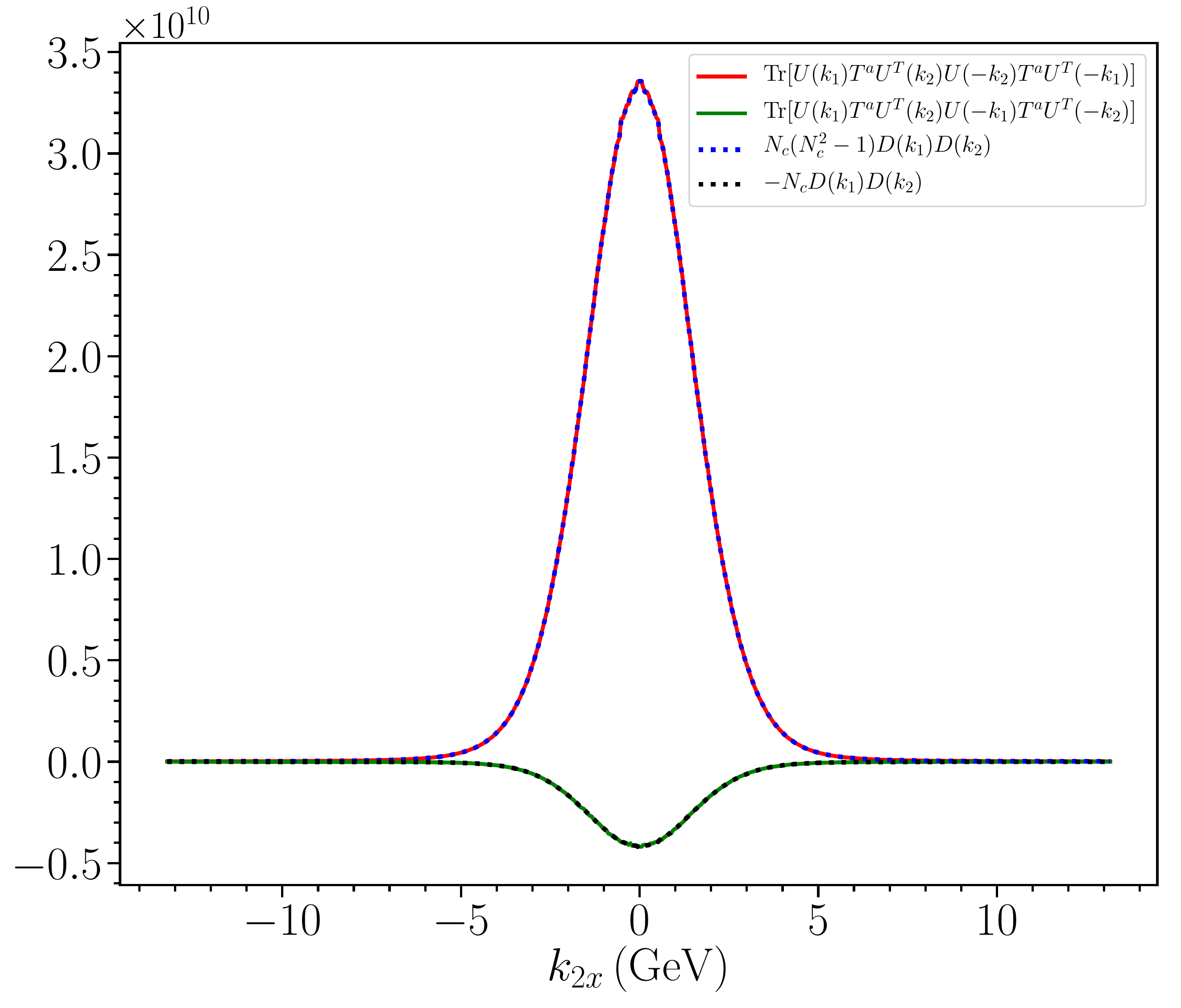}
\caption{check dipole approximation.}
\label{fig:factorizations}
\end{subfigure}
\begin{subfigure}[b]{0.65\textwidth}
\includegraphics[width=\textwidth]{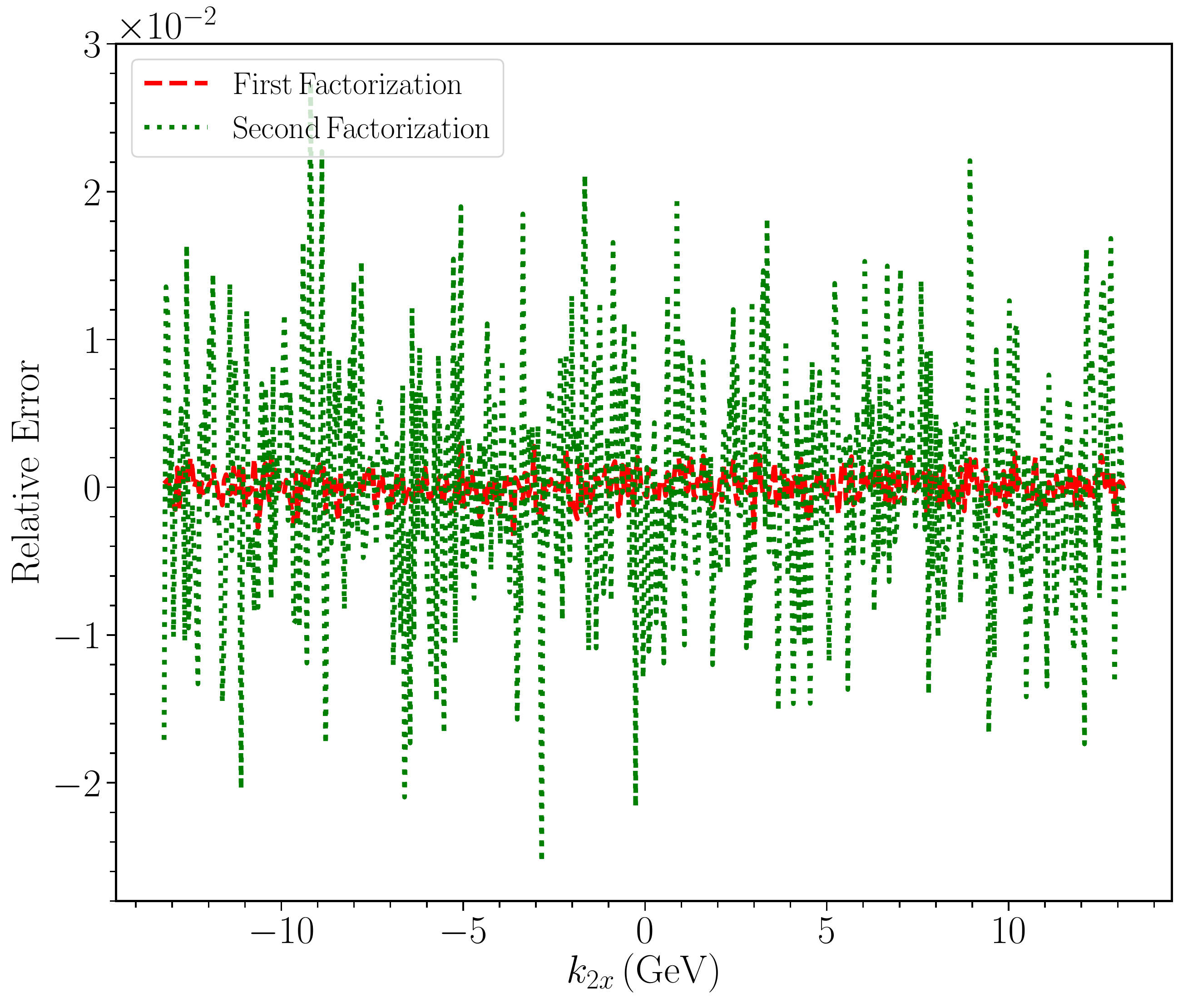}
\caption{Relative errors of the dipole approximation for the two factorization relations.}
\label{fig:errors_factorizations}
\end{subfigure}
\caption{Two factorization relations predicted by dipole approximation are checked.  The figures are plotted by choosing $\mathbf{k}_1 = (0, -0.3\,\mathrm{GeV})$ and varying $\mathbf{k}_2=(k_{2x},0)$.}
\label{fig:4U_two_factorizations_relations}
\end{figure} 
In Fig.~\ref{fig:4U_two_factorizations_relations}, the factorization  of eq.~\eqref{eq:factorization_one} and eq.~\eqref{eq:factorization_two} is demonstrated. We chose $\mathbf{k}_1 = (0, -0.3\,\mathrm{GeV})$ and vary $\mathbf{k}_2=(k_{2x},0)$.  For $N_{\mathrm{conf}} = 2000$ configurations, the errors are roughly $0.15 \%$ and $1.5\%$, respectively. We  checked that increasing the number of averaging configurations will reduce the errors.  Overall, the factorization relations hold surprising well even for regions $\mathbf{k}_1, \mathbf{k}_2\ll Q_{s}$ where possible violations might happen.

There are some subtleties when setting $\mathbf{k}_1 = \mathbf{k}_2$ because the left hand sides of the above two relations are the same while the right hand sides differ. It turns out that the correct one is
\begin{equation}
\langle\mathrm{Tr}[U(\mathbf{k}_1)T^aU^T(\mathbf{k}_1)U(-\mathbf{k}_1)T^aU^{T}(-\mathbf{k}_1)]\rangle = N_c(N_c^2-1)S_{\perp}^2[D(\mathbf{k}_1)]^2
\end{equation}
In other words, the first factorization relation eq.~\eqref{eq:factorization_one} applies to the case $\mathbf{k}_1=\mathbf{k}_2$ while the second factorization relation does not apply. Setting $\mathbf{k}_2 = -\mathbf{k}_1$ for both relations does not lead to any contradictions, as they have different expressions 
\begin{equation}
\begin{split}
&\langle\mathrm{Tr}[U(\mathbf{k}_1)T^aU^T(-\mathbf{k}_1)U(\mathbf{k}_1)T^aU^{T}(-\mathbf{k}_1)]\rangle = N_c(N_c^2-1)S_{\perp}^2[D(\mathbf{k}_1)]^2\\
&\langle\mathrm{Tr}[U(\mathbf{k}_1)T^aU^T(-\mathbf{k}_1)U(-\mathbf{k}_1)T^aU^{T}(\mathbf{k}_1)]\rangle  = -N_cS_{\perp}^2[D(\mathbf{k}_1)]^2\\
\end{split}
\end{equation}
To incorporate the problematic case   $\mathbf{k}_2 = \mathbf{k}_1$, one can modify  the second factorization relation as
\begin{equation}
\begin{split}
&\langle\mathrm{Tr}[U(\mathbf{k}_1)T^aU^T(\mathbf{k}_2)U(-\mathbf{k}_1)T^aU^{T}(-\mathbf{k}_2)]\rangle \\
= &-N_c S_{\perp}^2D(\mathbf{k}_1)D(\mathbf{k}_2) + S_{\perp}(2\pi)^2 \delta(\mathbf{k}_2-\mathbf{k}_1) N_c^3 D(\mathbf{k}_1)D(\mathbf{k}_2).\\
\end{split}
\end{equation} 
This makes sense from the perspective of large $N_c$ limit, in which only the first factorization relation contributes and no modification needed for $\mathbf{k}_1=\mathbf{k}_2$.


One can repeat similar numerical analysis for the second type of four-Wilson-line correlator and its dipole approximation in eq.~\eqref{eq:DA_result_4WL_one}. Again, the dipole approximation turns out to work surprisingly well. More detailed studies for various Wilson line correlators in momentum space and dipole approximation will be reported in future publications. The key parameter controlling the dipole approximation is $N_{\perp} = S_{\perp}Q_s^2$, for typically nucleus with radius $r\sim 5\, \mathrm{fm}$ and $Q_s \sim 1\, \mathrm{GeV}$, one obtains $N_{\perp} \sim 2\times 10^3$. As a result, configurations with four gluons within transverse area $1/Q_s^2$ would be suppressed by $1/N_{\perp}$. That probably is the reason why the dipole approximation predictions are surprisingly precise for the unrestricted momentum-space four-Wilson-line correlators. 

\subsection{When the dipole approximation fails.}
Mathematically, three-Wilson-line correlator is related to the four-Wilson-line correlator in coordinate space by
\begin{equation}\label{eq:3U_to_4U_coordinate}
\mathrm{Tr}[U(\mathbf{x}_1)T^eU^{\dagger}(\mathbf{x}_2)T^d]U^{de}(\mathbf{x}_3) = \mathrm{Tr}[U(\mathbf{x}_1)T^eU^{\dagger}(\mathbf{x}_2)U(\mathbf{x}_3)T^eU^{\dagger}(\mathbf{x}_3)].
\end{equation}
It should be pointed out that although expressed in terms of four Wilson lines, it still represents three gluons because two of the Wilson lines have the same transverse coordinate and thus represent one gluon. Only Wilson lines having different coordinates represent different gluons. For a system with three gluons, they cannot form separated gluon pairs and thus dipole approximation cannot be applied.  However, in momentum space, eq.~\eqref{eq:3U_to_4U_coordinate} becomes
\begin{equation}\label{eq:3U_related_to_4U}
\left\langle \mathrm{Tr}[U(\mathbf{k}_1)T^eU^T(\mathbf{k}_2)T^d]U^{de}(\mathbf{k}_3)\right\rangle =\int_{\mathbf{p}}\left\langle \mathrm{Tr}[U(\mathbf{k}_1) T^e U^T(\mathbf{k}_2)U(\mathbf{p})T^eU^T(\mathbf{k}_3-\mathbf{p})]\right\rangle \,.
\end{equation}
The right hand side involves four-Wilson line correlators we studied in the above section and we have confirmed numerically that the dipole approximation to the four-Wilson-line correlator is a very good approximation. Substituting the dipole approximation eq.~\eqref{eq:DA_result_4WL_two} into eq.~\eqref{eq:3U_related_to_4U}, one derives
\begin{equation}\label{eq:threeU_prediction}
\Big\langle \mathrm{Tr}[U(\mathbf{k}_1)T^eU^T(\mathbf{k}_2)T^d] U^{de}(\mathbf{k}_3)\Big\rangle  \simeq N_c(N_c^2-2) S_{\perp} D(\mathbf{k}_1)D(\mathbf{k}_2).
\end{equation}
(Recall the momentum constraint $\mathbf{k}_3 = -\mathbf{k}_1+\mathbf{k}_2$.)  However, numerical computations show that this prediction is incorrect. 
\begin{figure}[htp]
\centering 
\includegraphics[width = 0.7\textwidth]{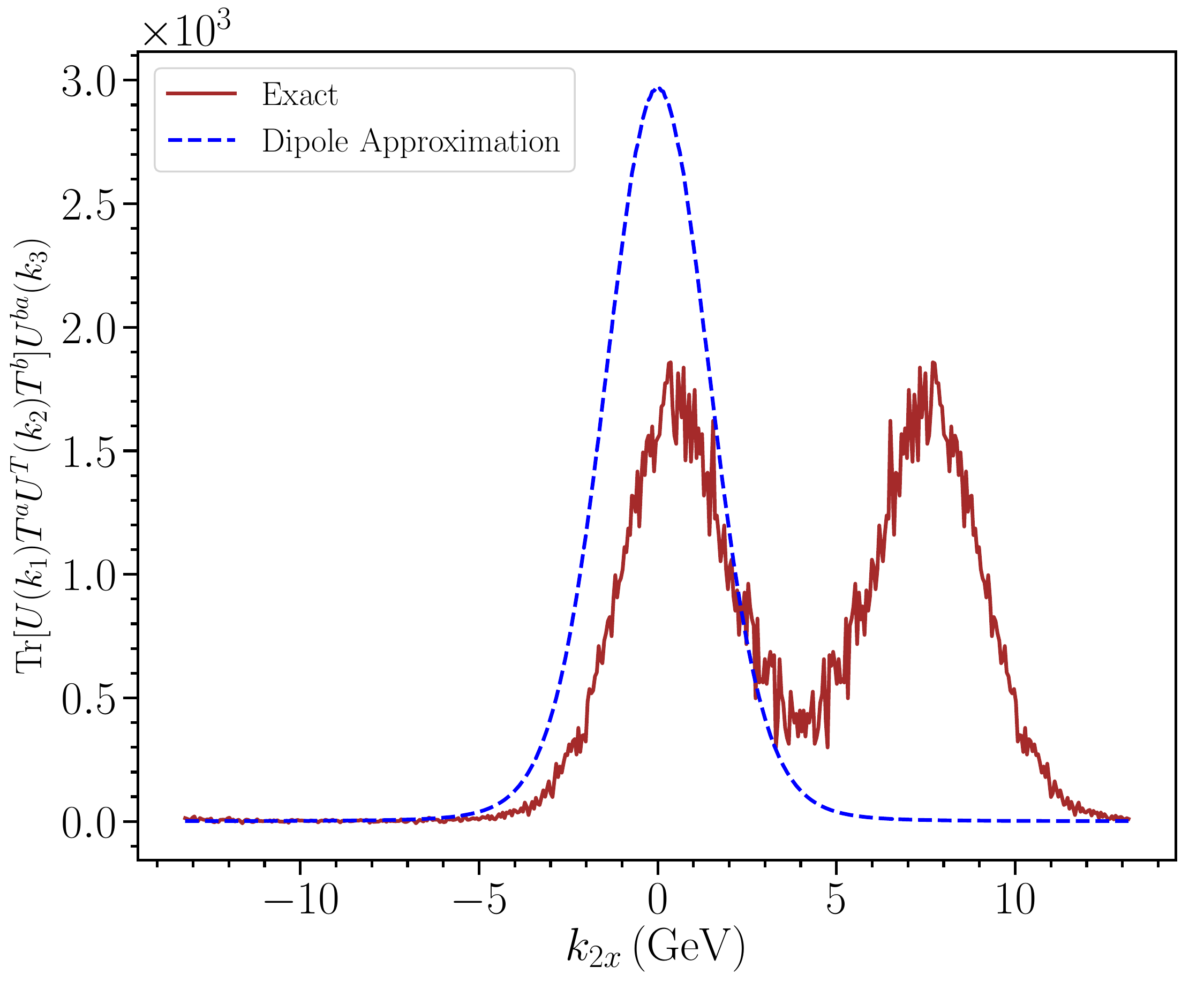}
\caption{ Dipole approximation gives incorrect prediction for the three-Wilson-line correlator. The figures are plotted by choosing $\mathbf{k}_1 = (-8.06\,\mathrm{GeV}, 0)$ and varying $\mathbf{k}_2=(k_{2x},0)$.}
\label{fig:3U_relations}
\end{figure} 
In Fig.~\ref{fig:3U_relations}, eq.~\eqref{eq:threeU_prediction} is checked by choosing $\mathbf{k}_1 = (-8.06\,\mathrm{GeV}, 0)$ and varying $\mathbf{k}_2=(k_{2x},0)$. The dipole approximation predictions one peak when $\mathbf{k}_{2}=0$ while the exact ensemble averaging shows that there are two equal magnitude peaks roughly at $\mathbf{k}_{2}=0$ and $\mathbf{k}_{2} = -\mathbf{k}_1$. This figure is from averaging over 2000 configurations. It is clear that the dipole approximation cannot be applied to the three-Wilson-line correlator.

To summarize, numerical comparison shows that for the four-Wilson-line correlators, dipole approximation works surprisingly well. For the three-Wilson-line correlator, we will directly compute it rather than using the dipole approximation. 

\section{First saturation correction to semi-hard gluon spectrum: Results}
\label{sec:final_results}

The first saturation correction to single inclusive semihard gluon production before averaging over the target is given in eq.~\eqref{eq:combined_M1M5_M3M3_average_rhoP}. The four-Wilson-line correlators present in this equation can be computed using the dipole approximation given in eq.~\eqref{eq:DA_result_4WL_one} and eq.~\eqref{eq:DA_result_4WL_two}. With the dipole approximation, after averaging over the target, eq.~\eqref{eq:combined_M1M5_M3M3_average_rhoP} becomes 
\begin{equation}\label{eq:FSC_ensemble_average}
\Big\langle \frac{dN}{d^2\mathbf{k}}\Big|_{\mathrm{FSC}}(\rho_P, \rho_T)\Big\rangle_{\rho_P, \rho_T}=\frac{1}{\alpha_s}\frac{S_{\perp}}{(2\pi)^3}\frac{(N_c^2-1)}{N_c}Q_{s,p}^4\Big[ N_{\mathrm{IS}}(\mathbf{k})+ N_{\mathrm{FS}}(\mathbf{k})\Big]\,.
\end{equation}
We have denoted the projectile saturation scale $Q^2_{s,p} = \frac{1}{2\pi}N_cg^4\bar{\mu}^2_P$ and $\alpha_s = g^2/4\pi$.  Here $N_{\mathrm{IS}}(\mathbf{k})$ and  $N_{\mathrm{FS}}(\mathbf{k})$ denote contributions coming from initial state interactions and final state interactions, respectively. 
We will separately discuss these two terms in the following sections. For comparison, the leading order single gluon production after ensemble averaging is
\begin{equation}\label{eq:LO_ensemble_average}
\Big\langle \frac{dN}{d^2\mathbf{k}} \Big|_{\mathrm{LO}}(\rho_P, \rho_T) \Big\rangle_{\rho_P, \rho_T}=\frac{1}{\alpha_s} \frac{S_{\perp}}{(2\pi)^3} \frac{(N_c^2-1)}{N_c}Q_{s,p}^2N_{\mathrm{LO}}(\mathbf{k}) 
\end{equation}
with 
\begin{equation}
N_{\mathrm{LO}}(\mathbf{k}) = \frac{1}{k_{\perp}^2} \int_{\mathbf{k}_1} \frac{k_1^2}{|\mathbf{k}-\mathbf{k}_1|^2}D(\mathbf{k}_1).
\end{equation}
The leading order result is linear in the target dipole operator.

\subsection{Initial State Interaction}
The initial state interaction within the first saturation correction is 
\begin{equation}\label{eq:FSC_IS_explicit}
N_{\mathrm{IS}}(\mathbf{k}) = (2\pi)\int_{\mathbf{p},\mathbf{k}_1} \mathsf{G}_1(\mathbf{p}, \mathbf{k}_1, \mathbf{k}) D(\mathbf{k}_1)+c.c.
\end{equation}
 Basically, it represents the next-to-leading order contributions from the WW field of the projectile.  In principle one can resum all order contributions from projectile's WW field \cite{Altinoluk:2014mta}. We derived the all order resummation in appendix.~\ref{appendixA} and the result is 
\begin{equation}\label{eq:master_eq_IS}
\frac{dN}{d^2\mathbf{k}}\Big|_{\mathrm{IS}} = \frac{1}{(2\pi)^2}\frac{S_{\perp}}{\pi k_{\perp}^2}\int\frac{d^2\mathbf{q}}{(2\pi)^2}[ (\mathbf{k}-\mathbf{q})^i(\mathbf{k}-\mathbf{q})^j +(k_{\perp}^2\delta^{ij} -\mathbf{k}^i\mathbf{k}^j)]  xG_{\mathrm{WW}}^{ij}(\mathbf{q}) D(\mathbf{k}-\mathbf{q})
\end{equation}
The all order initial state effects from the projectile are incorporated in the projectile WW field correlator $xG^{ij}_{\mathrm{WW}}(\mathbf{q})$.  The target dipole correlator $D(\mathbf{k})$ has a closed form expression; similarly the projectile WW field correlator has a simple integral form given in appendix.~\ref{appendixA}. Both expressions can be handled by a straightforward  numerical computation.  

\begin{figure}[htp]
\centering 
\includegraphics[width = 0.6\textwidth]{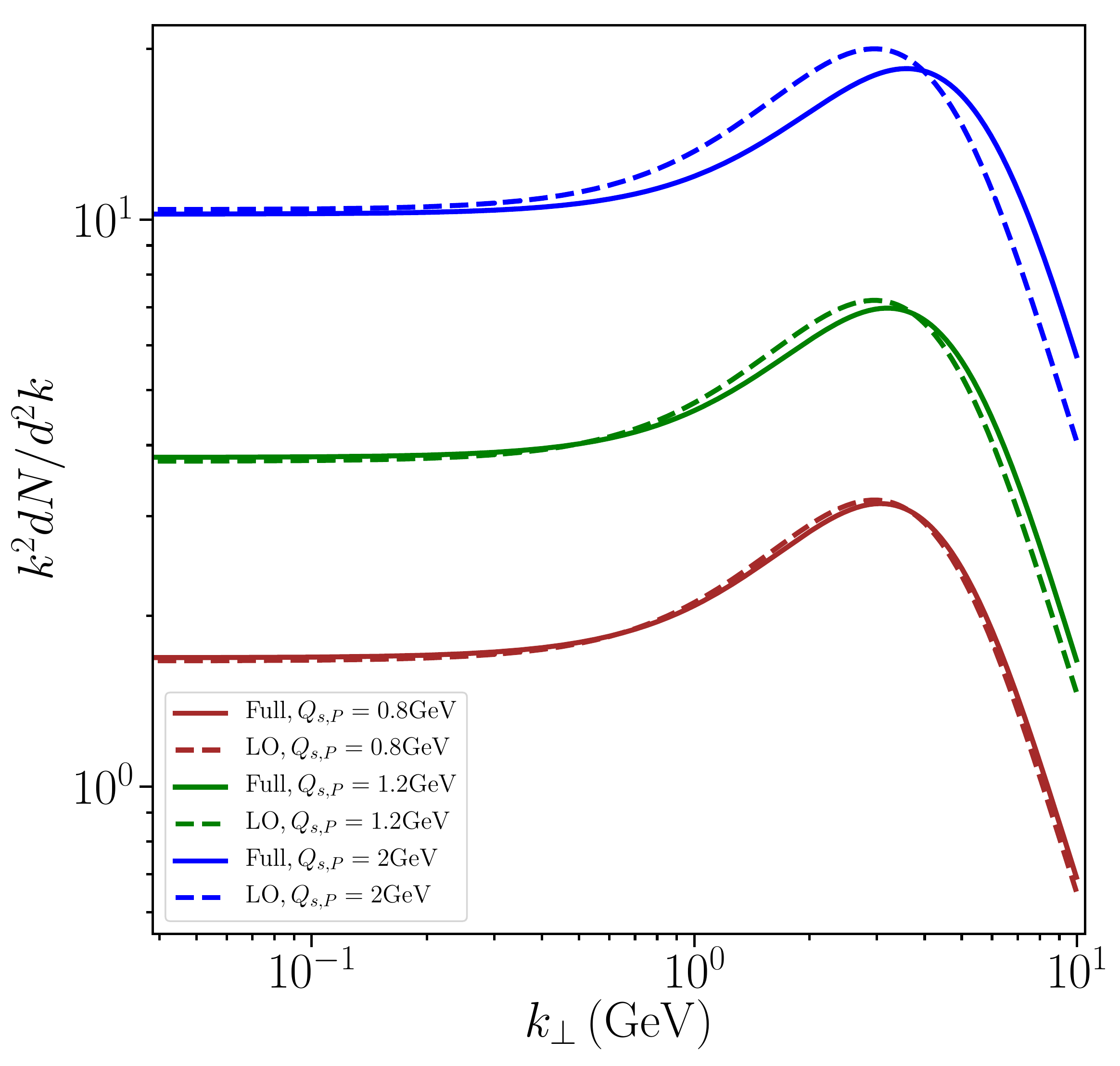}
\caption{Full result resumming all order initial state effect versus leading order result. The target saturation scale is chosen to be $Q_{s,T} = 2.0\,\mathrm{GeV}$. }
\label{fig:dNdk_fullvsLO_Qs}
\end{figure}

In Fig.~\ref{fig:dNdk_fullvsLO_Qs}, the full result eq.~\eqref{eq:master_eq_IS} which resums all order initial state effects is compared with leading order result eq.~\eqref{eq:LO_ensemble_average}. Instead of directly plotting $dN/d^2\mathbf{k}$, we plotted $k^2dN/d^2\mathbf{k}$. We also factored out the common factor $S_{\perp}/(2\pi)^3 \alpha_s$. With the target gluon saturation scale chosen as $Q_{s,T} = 2.0\,\mathrm{GeV}$, three different values of the projectile gluon saturation scale is plotted $Q_{s,P} = 0.8, 1.2, 2.0 \,\mathrm{GeV}$.  From the log-log plot, one can see that the leading order (and full) gluon distributions $dN/d^2\mathbf{k}$ behaves as $1/k_{\perp}^2$ for small momentum $k_{\perp}\ll Q_{s,P}$ but changes to $1/k_{\perp}^4$ behavior for large momentum $k_{\perp} \gg Q_{s,T}$. The transition region is around $\sim 1-2 Q_{s,T}$. 
This plot communicates  two main messages from. First, except in the extreme case that $Q_{s,P}$ is close to $Q_{s,T}$, higher order initial state effects are negligible compared to leading order effect. Second, the inclusion of higher order initial state effects cannot alleviate the $1/k_{\perp}^2$ divergent behavior at small $k_{\perp}$. In other words, the total number of gluons $N_{\mathrm{tot}} = \int d^2\mathbf{k} dN/d^2\mathbf{k}$ is still logarithmically divergent at the infrared even after including higher order initial state effects. Final state interactions must be included to have an infrared safe single gluon production spectrum.

\subsection{Final State Interaction}
The contribution from final state interactions is denoted by $N_{\mathrm{FS}}(\mathbf{k})$ in eq.~\eqref{eq:FSC_ensemble_average}. We separately denote terms from four-Wilson-line correlators as $N^{(4)}_{\mathrm{FS}}(\mathbf{k})$ and terms from three-Wilson-line correlators as $ N^{(3)}_{\mathrm{FS}}(\mathbf{k})$.
\begin{equation}\label{eq:FS_interaction}
N_{\mathrm{FS}}(\mathbf{k}) =N^{(4)}_{\mathrm{FS}}(\mathbf{k}) + N^{(3)}_{\mathrm{FS}}(\mathbf{k})
\end{equation}
with 
\begin{equation}\label{eq:FSC_n(4)}
\begin{split}
N^{(4)}_{\mathrm{FS}}(\mathbf{k}) 
=&\int_{\mathbf{p},\mathbf{k}_1, \mathbf{k}_2}\big[\mathsf{G}_5(\mathbf{p}, \mathbf{k}_1, \mathbf{k}_2,-\mathbf{k}_2, \mathbf{k}) - \frac{1}{N_c^2-1}\mathsf{G}_5(\mathbf{p}, \mathbf{k}_1, \mathbf{k}_2, -\mathbf{k}_1, \mathbf{k})\big] D(\mathbf{k}_1)D(\mathbf{k}_2)\\
&+\frac{1}{N_c^2-1}\int_{\mathbf{q},\mathbf{p}, \mathbf{k}_1}\big[\mathsf{G}_3(\mathbf{p}, \mathbf{q}, \mathbf{k}_1, -\mathbf{k}_1, \mathbf{k})-\mathsf{G}_3(\mathbf{p}, \mathbf{q}, \mathbf{k}_1, -\mathbf{q}+\mathbf{k}_1, \mathbf{k})\big] D(\mathbf{k}_1)D(\mathbf{q}-\mathbf{k}_1)\\
&+\frac{1}{N_c^2-1}\int_{\mathbf{q}, \mathbf{p}, \mathbf{k}_1}\big[\mathsf{G}_4(\mathbf{p}, \mathbf{q}, \mathbf{k}_1, -\mathbf{k}_1, \mathbf{k}) - \mathsf{G}_4(\mathbf{p}, \mathbf{q}, \mathbf{k}_1, -\mathbf{k}+\mathbf{k}_1, \mathbf{k})\big]D(\mathbf{k}_1)D(\mathbf{k}-\mathbf{k}_1)\\
&+c.c.\\
\end{split}
\end{equation}
and 
\begin{equation}\label{eq:FSC_n(3)}
\begin{split}
 N^{(3)}_{\mathrm{FS}}(\mathbf{k})
=&\frac{1}{S_{\perp}N_{\perp}(N_c^2-1)}\int_{\mathbf{k}_1,\mathbf{k}_2, \mathbf{q}} \mathsf{G}_2(\mathbf{q}, \mathbf{k}_1, \mathbf{k}_2, \mathbf{k})\big\langle \mathrm{Tr}\big[U(\mathbf{k}_1)T^eU^{T}(\mathbf{k}_2)T^{d}\big]U^{de}(-\mathbf{k}_1-\mathbf{k}_2)\big\rangle\\
&+c.c.\\
\end{split}
\end{equation}
The $N^{(4)}_{\mathrm{FS}}(\mathbf{k})$ is quadratic in the target dipole correlator as a result of applying the dipole approximation. On the other hand, $ N^{(3)}_{\mathrm{FS}}(\mathbf{k})$  cannot be simplified by dipole approximation and it has to be numerically computed directly by averaging over a large number of configurations. 

One can further simplify the first term in eq.~\eqref{eq:FSC_n(4)} because in the Dipole Approximation, Wilson line correlators are approximated by their real parts. An explicit expression for the integrand function $\mathsf{G}_5(\mathbf{p}, \mathbf{k}_1, \mathbf{k}_2, -\mathbf{k}_2, \mathbf{k})$ can be obtained. To be specific, only the imaginary parts of the six auxiliary functions in eq.~\eqref{eq:upsilon3parallel} and eq.~\eqref{eq:upsilon3perpendicular} contribute and one can carry out the angular integrations in these auxiliary functions analytically. As will be shown later, the term containing $\mathsf{G}_5$ is the dominant one in the final state interaction. It's convenient to have an explicit expression at hand.  

From the last term in eq.~\eqref{eq:M3M3_average_rhoP}, Dipole Approximation generates a momentum delta function $\delta(\mathbf{p}+\mathbf{q})$. Integrating out the momentum $\mathbf{q}$, one obtains
\begin{equation}
\begin{split}
&\widetilde{\mathcal{H}}_3^{(B)}(\mathbf{p}, \mathbf{q}=-\mathbf{p}, \mathbf{p}_1, \mathbf{p}_2, \mathbf{k}) \\
=&\frac{1}{k^2}\Bigg\{\left[\frac{(\mathbf{p}-\mathbf{p}_2)\cdot\mathbf{p}_2}{p_2^2} \frac{(\mathbf{k}-\mathbf{p})\cdot\mathbf{p}_1}{ |\mathbf{k}-\mathbf{p}|^2p_1^2}\right]^2+\left[\frac{(\mathbf{p} \times \mathbf{p}_2)}{p_2^2}  \frac{(\mathbf{k}-\mathbf{p})\cdot\mathbf{p}_1}{p_1^2}\frac{(\mathbf{k}\cdot\mathbf{p})}{p_{\perp}^2|\mathbf{k}-\mathbf{p}|^2}\right]^2\\
&\quad + \frac{1}{2} \left[\frac{(\mathbf{p}\cdot\mathbf{p}_2)}{p_2^2}  \frac{(\mathbf{k}-\mathbf{p})\cdot\mathbf{p}_1}{p_1^2} \frac{(\mathbf{k}\times\mathbf{p})}{p_{\perp}^2|\mathbf{k}-\mathbf{p}|^2}\right]^2 +\left[\frac{\mathbf{p} \times \mathbf{p}_2}{p_2^2}\frac{(\mathbf{k}-\mathbf{p}-\mathbf{p}_1)\cdot\mathbf{p}_1}{ p_1^2} \right]^2\frac{k^2}{p^4|\mathbf{k}-\mathbf{p}|^2}\\
&\quad+\frac{1}{2}\left[\frac{(\mathbf{p}-\mathbf{p}_2)\cdot\mathbf{p}_2}{p_2^2}\frac{(\mathbf{k}-\mathbf{p}-\mathbf{p}_1)\cdot\mathbf{p}_1}{p_1^2}\right]^2\frac{1}{p_{\perp}^2|\mathbf{k}-\mathbf{p}|^2} +\frac{1}{2}\left[\frac{\mathbf{p}\times \mathbf{p}_2}{p_2^2}\frac{(\mathbf{k}-\mathbf{p})\times \mathbf{p}_1}{p_1^2} \frac{k^2+p^2-\mathbf{k}\cdot\mathbf{p}}{p^2|\mathbf{k}-\mathbf{p}|^2}\right]^2\\
& \quad -\frac{(\mathbf{k}-\mathbf{p}-\mathbf{p}_1)\cdot\mathbf{p}_1}{p_1^2} \frac{(\mathbf{k}-\mathbf{p})\cdot\mathbf{p}_1}{ p_1^2}\frac{\mathbf{p}\cdot\mathbf{p}_2}{ p_2^2} \frac{(\mathbf{p}-\mathbf{p}_2)\cdot\mathbf{p}_2}{p_2^2} \left[\frac{\mathbf{p}\cdot(\mathbf{k}-\mathbf{p})}{p_{\perp}^2|\mathbf{k}-\mathbf{p}|^2}\right]^2\Bigg\}\\
\end{split}
\end{equation}
In deriving this expression, we have used the fact that the dipole correlators $D(\mathbf{p}-\mathbf{p}_2)$ and $D(\mathbf{k}-\mathbf{p}-\mathbf{p}_1)$ from the Dipole Approximation only depend on the magnitude of the arguments. As a result, any terms containing odd powers of cross products $\mathbf{p}\times \mathbf{p}_2$ and $(\mathbf{k}-\mathbf{p})\times \mathbf{p}_1$ vanish when integrating over the momenta $\mathbf{p}_2$ and $\mathbf{p}_1$, respectively. 

From the last term in eq.~\eqref{eq:M1M5_average_rhoP}, the Dipole Approximation generates a Dirac delta function $\delta(\mathbf{k}+\mathbf{p}-\mathbf{q})$. Integrating out the momentum $\mathbf{q}$ by setting $\mathbf{q} = \mathbf{k}+\mathbf{p}$, one obtains
 \begin{equation}\label{eq:explicit_F3B}
 \begin{split}
 &\widetilde{\mathcal{F}}_3^{(B)}(\mathbf{p}, \mathbf{q} = \mathbf{k}+\mathbf{p}, \mathbf{p}_3, \mathbf{p}_4, \mathbf{k})\\
=&\left[\frac{\mathbf{p}\times\mathbf{p}_3}{p_{\perp}^2p_3^2} \right]^2\left[\frac{(\mathbf{k}+\mathbf{p}_4)\cdot\mathbf{p}_4}{k^2p_4^2}\right]^2\left(\frac{(\mathbf{k}\times \mathbf{p})^2}{|\mathbf{k}+\mathbf{p}|^2}  - \frac{1}{4}p_{\perp}^2\right)  \left[ \mathrm{sgn}(k^2-p^2) + 1\right]\\
& +\frac{1}{4}\left[ \frac{(\mathbf{p}+\mathbf{p}_3)\cdot\mathbf{p}_3}{p^2_{\perp}p_3^2}\right]^2 \left[\frac{(\mathbf{k}+\mathbf{p}_4)\cdot\mathbf{p}_4}{k^2p_4^2}\right]^2\frac{ (k^2+p^2)}{|\mathbf{k}+\mathbf{p}|^2} \left[k^2-p^2- (k^2+p^2)\mathrm{sgn}(k^2-p^2)\right]\\
& +\frac{(\mathbf{p}+\mathbf{p}_3)\cdot\mathbf{p}_3}{p_3^2} \frac{\mathbf{p}\cdot \mathbf{p}_3}{p_{\perp}^2p_3^2} \frac{\mathbf{k}\cdot\mathbf{p}_4}{ k^2p_4^2}\frac{(\mathbf{k}+\mathbf{p}_4)\cdot\mathbf{p}_4}{k^2p_4^2}\frac{\mathbf{p}\cdot(\mathbf{k}+\mathbf{p})}{|\mathbf{k}+\mathbf{p}|^2}\frac{\mathbf{k}\cdot\mathbf{p}}{p_{\perp}^2}-\frac{1}{2}\frac{1}{k^2}\left[\frac{(\mathbf{k}+\mathbf{p}_4)^2p_4^2}{p_4^4}\right]\left[\frac{\mathbf{p}\cdot\mathbf{p}_3}{p_{\perp}^2p_3^2} \right]^2 \\
&-\left[\frac{\mathbf{p}\times \mathbf{p}_3}{p_{\perp}^2p_3^2} \right]^2\left[\frac{\mathbf{k}\times \mathbf{p}_4}{k^2p_4^2}\right]^2  \frac{(\mathbf{k}\cdot\mathbf{p})^2}{|\mathbf{k}+\mathbf{p}|^2}\mathrm{sgn}(k^2-p^2)\\
&-\left[\frac{\mathbf{k}\times \mathbf{p}_4}{k_{\perp}^2p_4^2}\right]^2 \left[\frac{(\mathbf{p}+\mathbf{p}_3)\cdot\mathbf{p}_3}{p_{\perp}^2p_3^2} \right]^2 \left(\frac{(\mathbf{k}\times \mathbf{p})^2}{|\mathbf{k}+\mathbf{p}|^2} - \frac{1}{4}k^2\right) \left[1-\mathrm{sgn}(k^2-p^2) \right]\\
 \end{split}
 \end{equation}
Again, we used the fact that any terms containing odd powers of cross products $\mathbf{k}\times \mathbf{p}_4$ and $\mathbf{p}\times \mathbf{p}_3$ vanish when integrating over the  momenta $\mathbf{p}_4$ and $\mathbf{p}_3$, respectively. The sign function $\mathrm{sgn}(x) $ comes from the angular integral
\begin{equation}
 \int_{-\pi}^{\pi}\frac{ d\phi}{2\pi}\frac{ 1}{w^2_{\perp} }= \frac{1}{|k^2-p^2|}   
\end{equation}
with $w_{\perp} = \sqrt{k^2+p^2-2kp\cos\phi}$.
Using these two explicit expressions, one then gets
\begin{equation}\label{eq:G5=F3+H3}
\begin{split}
\mathsf{G}_5(\mathbf{p}, \mathbf{k}_1, \mathbf{k}_2,-\mathbf{k}_2, \mathbf{k})=&\widetilde{\mathcal{F}}_3^{(B)}(\mathbf{p},\,\mathbf{k}+\mathbf{p},\,-\mathbf{k}_1+\mathbf{p},\,-\mathbf{k}-\mathbf{k}_2,\,\mathbf{k}) \\
&\qquad+\frac{1}{2} \widetilde{\mathcal{H}}_{3}^{(B)}(\mathbf{p},\,-\mathbf{p},\,\mathbf{k}-\mathbf{k}_1-\mathbf{p},\,\mathbf{p}-\mathbf{k}_2,\,\mathbf{k}).\\
\end{split}
\end{equation}
Unfortunately, one cannot obtain a simplified analytic expression for the integrand function $\mathsf{G}_3$ in which additional angular integrals have to be computed numerically. \\

We use Monte Carlo method to numerically compute these momentum integrals. The numerical integration is carried out using the Vegas algorithm \cite{Lepage:1977sw} implemented in the GSL  and the Cuba (a library for multidimensional numerical integration \cite{Hahn:2004fe}) libraries.  For the dipole operator either a closed from expression can be used to or it can be precomputed by directly averaging over a large number of configurations. There are additional angular integrals inside the integrand functions  $\mathsf{G}_3$. Maximal number of dimensions of the integrals in eq.~\eqref {eq:FS_interaction} is seven. For $N_{\mathrm{FS}}^{(3)}(\mathbf{k})$ in  eq.~\eqref {eq:FSC_n(3)}, numerical integration is performed for a given configuration and it is repeated for many configurations.

In the numerical integrations, the limits of momentum integration are chosen to be $k_{\mathrm{UV}} =30\, \mathrm{GeV}$. We also checked that results are insensitive to the limits by choosing  $15\,\mathrm{GeV}$ and $20\,\mathrm{GeV}$.  Without specifying otherwise, the infrared scale is chosen to be $m=0.2\mathrm{GeV}$. The sensitivity to infrared scales will be discussed further in later sections. The target gluon saturation scale is chosen as $Q_{s,T} = 2 \, \mathrm{GeV}$. 
The relative error for numerical integrations using Vegas algorithm is set to be  $1\times 10^{-3}$. Since the gluon spectrum is isotropic, without loss of generality, we choose the external momentum $\mathbf{k}$ along the $y$-axis, $k_x =0$. \\

\begin{figure}[htp]
\centering 
\includegraphics[width = 0.65\textwidth]{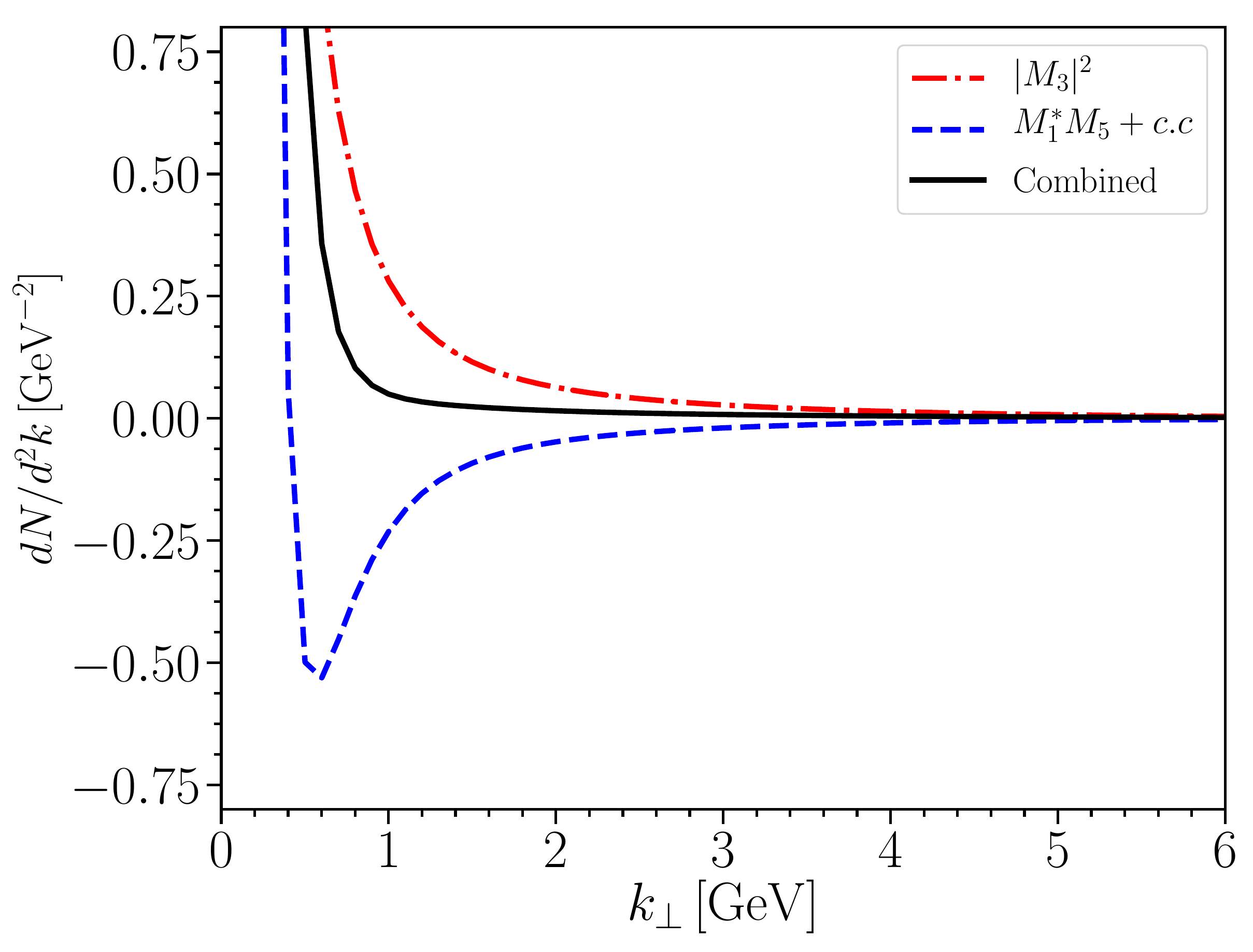}
\caption{Comparing the $|M_3|^2$ and $M_1^{\ast}M_5 +c.c$ terms inside the first saturation correction.}
\label{fig:dNdk_M1M5vsM3M3}
\end{figure}

From eq.~\eqref{eq:G5=F3+H3}, we know that the dominant part of the first saturation correction has two contributions. One comes from the order-$g^3$ amplitude square $|M_{3}(\mathbf{k})|^2$ and the other comes from the interference term of order-$g$ and order-$g^5$ amplitudes $M_1^{\ast}(\mathbf{k})M_5(\mathbf{k}) +c.c$, say also eqs.~\eqref{eq:M3M3_average_rhoP} and \eqref{eq:M1M5_average_rhoP}. They are plotted in Fig.~\ref{fig:dNdk_M1M5vsM3M3}. The contribution from $|M_3|^2$ is overall positive while the contribution from $M_1^{\ast}M_5+c.c$ is overall negative. The combined effect turns out to be positive. The sharply blowing up behavior in $M_1^{\ast}M_5+c.c.$ when $k_{\perp} \lesssim 0.6\,\rm{GeV}$ is not unexpected. Mathematically It is due to the second term in eq.~\eqref{eq:explicit_F3B}, which is related to the imaginary part the auxiliary function $\mathcal{I}_2$.

\begin{figure}[htp]
\centering 
\includegraphics[width = 0.65\textwidth]{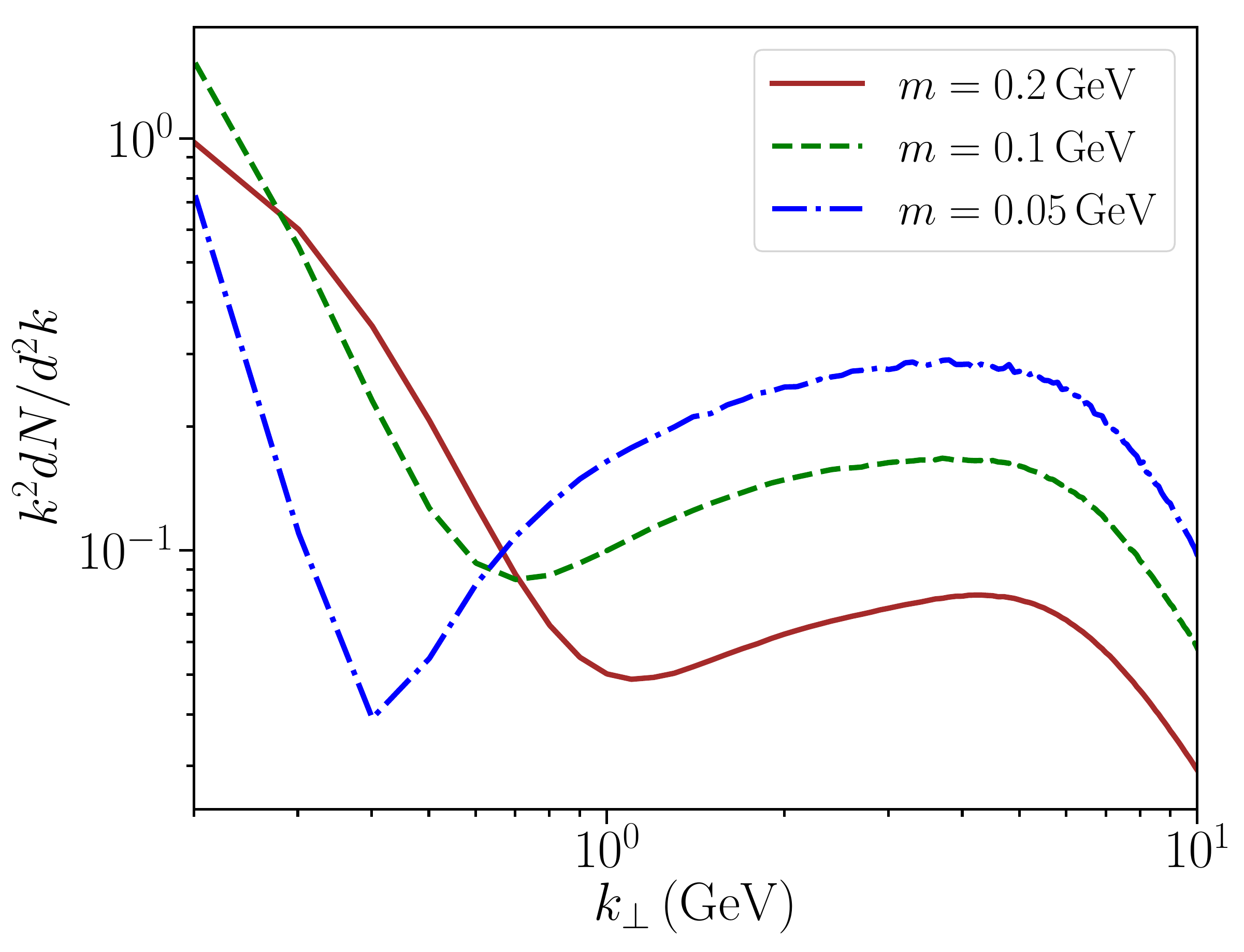}
\caption{Contributions from final state interactions within the first saturation corrections plotted for different infrared cutoff scales.}
\label{fig:dNdk_FS_differentm}
\end{figure}

In Fig.~\ref{fig:dNdk_FS_differentm}, contributions from the final state interaction part are plotted for different values of infrared cutoff scales. Depending on the infrared scales, the transitional regions $0.6\mathrm{Gev} \sim 6\mathrm{GeV}$ interpolating the small momentum and large momentum regions. For small momentum region, the gluon spectrum scales roughly $1/k^4$. For the large momentum region, it appears that the gluon spectrum also scales like $1/k^4$. However, this is an artifact of the Dipole Approximation which is supposed to be applicable for a dense target. To recover the $1/k^6$ scaling behavior at large momentum, one has to consider terms in the three-Wilson-line terms and terms from initial state interactions, essentially treating the target as dilute. Such a weak field expansion turns out rather involved in computation, we defer its study to a  future paper. 

\begin{figure}[htp]
\centering 
\includegraphics[width = 0.65\textwidth]{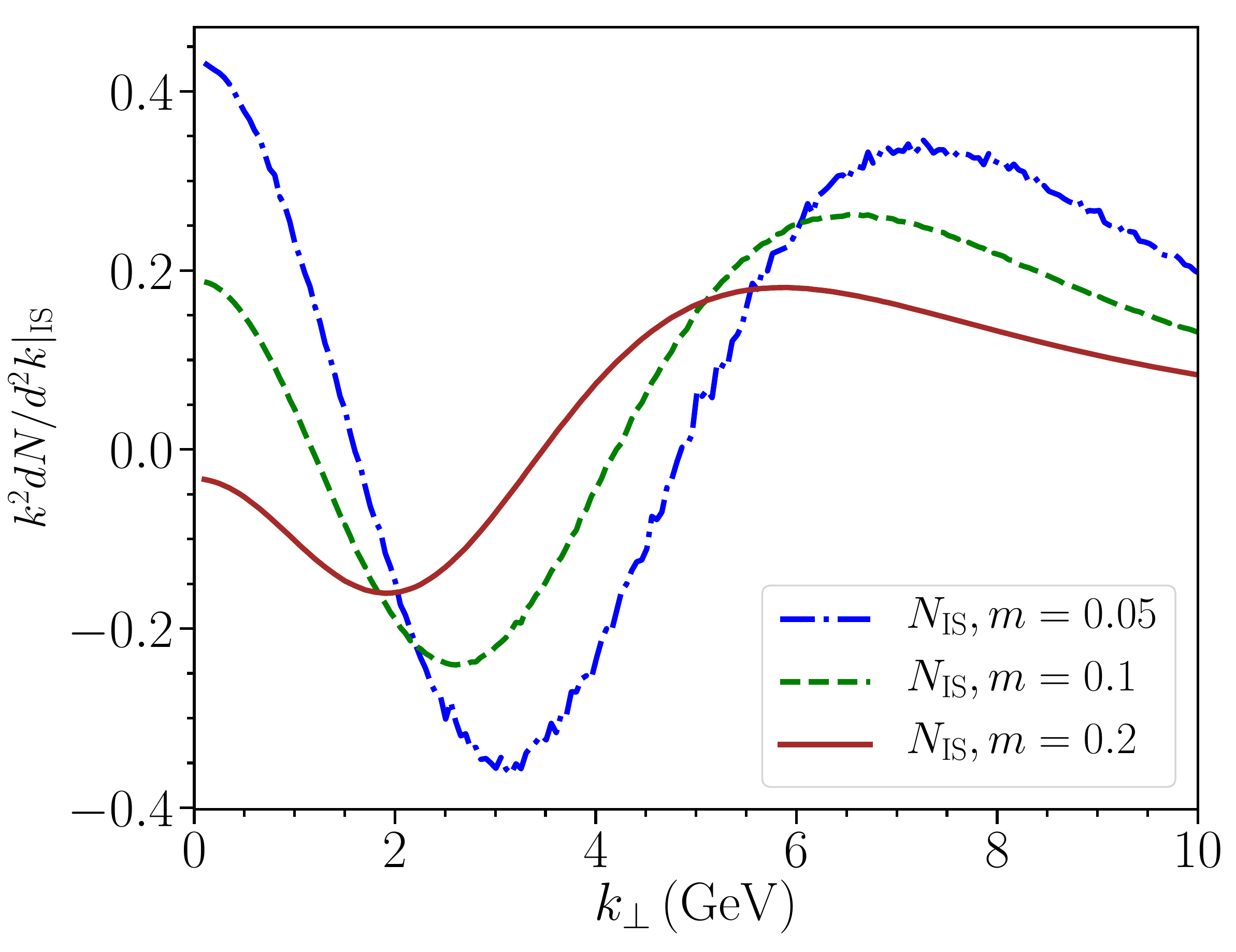}
\caption{Within the first saturation corrections, the initial state interactions are sensitive to the infrared scale cutoff $m$.}
\label{fig:dNdk_IS_m_dependence}
\end{figure} 


We would like to dwell a bit more on  the infrared cutoff sensitivity. In the numerical computations, the infrared cutoff enters by the replacement $1/p^2 \rightarrow 1/(p^2+m^2)$ for any momentum denominators in the integrand functions.   The leading order single gluon production given in eq.~\eqref{eq:LO_ensemble_average} turns out to be insensitive to the infrared cutoff (at most logarithmical dependence). This is also true for the all order resummed result of initial state interactions given in eq.~\eqref{eq:master_eq_IS}. However, expanding the all order resummed result to first order in saturation, eq.~\eqref{eq:FSC_IS_explicit} is sensitive to the infrared cutoff as shown in Fig.~\ref{fig:dNdk_IS_m_dependence}. The good news is that the first saturation correction from the initial state interaction has very small magnitude compared to the leading order result. The infrared scale sensitivity has negligible effects for initial state interactions. However, from a conceptual perspective, it seems an all order resummation is needed to alleviate the infrared scale sensitivity.  On the other hand, the infrared cutoff sensitivity for the final state interactions is more prominent as already demonstrated in Fig.~\ref{fig:dNdk_FS_differentm}. Unless there are more appropriate regularization schemes, it seems the only way to alleviate this infrared cutoff dependence is through an all order resummation.

\begin{figure}[htp]
\centering 
\includegraphics[width = 0.65\textwidth]{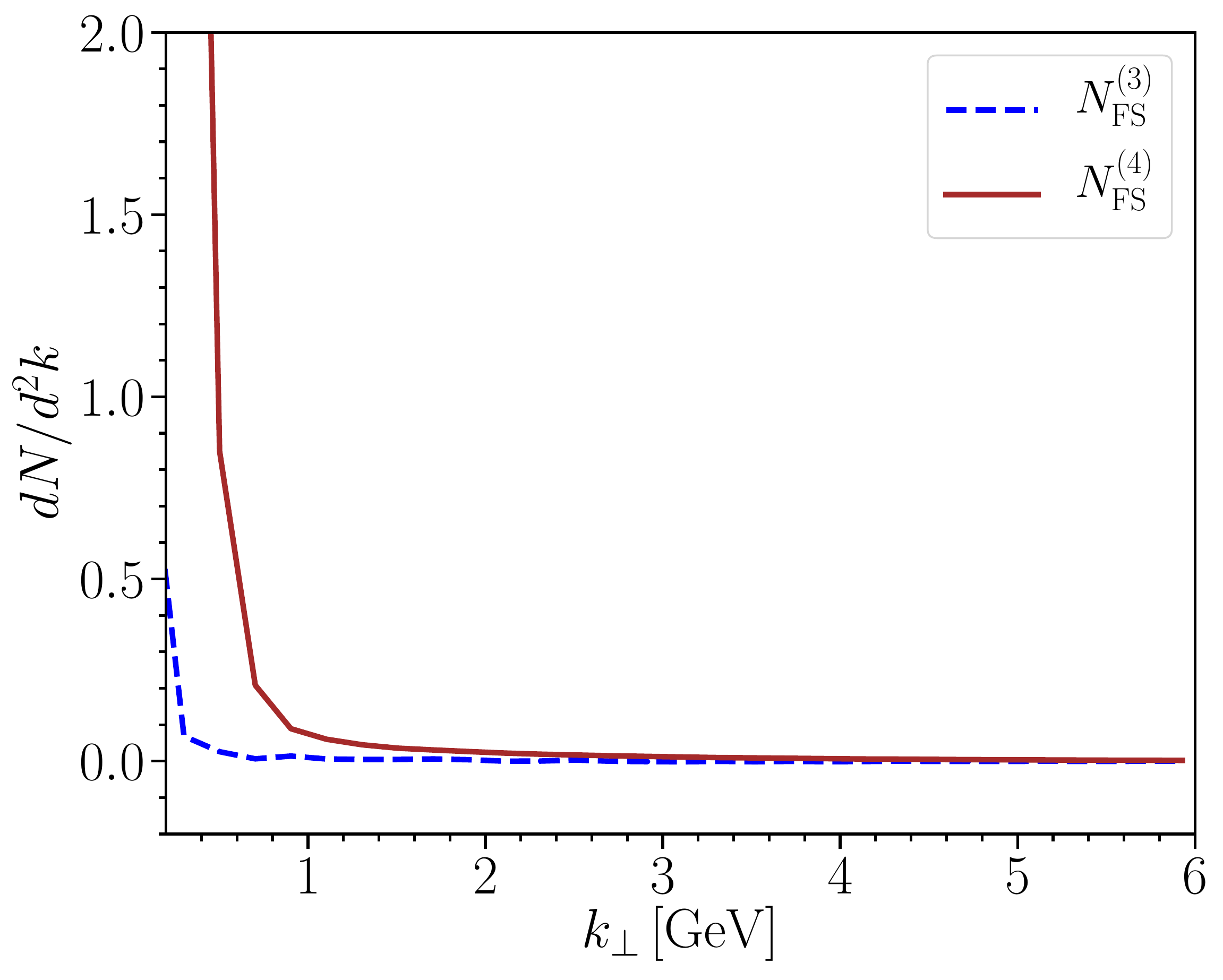}
\caption{Final state interactions from three-Wilson-line terms and four-Wilson-line terms. The target saturation scale is chosen to be $Q_{s,T} = 2\,\mathrm{GeV}$.}
\label{fig:dNdk_FS_3vs4}
\end{figure}

For completeness, we show evidence that the final state interactions is dominated by the $N_c$ leading four-Wilson-line correlator term, the first term in eq.~\eqref{eq:FSC_n(4)}. In Fig.~\ref{fig:dNdk_FS_3vs4}, $N^{(3)}_{\mathrm{FS}}(\mathbf{k})$ and $N^{(4)}_{\mathrm{FS}}(\mathbf{k})$ are plotted with the target saturation scale set to be $Q_{s,T} =2\, \mathrm{GeV}$. In this plot, when computing $N^{(3)}_{\mathrm{FS}}(\mathbf{k})$ and $N^{(4)}_{\mathrm{FS}}(\mathbf{k})$,  we explicitly averaged over $N_{\mathrm{conf}} =100$ configurations. The three-Wilson-line terms contribute positively to the gluon spectrum but is negligible compared to the four-Wilson-line terms (numerically only about $1\%$).  Furthermore, within the four-Wilson-line terms $N^{(4)}_{\mathrm{FS}}(\mathbf{k})$ given in eq.~\eqref{eq:FSC_n(4)}, the first term is the dominant one as other terms are large-$N_c$ suppressed. This observation is also confirmed by direct numerical computations for $N_c=3$. We therefore conclude that the final state interaction is dominated by the four-Wilson line terms, particularly the $N_c$ leading term. 

\section{Conclusions and Outlooks}
In this paper, we continued studying the first saturation correction to the single inclusive semi-hard gluon production in high energy proton-nucleus collisions. Based on the configuration-by-configuration expression, we  performed ensemble averaging over both the projectile and target color charge configurations using the MV model and the Dipole Approximation.  We demonstrated numerically that the Dipole Approximation, within its validity region, provides surprisingly good approximation to the four-Wilson-line correlators. Specifically, in momentum space, the four-Wilson-line correlators satisfy factorization relations in terms of the dipole operators.  Unlike the Glasma graph approximation \cite{Dusling:2012iga, Dusling:2012cg,Dusling:2012wy,Dusling:2013oia} or expansions based on large-$N_c$-closed form expression \cite{Kovchegov:2012nd},   which are related to  the dilute-dilute scattering regime,  the Dipole Approximation provides an avenue to isolate the dominant contributions analytically when the target is dense.

For the single inclusive gluon spectrum, the leading order result predicts scalings  $1/k_{\perp}^2$ at small $k_{\perp}$ and $1/k_{\perp}^4$ at large $k_{\perp}$. The first saturation correction in the projectile is expected to have the scalings  $1/k_{\perp}^4$ at small $k_{\perp}$ and $1/k_{\perp}^6$ at large $k_{\perp}$. The semi-hard momentum region $Q_{s,P}\ll k_{\perp} \ll Q_{s,T}$ interpolates between these two limits.  In the small momentum region $k_{\perp}\ll Q_{s,P}, Q_{s,T}$, higher order saturation corrections diverge even more severely at the infrared. No valuable information can be extracted at fixed order saturation corrections, requesting an all order resummation to obtain infrared finite result. For $Q_{s,P}\ll k_{\perp} \ll Q_{s,T}$, we demonstrated numerically that the the first saturation correction gives a positive contribution to  the gluon spectrum. We also found that within the first saturation correction, the major contribution comes from the final state interaction while the contribution due to the initial state interaction is negligible. We further isolated the dominant term in the final state interaction, which is consistent with expectations from the Dipole Approximation. Specifically, we showed that the terms with  a three-Wilson-line correlator are small compared to the terms with four-Wilson-line correlators. For the terms with four-Wilson-line correlators, the leading contribution coincides with the contribution obtained in the large-$N_c$ limit.

We also observed  that the first saturation correction is sensitive to the infrared cutoff when simply regularizing the momentum denominator. 
For initial state interactions, one could obtain an all order resummed result, which turns out to be insensitive to the infrared cutoff. Expanding it to the first order in saturation correction explicitly demonstrates the origin of the sensitivity to the infrared cutoff. We thus conclude that one has to perform an all order resummation to alleviate this problem. 

There are several avenues to pursue regarding the observable single inclusive gluon production. For example, to include small-x evolution and running coupling  effects \cite{Jalilian-Marian:2005ccm} as were done for the leading order result. However, the most urgent problem is to find some kind of resummation of all orders of the saturation corrections in eq.~\eqref{Eq:Expan}.  To discern  a possible resumption pattern, one would have to compute the second saturation correction. This would be extremely interesting and straightforward, but technically rather tedious.  

There are other interesting observables, e.g.  the double inclusive gluon productions, and   particularly the odd harmonics of two gluon correlations, in which the first saturation correction is the {\it leading} contribution~\cite{McLerran:2016snu, Kovchegov:2018jun}. It would be very interesting to analyze how the initial state interactions and final state interactions contribute to the  angular correlations.

\acknowledgments
We thank A. Kovner and Y. Kovchegov for insightful discussions. We thank Jack Featherstone for collaborating on establishing the region of validity of  the Dipole Approximation. We acknowledge computational resources from the high performance computing facilities at NC State University. 
We acknowledge support by the DOE Office of Nuclear Physics through Grant No. DE-SC0020081.\\

\appendix
\section{Initial state interaction: all order resummation}
\label{appendixA}
The initial conditions in the light-cone like subgauge are 
\begin{equation}
\begin{split}
&\beta^b(\tau=0, \mathbf{x}_{\perp}) = \frac{1}{2} \alpha_P^{a,i}(\mathbf{x}_{\perp}) \partial^i U^{ab}(\mathbf{x}_{\perp}),\\
&\beta^{b,i}(\tau=0, \mathbf{x}_{\perp}) = \alpha_P^{a,i}(\mathbf{x}_{\perp})U^{ab}(\mathbf{x}_{\perp}).\\
\end{split}
\end{equation}
The surface term for the two gluon polarizations, after performing the   Lehmann-Symanzik-Zimmermann reduction formula, are 
\begin{equation}
\mathfrak{S}_{\eta}(\mathbf{k}) = \frac{1}{\sqrt{\pi}k_{\perp}} 2\beta(\tau=0, \mathbf{k}), \quad \mathfrak{S}_{\perp}(\mathbf{k}) = \frac{1}{\sqrt{\pi}}\beta_{\perp}(\tau=0, \mathbf{k}).
\end{equation}
with 
\begin{equation}
\beta_{\perp}(\tau, \mathbf{k})=\frac{-i\epsilon_{ij}\mathbf{k}_i}{k_{\perp}} \beta_j(\tau, \mathbf{k}). 
\end{equation}
Then the single gluon production, originating from   initial state interaction, is computed as 
\begin{equation}
\begin{split}
\frac{dN}{d^2\mathbf{k}} = &\frac{1}{(2\pi)^2} \Big(|\mathfrak{S}_{\eta}(\mathbf{k})|^2 + |\mathfrak{S}_{\perp}(\mathbf{k})|^2\Big)\\
=&\frac{1}{(2\pi)^2}\frac{1}{\pi k^2_{\perp}}\int \frac{d^2\mathbf{p}}{(2\pi)^2}\frac{d^2\mathbf{q}}{(2\pi)^2}\Big[(\mathbf{k}-\mathbf{q})^i(\mathbf{k}+\mathbf{p})^j+k_{\perp}^2\delta^{ij}-\mathbf{k}^i\mathbf{k}^j\Big] \\
&\qquad \times\alpha_P^{a,i}(\mathbf{q})\alpha_P^{c,j}(\mathbf{p})U^{ab}(\mathbf{k}-\mathbf{q})U^{cb}(-\mathbf{k}-\mathbf{p})\\
\end{split}
\end{equation}
Averaging over the target Wilson line using the MV model gives
\begin{equation}
\langle U^{ab}(\mathbf{k}-\mathbf{q})U^{cb}(-\mathbf{k}-\mathbf{p})\rangle  = \delta^{ac} (2\pi)^2\delta(\mathbf{p}+\mathbf{q}) D(\mathbf{k}-\mathbf{q}).
\end{equation}
Here the dipole correlator in the adjoint representation is defined as 
\begin{equation}
D(\mathbf{k}) =\frac{1}{S_{\perp}} \frac{1}{N_c^2-1} \langle \mathrm{Tr}[U(\mathbf{k})U^T(-\mathbf{k})]\rangle . 
\end{equation}
Averaging over the projectile WW field correlator leads to 
\begin{equation}
xG^{ij}_{WW}(\mathbf{q}) = \frac{1}{S_{\perp}}\Big\langle \alpha_P^{a,i}(\mathbf{q})\alpha^{a,j}_P(-\mathbf{q})\Big\rangle. 
\end{equation}
As a result, the final result for the single gluon production from initial state interaction becomes 
\begin{equation}
\frac{dN}{d^2\mathbf{k}}\Big|_{\mathrm{IS}} = \frac{1}{(2\pi)^2}\frac{S_{\perp}}{\pi k_{\perp}^2}\int\frac{d^2\mathbf{q}}{(2\pi)^2}[ (\mathbf{k}-\mathbf{q})^i(\mathbf{k}-\mathbf{q})^j +(k_{\perp}^2\delta^{ij} -\mathbf{k}^i\mathbf{k}^j)]  xG_{WW}^{ij}(\mathbf{q}) D_T(\mathbf{k}-\mathbf{q})
\end{equation}
It is expressed as a convolution of  projectile WW field correlator and target dipole correlator.
It is straightforward  to verify that at the lowest order $\alpha_P^{a,i}(\mathbf{q}) = \frac{i\mathbf{q}^i}{q_{\perp}^2} \rho^a(\mathbf{q})$ and eq.~\eqref{eq:master_eq_IS} reduces to the well-known leading order single gluon production formula.\\

The WW field correlator has been extensively studied in the literature. 
Usually, $xG_{WW}^{ij}(\mathbf{q})$ is decomposed into finite trace and traceless part \cite{Metz:2011wb, Dominguez:2011br}
\begin{equation}
xG^{ij}_{WW}(\mathbf{q}) = \frac{1}{2} \delta^{ij} xG^{(1)}(\mathbf{q}) + \left(\frac{\mathbf{q}^i\mathbf{q}^j}{q_{\perp}^2} - \frac{1}{2}\delta^{ij}\right)xh^{(1)}(\mathbf{q})
\end{equation}
So that 
\begin{equation}
\begin{split}
&xG^{(1)}(\mathbf{q}) = \delta^{ij} xG^{ij}_{WW}(\mathbf{q}),\\
&xh^{(1)}(\mathbf{q}) =2 \left(\frac{\mathbf{q}^i\mathbf{q}^j}{q_{\perp}^2} - \frac{1}{2}\delta^{ij}\right) xG^{ij}_{WW}(\mathbf{q})
\end{split}
\end{equation}
In terms of the two distribution functions, the single gluon production becomes 
\begin{equation}
\begin{split}
\frac{dN}{d^2\mathbf{k}}\Big|_{\mathrm{IS}} = \frac{S_{\perp}}{(2\pi)^2\pi k_{\perp}^2}\int\frac{d^2\mathbf{q}}{(2\pi)^2}\left\{ \frac{1}{2}[(\mathbf{k}-\mathbf{q})^2 + k_{\perp}^2]xG^{(1)}(\mathbf{q}) + \frac{1}{2}[(\mathbf{k}-\mathbf{q})^2-k_{\perp}^2]xh^{(1)}(\mathbf{q})\right\}D_T(\mathbf{k}-\mathbf{q})
\end{split}
\end{equation}
The closed form expression of the WW field correlator is
\begin{equation}
\Big\langle \alpha_P^{a,i}(\mathbf{x}_{\perp}) \alpha_P^{b,j}( \mathbf{y}_{\perp}) \Big\rangle =\frac{-\delta^{ab}\partial^i_{\mathbf{r}}\partial^j_{\mathbf{r}} L(\mathbf{r})}{\frac{1}{2} C_A g^2 \Gamma(\mathbf{r})}  \left[e^{\frac{1}{2}C_A g^4\bar{\mu}^2 \Gamma(\mathbf{r})} -1\right] 
\end{equation}
with the functions 
\begin{equation}
-\partial^i_{\mathbf{r}}\partial^j_{\mathbf{r}} L(\mathbf{r}) = \int \frac{d^2\mathbf{p}}{(2\pi)^2} e^{-i\mathbf{p}\cdot\mathbf{r}} \frac{\mathbf{p}^i\mathbf{p}^j}{p_{\perp}^4}. 
\end{equation}
and 
\begin{equation}
\Gamma(\mathbf{r}) = 2L(\mathbf{r})-2L(0) = 2\int \frac{d^2\mathbf{p}}{(2\pi)^2} [e^{-i\mathbf{p}\cdot\mathbf{r}}-1] \frac{1}{p_{\perp}^4}=\frac{1}{2\pi} \frac{1}{m^2} \Big[-1 + (mr) K_1(mr)\Big]
\end{equation}
Here the IR scale $m$ enters by replacing the propagator $1/p^2$ with $1/(p^2+m^2)$. $K_1(x)$ is the modified Bessel function of second kind.  The definition of saturation scale $Q_s^2 = \frac{1}{2\pi} N_c g^4 \bar{\mu}^2$. 
The two distribution functions can be computed by the formula 
\begin{equation}
\begin{split}
&xG^{(1)}(q) 
=\frac{4\pi(N_c^2-1)}{N_c g^2} \int_0^{\infty}rdr J_0(qr) \frac{K_0(mr)}{\frac{1}{m^2}[1-(mr)K_1(mr)]} \Big[ 1 -e^{-\frac{1}{2}\frac{Q_s^2}{m^2}[1 - (mr)K_1(mr)]}\Big],\\
&xh^{(1)}(q) 
=\frac{4\pi (N_c^2-1)}{N_c g^2}\int_0^{\infty} rdr J_2(qr) \frac{ \frac{2}{(mr)^2} -K_2(mr)}{\frac{1}{m^2}[1 - (mr)K_1(mr)]} \Big[ 1 -e^{-\frac{1}{2}\frac{Q_s^2}{m^2}[1 - (mr)K_1(mr)]}\Big].\\
\end{split}
\end{equation}
One can expand the modified Bessel functions in the regime $mr\ll 1$ and recover the expressions derived in the literature \cite{Metz:2011wb, Dominguez:2011br}.\\

We would like to expand perturbatively in terms of projectile saturation scale $Q_{s,P}^2$. 
\begin{equation}\label{eq:WW_expansion}
\begin{split}
xG_{\mathrm{WW}}^{ij}(\mathbf{q}) = &\int d^2\mathbf{r}e^{i\mathbf{q}\cdot\mathbf{r}}\Big\langle \alpha_P^{a,i}(\mathbf{x}_{\perp}) A^{a,j}_P( \mathbf{y}_{\perp}) \Big\rangle\\
=&\frac{2(N_c^2-1)}{N_cg^2} \int d^2\mathbf{r}e^{i\mathbf{q}\cdot\mathbf{r}}\frac{-\partial^i_{\mathbf{r}}\partial^j_{\mathbf{r}} L(\mathbf{r})}{\Gamma(\mathbf{r})} \Big[ e^{\pi Q_{s,P}^2 \Gamma(\mathbf{r})}-1\Big]\\
=&\frac{2(N_c^2-1)}{N_cg^2} \int d^2\mathbf{r}e^{i\mathbf{q}\cdot\mathbf{r}} \int \frac{d^2\mathbf{p}}{(2\pi)^2} e^{-i\mathbf{p}\cdot\mathbf{r}} \frac{\mathbf{p}^i\mathbf{p}^j}{p_{\perp}^4}\Big[ \pi Q_{s,P}^2 + \frac{1}{2}\pi^2Q_{s,P}^4 \Gamma(\mathbf{r})  + \ldots\Big]\,.
\end{split}
\end{equation}
The leading order WW field correlator is 
\begin{equation}
xG_{\mathrm{WW}}^{ij}(\mathbf{q})\Big|_{\mathrm{LO}} = \frac{(N_c^2-1)}{2N_c \alpha_s} Q_{s,P}^2 \frac{\mathbf{q}^i\mathbf{q}^j}{q^4}\,.
\end{equation}
From it one reproduce the leading order single gluon production
\begin{equation}
\frac{dN}{d^2\mathbf{k}}\Big|_{\mathrm{LO}} 
=\frac{1}{\alpha_s}\frac{S_{\perp}}{(2\pi)^3} \frac{N_c^2-1}{N_c} \frac{Q_{s,P}^2}{k^2} \int\frac{d^2\mathbf{q}}{(2\pi)^2}\frac{(\mathbf{k}-\mathbf{q})^2}{q^2} D_T(\mathbf{k}-\mathbf{q}).
\end{equation}
The next-to-leading order term in the WW field correlator is 
\begin{equation}
xG_{\mathrm{WW}}^{ij}(\mathbf{q})\Big|_{\mathrm{NLO}} = 
\frac{\pi (N_c^2-1)}{2N_c \alpha_s} Q_{s,P}^4\int \frac{d^2\mathbf{q}_1}{(2\pi)^2}\Big[\frac{(\mathbf{q}-\mathbf{q}_1)^i(\mathbf{q}-\mathbf{q}_1)^j}{|\mathbf{q}-\mathbf{q}_1|^4q_1^4} - \frac{\mathbf{q}^i\mathbf{q}^j}{q^4q_1^4}\Big] \,.
\end{equation}
And the next-to-leading order single gluon production is 
\begin{equation}
\frac{dN}{d^2\mathbf{k}}\Big|_{\mathrm{NLO}} 
=\frac{1}{\alpha_s}\frac{S_{\perp}}{(2\pi)^3} \frac{N_c^2-1}{N_c} \frac{Q_{s,P}^4}{k^2} (2\pi) \int \frac{d^2\mathbf{q}}{(2\pi)^2} \frac{d^2\mathbf{q}_1}{(2\pi)^2} \mathsf{G}_1(\mathbf{q}, \mathbf{q}_1, \mathbf{k}) D_T(\mathbf{k}-\mathbf{q}).
\end{equation}
with 
\begin{equation}
 \mathsf{G}_1(\mathbf{q}, \mathbf{q}_1, \mathbf{k})  = \frac{1}{2} \left(\frac{[(\mathbf{k}-\mathbf{q})\cdot(\mathbf{q}-\mathbf{q}_1)]^2 + [\mathbf{k} \times (\mathbf{q}-\mathbf{q}_1)]^2}{|\mathbf{q}-\mathbf{q}_1|^4q_1^4} - \frac{(\mathbf{k}-\mathbf{q})^2}{q^2q_1^4}\right). 
\end{equation}

\section{Perturbative Solutions of WW Field}
\label{appendixB}
In this appendix, we compute the perturbative solutions of WW field keeping the longitudinal coordinate explicit. Path ordering along the longitudinal direction is important for computations involving initial state interactions. The light-like Wilson line in the fundamental representation is 
\begin{equation}
V(x^-, \mathbf{x}) = \mathcal{P} \mathrm{exp}\left\{ig^2 \int_{-\infty}^{x^-} dz^-\Phi^a(z^-, \mathbf{x})t^a\right\}\,.
\end{equation}
Here the gauge potential field is related to the covariant gauge color charge density by $\partial^2 \Phi^a(x^-, \mathbf{x}) = \rho^a_P(x^-, \mathbf{x})$. The WW field is defined as 
\begin{equation}
\alpha_P^i(x^-, \mathbf{x})  = \frac{i}{g} V(x^-, \mathbf{x})  \partial^i V^{\dagger}(x^-, \mathbf{x})\,.
\end{equation}
We would like to perturbatively expand the light-like Wilson line and thus obtain the $\alpha_P^i$ perturbatively.
\begin{equation}\label{eq:alphaP(1)}
\alpha_{P, (1)}^i(x^-, \mathbf{x}) =  \int_{-\infty}^{x^-} dy_1^- \partial^i\Phi^{b}(y_1^-, \mathbf{x}) t^b. 
\end{equation}
\begin{equation}
\alpha_{P, (3)}^i(x^-, \mathbf{x}) 
=i\int_{-\infty}^{x^-} dy_1^- \int_{-\infty}^{y_1^-}dy_2^- \partial^i\Phi^{b_1}(y_1^-, \mathbf{x}) \Phi^{b_2}(y_2^-,\mathbf{x}) if^{b_2b_1b}t^b.
\end{equation}
The fifth order solution 
\begin{equation}\label{eq:alphaP(5)}
\begin{split}
&\alpha_{P, (5)}^i(x^-, \mathbf{x})\\
=& \int_{-\infty}^{x^-} dy_1^-\int_{-\infty}^{y_1^-} dy_2^-\int_{-\infty}^{y_2^-} dy_3^-\partial^i[\Phi^{b_1}(y_1^-, \mathbf{x})\Phi^{b_2}(y_2^-, \mathbf{x})\Phi^{b_3}(y_3^-, \mathbf{x})]t^{b_1}t^{b_2}t^{b_3}\\
&- \int_{-\infty}^{x^-} dy_1^- \int_{-\infty}^{x^-}dy_2^- \int_{-\infty}^{y_2^-}dy_3^- \Phi^{b_1}(y_1^-, \mathbf{x}) \partial^i [\Phi^{b_2}(y_2^-,\mathbf{x})\Phi^{b_1}(y_3^-, \mathbf{x})] t^{b_1}t^{b_2}t^{b_3}\\
&+ \int_{-\infty}^{x^-} dy_1^- \int_{-\infty}^{y_1^-} dy_2^- \int_{-\infty}^{x^-} dy_3^- \Phi^{b_1}(y_1^-, \mathbf{x}) \Phi^{b_2}(y_2^-, \mathbf{x}) \partial^i \Phi^{b_3}(y_3^-,\mathbf{x}) t^{b_1}t^{b_2}t^{b_3}.
\end{split}
\end{equation}

Using these explicit expressions, one can compute the WW field correlator order-by-order.
\begin{equation}\label{eq:alphaP(3)alphaP(3)}
\begin{split}
&\left\langle \alpha_{P,(3)}^{a,i}(x^-, \mathbf{x}) \alpha_{P, (3)}^{a,j}(y^-, \mathbf{y})\right\rangle \\
=& -  \int_{-\infty}^{x^-} dy_1^- \int_{-\infty}^{y_1^-}dy_2^-\int_{-\infty}^{y^-} dz_1^- \int_{-\infty}^{z_1^-}dz_2^-\langle \partial^i\Phi^{b_1}(y_1^-, \mathbf{x}) \Phi^{b_2}(y_2^-,\mathbf{x})\partial^j\Phi^{c_1}(z_1^-, \mathbf{y}) \Phi^{c_2}(z_2^-,\mathbf{y})\rangle \\
&\qquad \times if^{b_1b_2b}if^{c_1c_2c}2\mathrm{Tr}(t^bt^c)\\
=&N_c(N_c^2-1) \int_{-\infty}^{x^-} dy_1^- \int_{-\infty}^{y_1^-}dy_2^-\mu^2(y_1^-)\mu^2(y_2^-)\partial^i_{\mathbf{x}}\partial^j_{\mathbf{y}}L(\mathbf{x}, \mathbf{y})L(\mathbf{x}, \mathbf{y})\\
=&N_c(N_c^2-1) \frac{1}{2}\bar{\mu}^4 \partial^i_{\mathbf{x}}\partial^j_{\mathbf{y}}L(\mathbf{x}, \mathbf{y})L(\mathbf{x}, \mathbf{y}).
\end{split}
\end{equation}
In obtaining the second equality, there is only one possible contraction allowed in the ensemble averaging due to path orderings and we used the expression
\begin{equation}
\begin{split}
&\langle \partial^i\Phi^{b_1}(y_1^-, \mathbf{x}) \Phi^{b_2}(y_2^-,\mathbf{x})\partial^j\Phi^{c_1}(z_1^-, \mathbf{y}) \Phi^{c_2}(z_2^-,\mathbf{y})\rangle \\
=&\langle \partial^i\Phi^{b_1}(y_1^-, \mathbf{x}) \partial^j\Phi^{c_1}(z_1^-, \mathbf{y}) \rangle \langle \Phi^{b_2}(y_2^-,\mathbf{x})\Phi^{c_2}(z_2^-,\mathbf{y})\rangle\\
=&\delta^{b_1c_1} \delta(y_1^--z_1^-) \mu^2(y^-_1) \partial^i_{\mathbf{x}}\partial^j_{\mathbf{y}}L(\mathbf{x}, \mathbf{y})\delta^{b_2c_2}\delta(y_2^--z_2^-) \mu^2(y_2^-) L(\mathbf{x}, \mathbf{y}).\\
\end{split}
\end{equation}

The other correlator at order-$g^6$ is 
\begin{equation}
\begin{split}
&\left\langle \alpha_{P, (1)}^{a, j}(y^-, \mathbf{y}) \alpha_{P,(5)}^{a, i}(x^-, \mathbf{x})\right\rangle .\\
\end{split}
\end{equation}
Combining eq.~\eqref{eq:alphaP(1)} and eq.~\eqref{eq:alphaP(5)}, there are  three terms, we analyze each one separately. For the first one 
\begin{equation}
\begin{split}
&i \int_{-\infty}^{y^-} dz^- \partial^j \Phi^{b}(z^-, \mathbf{y})\int_{-\infty}^{x^-} dy_1^-\int_{-\infty}^{y_1^-} dy_2^-\int_{-\infty}^{y_2^-} dy_3^-\partial^i[\Phi^{b_1}(y_1^-, \mathbf{x})\Phi^{b_2}(y_2^-, \mathbf{x})\Phi^{b_3}(y_3^-, \mathbf{x})]\\
&\qquad \times 2\mathrm{Tr}[t^bt^{b_1}t^{b_2}t^{b_3}]\\
=&i  \frac{(N_c^2-1)^2}{N_c} \partial^j_{\mathbf{y}}\partial^i_{\mathbf{x}} L(\mathbf{x},\mathbf{y}) L(\mathbf{x},\mathbf{x})\frac{1}{4} \bar{\mu}^4
\end{split}
\end{equation}
In the ensemble averaging, there are only two possible contractions allowed due to path orderings 
\begin{equation}
\begin{split}
&\langle\partial^j \Phi^{b}(z^-, \mathbf{y}) \partial^i[\Phi^{b_1}(y_1^-, \mathbf{x})\Phi^{b_2}(y_2^-, \mathbf{x})\Phi^{b_3}(y_3^-, \mathbf{x})] \rangle \\
=&\partial^j_{\mathbf{y}} \partial^i_{\mathbf{x}}\Big[ \delta^{bb_1} \delta(z^--y_1^-) \mu^2(y_1^-) L(\mathbf{x},\mathbf{y}) \delta^{b_2b_3}\delta(y_2^--y_3^-)\mu^2(y_2^-) L(\mathbf{x}, \mathbf{x}) \\
&+ \delta^{bb_3}\delta(z^--y_3^-)\mu^2(y_3^-) L(\mathbf{x}, \mathbf{y}) \delta^{b_1b_2}\delta(y_1^--y_2^-) \mu^2(y_2^-)L(\mathbf{x}, \mathbf{x})\Big].
\end{split}
\end{equation}
The second term is 
\begin{equation}
\begin{split}
&-i\int_{-\infty}^{y^-} dz^- \partial^j \Phi^{b}(z^-, \mathbf{y})  \int_{-\infty}^{x^-} dy_1^- \int_{-\infty}^{x^-}dy_2^- \int_{-\infty}^{y_2^-}dy_3^- \Phi^{b_1}(y_1^-, \mathbf{x}) \partial^i [\Phi^{b_2}(y_2^-,\mathbf{x})\Phi^{b_3}(y_3^-, \mathbf{x})]\\
&\qquad\times  2\mathrm{Tr}[t^bt^{b_1}t^{b_2}t^{b_3}]\\
=&i \frac{(N_c^2-1)}{2N_c} \partial^j_{\mathbf{y}}\partial^i_{\mathbf{x}} L(\mathbf{x},\mathbf{y}) L(\mathbf{x},\mathbf{x})\frac{1}{2} \bar{\mu}^4 -i  \frac{(N_c^2-1)^2}{2N_c} \partial^j_{\mathbf{y}}\partial^i_{\mathbf{x}} L(\mathbf{x},\mathbf{y}) L(\mathbf{x},\mathbf{x})\frac{1}{2} \bar{\mu}^4\,.
\end{split}
\end{equation}
Again, there are only two contractions allowed in obtaining the above expression
\begin{equation}
\begin{split}
&\langle \partial^j \Phi^{b}(z^-, \mathbf{y})\Phi^{b_1}(y_1^-, \mathbf{x}) \partial^i [\Phi^{b_2}(y_2^-,\mathbf{x})\Phi^{b_3}(y_3^-, \mathbf{x})]\rangle \\
=&\langle \partial^j \Phi^{b}(z^-, \mathbf{y})\partial^j \Phi^{b_2}(y_2^-,\mathbf{x})\rangle \langle \Phi^{b_1}(y_1^-, \mathbf{x}) \Phi^{b_3}(y_3^-, \mathbf{x})\rangle\\
&+\langle \partial^i \Phi^{b}(z^-, \mathbf{y})\partial^j\Phi^{b_3}(y_3^-, \mathbf{x})\rangle \rangle \langle \Phi^{b_1}(y_1^-, \mathbf{x}) \Phi^{b_2}(y_2^-,\mathbf{x})\rangle \\
=&\partial_{\mathbf{y}}^j\partial^i_{\mathbf{x}} L(\mathbf{x},\mathbf{y})L(\mathbf{x},\mathbf{x}) \Big[ \delta^{bb_2} \delta(z^--y_2^-) \mu^2(z^-) \delta^{b_1b_3}\delta(y_1^--y_3^-)\mu^2(y_1^-) \\
&+ \delta^{bb_3}\delta(z^--y_3^-)\mu^2(y_3^-) \delta^{b_1b_2}\delta(y_1^--y_2^-) \mu^2(y_2^-)\Big]\,.
\end{split}
\end{equation}
The third term is 
\begin{equation}
\begin{split}
&i \int_{-\infty}^{y^-} dz^- \partial^j \Phi^{b}(z^-, \mathbf{y}) t^b  \int_{-\infty}^{x^-} dy_1^- \int_{-\infty}^{y_1^-} dy_2^- \int_{-\infty}^{x^-} dy_3^- \Phi^{b_1}(y_1^-, \mathbf{x}) \Phi^{b_2}(y_2^-, \mathbf{x}) \partial^i \Phi^{b_3}(y_3^-,\mathbf{x})\\
&\qquad\times 2\mathrm{Tr}[t^b t^{b_1}t^{b_2}t^{b_3}]\\
=&i  \frac{(N_c^2-1)^2}{2N_c} \partial^i_{\mathbf{y}}\partial^j_{\mathbf{x}} L(\mathbf{x},\mathbf{y}) L(\mathbf{x},\mathbf{x})\frac{1}{2} \bar{\mu}^4\,.
\end{split}
\end{equation}
This time there is only one contraction that is non-vanishing 
\begin{equation}
\begin{split}
&\langle \partial^j \Phi^{b}(z^-, \mathbf{y})\Phi^{b_1}(y_1^-, \mathbf{x}) \Phi^{b_2}(y_2^-, \mathbf{x}) \partial^i \Phi^{b_3}(y_3^-,\mathbf{x})\rangle\\
=&\langle \partial^j \Phi^{b}(z^-, \mathbf{y})\partial^i \Phi^{b_3}(y_3^-,\mathbf{x})\rangle\langle \Phi^{b_1}(y_1^-, \mathbf{x}) \Phi^{b_2}(y_2^-, \mathbf{x}) \rangle \\
=&\partial^j_{\mathbf{y}}\partial^i_{\mathbf{x}}L(\mathbf{x},\mathbf{y}) L(\mathbf{x},\mathbf{x}) \delta^{bb_3}\delta^{b_1b_2}\delta(z^--y_3^-)\mu^2(y_3^-) \delta(y_1^--y_2^-)\mu^2(y_1^-)\,.
\end{split}
\end{equation}
Adding these three terms and taking into account the complex conjugate part
\begin{equation}\label{eq:alphaP(1)alphaP(5)}
\begin{split}
&\left\langle \alpha_{P, (1)}^{a, j}(y^-, \mathbf{y}) \alpha_{P,(5)}^{a, i}(x^-, \mathbf{x})\right\rangle + \left\langle \alpha_{P, (5)}^{a, j}(y^-, \mathbf{y}) \alpha_{P,(1)}^{a, i}(x^-, \mathbf{x})\right\rangle\\
=&-  N_c(N_c^2-1)\partial^j_{\mathbf{y}}\partial^i_{\mathbf{x}} L(\mathbf{x},\mathbf{y}) L(\mathbf{x},\mathbf{x})\frac{1}{2} \bar{\mu}^4.\\
\end{split}
\end{equation}
Combining eq.~\eqref{eq:alphaP(3)alphaP(3)} and eq.~\eqref{eq:alphaP(1)alphaP(5)}, one obtains 
\begin{equation}
\frac{1}{4}g^6\bar{\mu}^4N_c(N_c^2-1) \partial^i_{\mathbf{x}}\partial^j_{\mathbf{y}} L(\mathbf{x},\mathbf{y}) \Gamma(\mathbf{r})\,.
\end{equation}
This direct computation precisely reproduces the expanded expression in eq.~\eqref{eq:WW_expansion}.

\bibliography{spires}

\providecommand{\href}[2]{#2}\begingroup\raggedright\begin{thebibliography}{10}

\bibitem{Li:2021zmf}
M.~Li and V.V.~Skokov, \emph{{First saturation correction in high energy
  proton-nucleus collisions. Part I. Time evolution of classical Yang-Mills
  fields beyond leading order}},
  \href{https://doi.org/10.1007/JHEP06(2021)140}{\emph{JHEP} {\bfseries 06}
  (2021) 140} [\href{https://arxiv.org/abs/2102.01594}{{\ttfamily
  2102.01594}}].

\bibitem{Li:2021yiv}
M.~Li and V.V.~Skokov, \emph{{First saturation correction in high energy
  proton-nucleus collisions. Part II. Single inclusive semi-hard gluon
  production}}, \href{https://doi.org/10.1007/JHEP06(2021)141}{\emph{JHEP}
  {\bfseries 06} (2021) 141}
  [\href{https://arxiv.org/abs/2104.01879}{{\ttfamily 2104.01879}}].

\bibitem{McLerran:2001sr}
L.D.~McLerran, \emph{{The Color glass condensate and small x physics: Four
  lectures}}, \href{https://doi.org/10.1007/3-540-45792-5_8}{\emph{Lect. Notes
  Phys.} {\bfseries 583} (2002) 291}
  [\href{https://arxiv.org/abs/hep-ph/0104285}{{\ttfamily hep-ph/0104285}}].

\bibitem{Iancu:2003xm}
E.~Iancu and R.~Venugopalan, \emph{{The Color glass condensate and high-energy
  scattering in QCD}},  in \emph{{Quark-gluon plasma 4}}, R.C.~Hwa and
  X.-N.~Wang, eds., pp.~249--3363 (2003),
  \href{https://doi.org/10.1142/9789812795533_0005}{DOI}
  [\href{https://arxiv.org/abs/hep-ph/0303204}{{\ttfamily hep-ph/0303204}}].

\bibitem{Kovchegov:2012mbw}
Y.V.~Kovchegov and E.~Levin, \emph{{Quantum chromodynamics at high energy}},
  vol.~33, Cambridge University Press (8, 2012),
  \href{https://doi.org/10.1017/CBO9781139022187}{10.1017/CBO9781139022187}.

\bibitem{Kovner:2010xk}
A.~Kovner and M.~Lublinsky, \emph{{Angular Correlations in Gluon Production at
  High Energy}}, \href{https://doi.org/10.1103/PhysRevD.83.034017}{\emph{Phys.
  Rev. D} {\bfseries 83} (2011) 034017}
  [\href{https://arxiv.org/abs/1012.3398}{{\ttfamily 1012.3398}}].

\bibitem{McLerran:1993ni}
L.D.~McLerran and R.~Venugopalan, \emph{{Computing quark and gluon distribution
  functions for very large nuclei}},
  \href{https://doi.org/10.1103/PhysRevD.49.2233}{\emph{Phys. Rev. D}
  {\bfseries 49} (1994) 2233}
  [\href{https://arxiv.org/abs/hep-ph/9309289}{{\ttfamily hep-ph/9309289}}].

\bibitem{McLerran:1993ka}
L.D.~McLerran and R.~Venugopalan, \emph{{Gluon distribution functions for very
  large nuclei at small transverse momentum}},
  \href{https://doi.org/10.1103/PhysRevD.49.3352}{\emph{Phys. Rev. D}
  {\bfseries 49} (1994) 3352}
  [\href{https://arxiv.org/abs/hep-ph/9311205}{{\ttfamily hep-ph/9311205}}].

\bibitem{Dumitru:2020fdh}
A.~Dumitru, V.~Skokov and T.~Stebel, \emph{{Subfemtometer scale color charge
  correlations in the proton}},
  \href{https://doi.org/10.1103/PhysRevD.101.054004}{\emph{Phys. Rev. D}
  {\bfseries 101} (2020) 054004}
  [\href{https://arxiv.org/abs/2001.04516}{{\ttfamily 2001.04516}}].

\bibitem{Dumitru:2020gla}
A.~Dumitru and R.~Paatelainen, \emph{{Sub-femtometer scale color charge
  fluctuations in a proton made of three quarks and a gluon}},
  \href{https://doi.org/10.1103/PhysRevD.103.034026}{\emph{Phys. Rev. D}
  {\bfseries 103} (2021) 034026}
  [\href{https://arxiv.org/abs/2010.11245}{{\ttfamily 2010.11245}}].

\bibitem{Dumitru:2021tvw}
A.~Dumitru, H.~M\"antysaari and R.~Paatelainen, \emph{{Color charge
  correlations in the proton at NLO: Beyond geometry based intuition}},
  \href{https://doi.org/10.1016/j.physletb.2021.136560}{\emph{Phys. Lett. B}
  {\bfseries 820} (2021) 136560}
  [\href{https://arxiv.org/abs/2103.11682}{{\ttfamily 2103.11682}}].

\bibitem{Kovner:2017ssr}
A.~Kovner and A.H.~Rezaeian, \emph{{Double parton scattering in the CGC: Double
  quark production and effects of quantum statistics}},
  \href{https://doi.org/10.1103/PhysRevD.96.074018}{\emph{Phys. Rev. D}
  {\bfseries 96} (2017) 074018}
  [\href{https://arxiv.org/abs/1707.06985}{{\ttfamily 1707.06985}}].

\bibitem{Krasnitz:2000gz}
A.~Krasnitz and R.~Venugopalan, \emph{{The Initial gluon multiplicity in heavy
  ion collisions}},
  \href{https://doi.org/10.1103/PhysRevLett.86.1717}{\emph{Phys. Rev. Lett.}
  {\bfseries 86} (2001) 1717}
  [\href{https://arxiv.org/abs/hep-ph/0007108}{{\ttfamily hep-ph/0007108}}].

\bibitem{Krasnitz:2001qu}
A.~Krasnitz, Y.~Nara and R.~Venugopalan, \emph{{Coherent gluon production in
  very high-energy heavy ion collisions}},
  \href{https://doi.org/10.1103/PhysRevLett.87.192302}{\emph{Phys. Rev. Lett.}
  {\bfseries 87} (2001) 192302}
  [\href{https://arxiv.org/abs/hep-ph/0108092}{{\ttfamily hep-ph/0108092}}].

\bibitem{Lappi:2003bi}
T.~Lappi, \emph{{Production of gluons in the classical field model for heavy
  ion collisions}},
  \href{https://doi.org/10.1103/PhysRevC.67.054903}{\emph{Phys. Rev. C}
  {\bfseries 67} (2003) 054903}
  [\href{https://arxiv.org/abs/hep-ph/0303076}{{\ttfamily hep-ph/0303076}}].

\bibitem{Blaizot:2010kh}
J.P.~Blaizot, T.~Lappi and Y.~Mehtar-Tani, \emph{{On the gluon spectrum in the
  glasma}}, \href{https://doi.org/10.1016/j.nuclphysa.2010.06.009}{\emph{Nucl.
  Phys. A} {\bfseries 846} (2010) 63}
  [\href{https://arxiv.org/abs/1005.0955}{{\ttfamily 1005.0955}}].

\bibitem{Jalilian-Marian:2005ccm}
J.~Jalilian-Marian and Y.V.~Kovchegov, \emph{{Saturation physics and
  deuteron-Gold collisions at RHIC}},
  \href{https://doi.org/10.1016/j.ppnp.2005.07.002}{\emph{Prog. Part. Nucl.
  Phys.} {\bfseries 56} (2006) 104}
  [\href{https://arxiv.org/abs/hep-ph/0505052}{{\ttfamily hep-ph/0505052}}].

\bibitem{Kovchegov:1998bi}
Y.V.~Kovchegov and A.H.~Mueller, \emph{{Gluon production in current nucleus and
  nucleon - nucleus collisions in a quasiclassical approximation}},
  \href{https://doi.org/10.1016/S0550-3213(98)00384-8}{\emph{Nucl. Phys. B}
  {\bfseries 529} (1998) 451}
  [\href{https://arxiv.org/abs/hep-ph/9802440}{{\ttfamily hep-ph/9802440}}].

\bibitem{Kopeliovich:1998nw}
B.Z.~Kopeliovich, A.V.~Tarasov and A.~Schafer, \emph{{Bremsstrahlung of a quark
  propagating through a nucleus}},
  \href{https://doi.org/10.1103/PhysRevC.59.1609}{\emph{Phys. Rev. C}
  {\bfseries 59} (1999) 1609}
  [\href{https://arxiv.org/abs/hep-ph/9808378}{{\ttfamily hep-ph/9808378}}].

\bibitem{Kovner:2001vi}
A.~Kovner and U.A.~Wiedemann, \emph{{Eikonal evolution and gluon radiation}},
  \href{https://doi.org/10.1103/PhysRevD.64.114002}{\emph{Phys. Rev. D}
  {\bfseries 64} (2001) 114002}
  [\href{https://arxiv.org/abs/hep-ph/0106240}{{\ttfamily hep-ph/0106240}}].

\bibitem{Dumitru:2001ux}
A.~Dumitru and L.D.~McLerran, \emph{{How protons shatter colored glass}},
  \href{https://doi.org/10.1016/S0375-9474(01)01301-X}{\emph{Nucl. Phys. A}
  {\bfseries 700} (2002) 492}
  [\href{https://arxiv.org/abs/hep-ph/0105268}{{\ttfamily hep-ph/0105268}}].

\bibitem{Blaizot:2004wu}
J.P.~Blaizot, F.~Gelis and R.~Venugopalan, \emph{{High-energy pA collisions in
  the color glass condensate approach. 1. Gluon production and the Cronin
  effect}}, \href{https://doi.org/10.1016/j.nuclphysa.2004.07.005}{\emph{Nucl.
  Phys. A} {\bfseries 743} (2004) 13}
  [\href{https://arxiv.org/abs/hep-ph/0402256}{{\ttfamily hep-ph/0402256}}].

\bibitem{Kowalski:2003hm}
H.~Kowalski and D.~Teaney, \emph{{An Impact parameter dipole saturation
  model}}, \href{https://doi.org/10.1103/PhysRevD.68.114005}{\emph{Phys. Rev.
  D} {\bfseries 68} (2003) 114005}
  [\href{https://arxiv.org/abs/hep-ph/0304189}{{\ttfamily hep-ph/0304189}}].

\bibitem{Mantysaari:2016ykx}
H.~M\"antysaari and B.~Schenke, \emph{{Evidence of strong proton shape
  fluctuations from incoherent diffraction}},
  \href{https://doi.org/10.1103/PhysRevLett.117.052301}{\emph{Phys. Rev. Lett.}
  {\bfseries 117} (2016) 052301}
  [\href{https://arxiv.org/abs/1603.04349}{{\ttfamily 1603.04349}}].

\bibitem{Fujii:2002vh}
H.~Fujii, \emph{{Penetration of a high-energy Q anti-Q bound state through
  SU(N) color background}},
  \href{https://doi.org/10.1016/S0375-9474(02)01072-2}{\emph{Nucl. Phys. A}
  {\bfseries 709} (2002) 236}
  [\href{https://arxiv.org/abs/nucl-th/0205066}{{\ttfamily nucl-th/0205066}}].

\bibitem{Blaizot:2004wv}
J.P.~Blaizot, F.~Gelis and R.~Venugopalan, \emph{{High-energy pA collisions in
  the color glass condensate approach. 2. Quark production}},
  \href{https://doi.org/10.1016/j.nuclphysa.2004.07.006}{\emph{Nucl. Phys. A}
  {\bfseries 743} (2004) 57}
  [\href{https://arxiv.org/abs/hep-ph/0402257}{{\ttfamily hep-ph/0402257}}].

\bibitem{Fukushima:2007dy}
K.~Fukushima and Y.~Hidaka, \emph{{Light projectile scattering off the color
  glass condensate}},
  \href{https://doi.org/10.1088/1126-6708/2007/06/040}{\emph{JHEP} {\bfseries
  06} (2007) 040} [\href{https://arxiv.org/abs/0704.2806}{{\ttfamily
  0704.2806}}].

\bibitem{Fukushima:2017mko}
K.~Fukushima and Y.~Hidaka, \emph{{General formulae for dipole Wilson line
  correlators with the Color Glass Condensate}},
  \href{https://doi.org/10.1007/JHEP11(2017)114}{\emph{JHEP} {\bfseries 11}
  (2017) 114} [\href{https://arxiv.org/abs/1708.03051}{{\ttfamily
  1708.03051}}].

\bibitem{Albacete:2010sy}
J.L.~Albacete, N.~Armesto, J.G.~Milhano, P.~Quiroga-Arias and C.A.~Salgado,
  \emph{{AAMQS: A non-linear QCD analysis of new HERA data at small-x including
  heavy quarks}},
  \href{https://doi.org/10.1140/epjc/s10052-011-1705-3}{\emph{Eur. Phys. J. C}
  {\bfseries 71} (2011) 1705}
  [\href{https://arxiv.org/abs/1012.4408}{{\ttfamily 1012.4408}}].

\bibitem{Albacete:2012xq}
J.L.~Albacete, A.~Dumitru, H.~Fujii and Y.~Nara, \emph{{CGC predictions for p +
  Pb collisions at the LHC}},
  \href{https://doi.org/10.1016/j.nuclphysa.2012.09.012}{\emph{Nucl. Phys. A}
  {\bfseries 897} (2013) 1} [\href{https://arxiv.org/abs/1209.2001}{{\ttfamily
  1209.2001}}].

\bibitem{Lappi:2013zma}
T.~Lappi and H.~M\"antysaari, \emph{{Single inclusive particle production at
  high energy from HERA data to proton-nucleus collisions}},
  \href{https://doi.org/10.1103/PhysRevD.88.114020}{\emph{Phys. Rev. D}
  {\bfseries 88} (2013) 114020}
  [\href{https://arxiv.org/abs/1309.6963}{{\ttfamily 1309.6963}}].

\bibitem{JalilianMarian:2004da}
J.~Jalilian-Marian and Y.V.~Kovchegov, \emph{{Inclusive two-gluon and valence
  quark-gluon production in DIS and pA}},
  \href{https://doi.org/10.1103/PhysRevD.71.079901}{\emph{Phys. Rev. D}
  {\bfseries 70} (2004) 114017}
  [\href{https://arxiv.org/abs/hep-ph/0405266}{{\ttfamily hep-ph/0405266}}].

\bibitem{Dominguez:2012ad}
F.~Dominguez, C.~Marquet, A.M.~Stasto and B.-W.~Xiao, \emph{{Universality of
  multiparticle production in QCD at high energies}},
  \href{https://doi.org/10.1103/PhysRevD.87.034007}{\emph{Phys. Rev. D}
  {\bfseries 87} (2013) 034007}
  [\href{https://arxiv.org/abs/1210.1141}{{\ttfamily 1210.1141}}].

\bibitem{Kovchegov:2012nd}
Y.V.~Kovchegov and D.E.~Wertepny, \emph{{Long-Range Rapidity Correlations in
  Heavy-Light Ion Collisions}},
  \href{https://doi.org/10.1016/j.nuclphysa.2013.03.006}{\emph{Nucl. Phys. A}
  {\bfseries 906} (2013) 50} [\href{https://arxiv.org/abs/1212.1195}{{\ttfamily
  1212.1195}}].

\bibitem{Altinoluk:2018ogz}
T.~Altinoluk, N.~Armesto, A.~Kovner and M.~Lublinsky, \emph{{Double and triple
  inclusive gluon production at mid rapidity: quantum interference in p-A
  scattering}},
  \href{https://doi.org/10.1140/epjc/s10052-018-6186-1}{\emph{Eur. Phys. J. C}
  {\bfseries 78} (2018) 702}
  [\href{https://arxiv.org/abs/1805.07739}{{\ttfamily 1805.07739}}].

\bibitem{Agostini:2021xca}
P.~Agostini, T.~Altinoluk and N.~Armesto, \emph{{Multi-particle production in
  proton-nucleus collisions in the color glass condensate}},
  \href{https://doi.org/10.1140/epjc/s10052-021-09475-0}{\emph{Eur. Phys. J. C}
  {\bfseries 81} (2021) 760}
  [\href{https://arxiv.org/abs/2103.08485}{{\ttfamily 2103.08485}}].

\bibitem{Lappi:2007ku}
T.~Lappi, \emph{{Wilson line correlator in the MV model: Relating the glasma to
  deep inelastic scattering}},
  \href{https://doi.org/10.1140/epjc/s10052-008-0588-4}{\emph{Eur. Phys. J. C}
  {\bfseries 55} (2008) 285} [\href{https://arxiv.org/abs/0711.3039}{{\ttfamily
  0711.3039}}].

\bibitem{Schlichting:2019bvy}
S.~Schlichting and V.~Skokov, \emph{{Saturation corrections to dilute-dense
  particle production and azimuthal correlations in the Color Glass
  Condensate}},
  \href{https://doi.org/10.1016/j.physletb.2020.135511}{\emph{Phys. Lett. B}
  {\bfseries 806} (2020) 135511}
  [\href{https://arxiv.org/abs/1910.12496}{{\ttfamily 1910.12496}}].

\bibitem{Schenke:2012wb}
B.~Schenke, P.~Tribedy and R.~Venugopalan, \emph{{Fluctuating Glasma initial
  conditions and flow in heavy ion collisions}},
  \href{https://doi.org/10.1103/PhysRevLett.108.252301}{\emph{Phys. Rev. Lett.}
  {\bfseries 108} (2012) 252301}
  [\href{https://arxiv.org/abs/1202.6646}{{\ttfamily 1202.6646}}].

\bibitem{Schenke:2012hg}
B.~Schenke, P.~Tribedy and R.~Venugopalan, \emph{{Event-by-event gluon
  multiplicity, energy density, and eccentricities in ultrarelativistic
  heavy-ion collisions}},
  \href{https://doi.org/10.1103/PhysRevC.86.034908}{\emph{Phys. Rev. C}
  {\bfseries 86} (2012) 034908}
  [\href{https://arxiv.org/abs/1206.6805}{{\ttfamily 1206.6805}}].

\bibitem{Altinoluk:2014mta}
T.~Altinoluk, A.~Kovner, E.~Levin and M.~Lublinsky, \emph{{Reggeon Field Theory
  for Large Pomeron Loops}},
  \href{https://doi.org/10.1007/JHEP04(2014)075}{\emph{JHEP} {\bfseries 04}
  (2014) 075} [\href{https://arxiv.org/abs/1401.7431}{{\ttfamily 1401.7431}}].

\bibitem{Lepage:1977sw}
G.P.~Lepage, \emph{{A New Algorithm for Adaptive Multidimensional
  Integration}}, \href{https://doi.org/10.1016/0021-9991(78)90004-9}{\emph{J.
  Comput. Phys.} {\bfseries 27} (1978) 192}.

\bibitem{Hahn:2004fe}
T.~Hahn, \emph{{CUBA: A Library for multidimensional numerical integration}},
  \href{https://doi.org/10.1016/j.cpc.2005.01.010}{\emph{Comput. Phys. Commun.}
  {\bfseries 168} (2005) 78}
  [\href{https://arxiv.org/abs/hep-ph/0404043}{{\ttfamily hep-ph/0404043}}].

\bibitem{Dusling:2012iga}
K.~Dusling and R.~Venugopalan, \emph{{Azimuthal collimation of long range
  rapidity correlations by strong color fields in high multiplicity
  hadron-hadron collisions}},
  \href{https://doi.org/10.1103/PhysRevLett.108.262001}{\emph{Phys. Rev. Lett.}
  {\bfseries 108} (2012) 262001}
  [\href{https://arxiv.org/abs/1201.2658}{{\ttfamily 1201.2658}}].

\bibitem{Dusling:2012cg}
K.~Dusling and R.~Venugopalan, \emph{{Evidence for BFKL and saturation dynamics
  from dihadron spectra at the LHC}},
  \href{https://doi.org/10.1103/PhysRevD.87.051502}{\emph{Phys. Rev. D}
  {\bfseries 87} (2013) 051502}
  [\href{https://arxiv.org/abs/1210.3890}{{\ttfamily 1210.3890}}].

\bibitem{Dusling:2012wy}
K.~Dusling and R.~Venugopalan, \emph{{Explanation of systematics of CMS p+Pb
  high multiplicity di-hadron data at $\sqrt{s}_{\rm NN} = 5.02$ TeV}},
  \href{https://doi.org/10.1103/PhysRevD.87.054014}{\emph{Phys. Rev. D}
  {\bfseries 87} (2013) 054014}
  [\href{https://arxiv.org/abs/1211.3701}{{\ttfamily 1211.3701}}].

\bibitem{Dusling:2013oia}
K.~Dusling and R.~Venugopalan, \emph{{Comparison of the color glass condensate
  to dihadron correlations in proton-proton and proton-nucleus collisions}},
  \href{https://doi.org/10.1103/PhysRevD.87.094034}{\emph{Phys. Rev. D}
  {\bfseries 87} (2013) 094034}
  [\href{https://arxiv.org/abs/1302.7018}{{\ttfamily 1302.7018}}].

\bibitem{McLerran:2016snu}
L.~McLerran and V.~Skokov, \emph{{Odd Azimuthal Anisotropy of the Glasma for pA
  Scattering}},
  \href{https://doi.org/10.1016/j.nuclphysa.2016.12.011}{\emph{Nucl. Phys. A}
  {\bfseries 959} (2017) 83}
  [\href{https://arxiv.org/abs/1611.09870}{{\ttfamily 1611.09870}}].

\bibitem{Kovchegov:2018jun}
Y.V.~Kovchegov and V.V.~Skokov, \emph{{How classical gluon fields generate odd
  azimuthal harmonics for the two-gluon correlation function in high-energy
  collisions}}, \href{https://doi.org/10.1103/PhysRevD.97.094021}{\emph{Phys.
  Rev. D} {\bfseries 97} (2018) 094021}
  [\href{https://arxiv.org/abs/1802.08166}{{\ttfamily 1802.08166}}].

\bibitem{Metz:2011wb}
A.~Metz and J.~Zhou, \emph{Distribution of linearly polarized gluons inside a
  large nucleus}, \href{https://doi.org/10.1103/PhysRevD.84.051503}{\emph{Phys.
  Rev.} {\bfseries D84} (2011) 051503}
  [\href{https://arxiv.org/abs/1105.1991}{{\ttfamily 1105.1991}}].

\bibitem{Dominguez:2011br}
F.~Dominguez, J.-W.~Qiu, B.-W.~Xiao and F.~Yuan, \emph{On the linearly
  polarized gluon distributions in the color dipole model},
  \href{https://doi.org/10.1103/PhysRevD.85.045003}{\emph{Phys. Rev.}
  {\bfseries D85} (2012) 045003}
  [\href{https://arxiv.org/abs/1109.6293}{{\ttfamily 1109.6293}}].

\end{thebibliography}\endgroup
\end{document}